\documentclass[12pt,a4paper]{article}
\usepackage[cp850]{inputenc}
\usepackage{amssymb}
\usepackage{latexsym}
\usepackage[dvips]{epsfig}
\usepackage{pstricks,pst-node}
\renewcommand{\baselinestretch}{1.5}
\begin{document}
\begin{titlepage} 
\renewcommand{\baselinestretch}{1}
\small\normalsize
\begin{flushright}
hep-th/0205062\\
MZ-TH/02-07     \\
\end{flushright}

\vspace{0.1cm}

\begin{center}   

{\LARGE \sc Flow Equation of\\[3mm] Quantum Einstein Gravity in a\\[5mm]
Higher-Derivative Truncation}

\vspace{1.4cm}
{\large O. Lauscher and M. Reuter}\\

\vspace{0.7cm}
\noindent
{\it Institute of Physics, University of Mainz\\
Staudingerweg 7, D-55099 Mainz, Germany}\\

\end{center}   
 
\vspace*{0.6cm}
\begin{abstract}
Motivated by recent evidence indicating that Quantum Einstein Gravity (QEG)
might be nonperturbatively renormalizable, the exact renormalization group
equation of QEG is evaluated in a truncation of theory space which generalizes
the Einstein-Hilbert truncation by the inclusion of a higher-derivative term
$(R^2)$. The beta-functions describing the renormalization group flow of the
cosmological constant, Newton's constant, and the $R^2$-coupling are computed
explicitly. The fixed point properties of the 3-dimensional flow are
investigated, and they are confronted with those of the 2-dimensional
Einstein-Hilbert flow. The non-Gaussian fixed point predicted by the latter is
found to generalize to a fixed point on the enlarged theory space. In order 
to test
the reliability of the $R^2$-truncation near this fixed point we analyze the
residual scheme dependence of various universal quantities; it turns out to
be very weak. The two truncations are compared in detail, and their numerical
predictions are found to agree with a suprisingly high precision. Due to the
consistency of the results it appears increasingly unlikely that the
non-Gaussian fixed point is an artifact of the truncation.
If it is present in the exact theory QEG is probably nonperturbatively
renormalizable and ``asymptotically safe''. We discuss how the
conformal factor problem of Euclidean gravity manifests itself in the exact
renormalization group approach and show that, in the $R^2$-truncation, the
investigation of the fixed point is not afflicted with this problem. Also the
Gaussian fixed point of the Einstein-Hilbert truncation is analyzed; it turns
out that it does not generalize to a corresponding fixed point on the
enlarged theory space.
\end{abstract}
\end{titlepage}
%
\section{Introduction}
\renewcommand{\theequation}{1.\arabic{equation}}
\setcounter{equation}{0}
\label{3intro}
Recently a lot of work on Quantum Einstein Gravity (QEG) went into
constructing an appropriate exact renormalization group (RG) equation
\cite{Reu96,LR1}, finding approximate solutions to it
\cite{LR2,LR4,frank,frank2,DP97}, and exploring their implications for black 
hole physics \cite{bh1,bh2} and cosmology \cite{cosmo}. In particular, strong
indications were found that QEG might be nonperturbatively renormalizable.
If so, it could have the status of a fundamental, microscopic quantum theory of
gravity.

The basic tool used in these investigations is the effective average action
and its exact RG equation \cite{ber}. It is a continuum analog of Wilson's
lattice renormalization group of iterated block spin transformations
\cite{wilson}. Both in quantum field theory and statistical mechanics the
idea is to integrate out all fluctuation modes which have momenta larger than
a certain infrared (IR) cutoff $k$ (``fast degrees of freedom''), and to take
account of those modes in an implicit way by the modified dynamics which they
induce for the remaining fluctuations with momenta smaller than $k$ (the
``slow degrees of freedom''). In field theory this ``renormalized'' dynamics
is encoded in a scale dependent effective action, $\Gamma_k$, whose dependence
on the cutoff scale $k$ is governed by a functional differential equation
referred to as the ``exact RG equation'' \cite{bag}. This equation gives rise
to a flow on the space of all actions (``theory space''). The functional
$\Gamma_k$
defines an effective field theory valid near the scale $k$; evaluated at tree
level, it describes all loop effects due to the high-momentum modes. The
effective average action can be thought of as a kind of microscope with a
variable resolution. At large $k$, the physics at short distances $\ell=1/k$
can be read off directly from $\Gamma_k$; at small $k$ we see a coarse-grained
picture suitable for a simple description of structures with a large
characteristic length scale $\ell=1/k$ \cite{ber}.

The effective average action $\Gamma_k$, regarded as a function of $k$,
interpolates between the ordinary effective action
$\Gamma=\lim_{k\rightarrow 0}\Gamma_k$ and the bare (classical) action $S$
which is approached for $k\rightarrow\infty$. The construction of $\Gamma_k$
begins by adding a IR cutoff term $\Delta_k S$ to the classical action
entering the standard Euclidean functional integral for the generating
functional $W$ of the connected Green's functions. The new piece $\Delta_k S$
introduces a momentum dependent $(\mbox{mass})^2$-term ${\cal R}_k(p^2)$ for
each mode of the quantum field with momentum $p$. For $p^2\gg k^2$, the
cutoff function ${\cal R}_k(p^2)$ is assumed to vanish so that the
high-momentum modes get integrated out unsuppressed. For $p^2\ll k^2$, it
behaves as ${\cal R}_k(p^2)\propto k^2$; hence the small-momentum modes are
suppressed in the path integral by a mass term $\propto k^2$. Apart from a
correction term which is known explicitly \cite{ber}, the effective average
action $\Gamma_k$ is given by the Legendre transform of the modified
generating functional $W_k$.

From this definition one can derive the exact RG equation obeyed by
$\Gamma_k$. In a slightly symbolic notation it is of the form
\begin{eqnarray}
\label{3in1}
k\,\partial_k\Gamma_k=\frac{1}{2}\;{\rm Tr}\left[\left(\Gamma_k^{(2)}
+{\cal R}_k(-\Delta)\right)^{-1}\,k\,\partial_k{\cal R}_k(-\Delta)\right]\;.
\end{eqnarray}
The RHS of this equation is a kind of ``beta-functional'' which summarizes
the beta-functions for infinitely many running couplings. Geometrically, it
defines a vector field on theory space, the corresponding
flow lines being the RG trajectories $k\mapsto\Gamma_k$.

The functional $\Gamma_k$ enters this vector field via its Hessian 
$\Gamma_k^{(2)}$, i.e.
the infinite-dimensional matrix of all second functional derivatives of
$\Gamma_k$ with respect to the dynamical, i.e., non-background fields.

In eq. (\ref{3in1}) the c-number argument of ${\cal R}_k$ is replaced with the
operator $-\Delta$. The discrimination of high-``momentum'' vs.
low-``momentum'' modes is performed according to the spectrum of this operator,
i.e. $p^2$ is an eigenvalue of $-\Delta$. In simple theories where no gauge
or diffeomorphism invariance needs to be respected, $\Delta$ is the free
Laplacian, $\Delta=\partial_\mu\partial^\mu$, whose eigenmodes are momentum
eigenstates in the usual sense of the word. In Yang-Mills theory \cite{ym,ym2}
it has proven convenient to use the background field formalism \cite{Ab81,back}
and to set $\Delta=\bar{D}_\mu\bar{D}^\mu$ where $\bar{D}_\mu$ is the covariant
derivative in the background field. The background field technique plays
a dual role in this context. Using a background gauge fixing term makes
$\Gamma_k$ a gauge invariant functional of its argument, and using
$\bar{D}_\mu$ in the cutoff leads to a flow equation of the relatively simple
type (\ref{3in1}), similar to non-gauge theories. (The RG equation resulting
from $\Delta=D_\mu D^\mu$ with $D_\mu$ constructed from the {\it dynamical}
gauge field is quite unwieldy.)

Along similar lines an effective average action for $d$-dimensional Euclidean
QEG has been constructed in ref. \cite{Reu96}, and the corresponding flow
equation has been derived. Leaving the Faddeev-Popov ghosts aside for a moment,
the gravitational effective average action,
$\Gamma_k[g_{\mu\nu},\bar{g}_{\mu\nu}]$, is a functional of two different
metrics, the ``ordinary'' dynamical metric $g_{\mu\nu}$ and the background
metric $\bar{g}_{\mu\nu}$. The usual effective action $\Gamma[g_{\mu\nu}]$
is recovered by taking the limit $k\rightarrow 0$ of the functional
$\Gamma_k[g_{\mu\nu}]\equiv\Gamma_k[g_{\mu\nu},\bar{g}_{\mu\nu}=g_{\mu\nu}]$ in
which the two metrics are taken equal. Thanks to the background gauge fixing
condition this construction leads to a functional $\Gamma_k[g_{\mu\nu}]$
which is invariant under general coordinate transformations.

One of the many advantages which the exact RG approach has in comparison to the
standard canonical or path integral quantization is that it offers a very
natural and intuitive nonperturbative approximation scheme. By truncating the
theory space one can obtain approximate solutions to the RG equation which do
not need a small expansion parameter. The idea is to project the RG flow from
the ``huge'' infinite-dimensional space of all actions onto some smaller,
typically finite dimensional subspace which is easier to handle. In this way
the functional RG equation for $\Gamma_k$ becomes a system of ordinary
differential equations for a (finite) set of coupling constants which have the
geometrical interpretation of coordinates on the subspace. It is clear that
in applying this strategy the key problem is finding the ``relevant'' subspace
which contains the essential physics.

In a first attempt at solving the gravitational RG equation \cite{Reu96}
the flow has been projected onto the 2-dimensional subspace of theory space
which is spanned by the invariants $\int d^dx\,\sqrt{g}$ and
$\int d^dx\,\sqrt{g}R$. This is the so-called Einstein-Hilbert truncation
defined by the ansatz
\begin{eqnarray}
\label{3in2}
\Gamma_k[g,\bar{g}]=\left(16\pi G_k\right)^{-1}\int d^dx\,\sqrt{g}\left\{
-R(g)+2\bar{\lambda}_k\right\}+\mbox{classical gauge fixing}.
\end{eqnarray}
The two running couplings involved are the running Newton constant $G_k$
and the running cosmological constant $\bar{\lambda}_k$. The similarity of
(\ref{3in2}) to the action of classical General Relativity is accidental in a
sense; improved truncations would include both higher powers of the curvature
and nonlocal terms \cite{nonloc,frank2}. In (\ref{3in2}) only the gauge
fixing term depends on $\bar{g}_{\mu\nu}$; it vanishes when we set
$\bar{g}_{\mu\nu}=g_{\mu\nu}$. 

The scale dependence of $G_k$ and
$\bar{\lambda}_k$ is most conveniently visualized as a flow in the
$\lambda$-$g-$plane where $g_k\equiv k^{d-2}G_k$ and $\lambda_k\equiv 
\bar{\lambda}_k/k^2$ are the dimensionless Newton constant and cosmological
constant, respectively.
Using the original cutoff of ``type A'' \cite{Reu96} the system of equations
for $g_k$ and $\lambda_k$ was derived in \cite{Reu96} and solved numerically
in \cite{frank}. In \cite{LR1} a new cutoff of ``type B'' was introduced and
the corresponding flow equations in the Einstein-Hilbert truncation were
derived. (The ``type B'' cutoff is convenient if one uses the TT-decomposition
of the metric \cite{York}.) The fixed point properties of these equations
were first discussed in \cite{souma1} and \cite{bh2}, and analyzed in detail in
\cite{LR1} and \cite{frank}. One finds that the RG flow in the
$\lambda$-$g-$plane is governed by two fixed points $(\lambda_*,g_*)$: a
trivial or ``Gaussian'' fixed point at $(\lambda_*,g_*)=(0,0)$, and a
non-Gaussian fixed point with $\lambda_*\neq 0$ and $g_*\neq 0$.

In order to appreciate the importance of the non-Gaussian fixed point we
recall what it means to ``quantize'' a theory in the average action approach.
One picks a bare action $S$ and imposes the initial condition
$\Gamma_{\widehat{k}}=S$ at the ultraviolet (UV) cutoff scale $\widehat{k}$,
uses the RG equation to find $\Gamma_k$ at all lower scales $k\le\widehat{k}$,
and finally sends $k\rightarrow 0$ and $\widehat{k}\rightarrow\infty$.
A {\it fundamental} theory has the property that the ``continuum'' limit
$\widehat{k}\rightarrow\infty$ actually exists after redefining only
finitely many parameters in the action. This is the case in perturbatively
renormalizable theories \cite{polch}, but there are also examples of
perturbatively nonrenormalizable theories which possess a limit
$\widehat{k}\rightarrow\infty$ \cite{parisi}. The continuum limit of those
``nonperturbatively renormalizable'' theories is taken at a non-Gaussian
fixed point, i.e. the theory is defined by the set of RG trajectories which
leave the fixed point when we lower $k$. These trajectories span the
UV critical hypersurface of the fixed point, ${\cal S}_{\rm UV}$. If it is
finite dimensional, the quantum theory thus constructed has only finitely
many free parameters and therefore keeps its predictive power even at
arbitrarily large momentum scales. This behavior is to be contrasted with
an {\it effective} field theory which, at high energies, typically contains an
increasing number of free parameters which must be taken from the experiment.

In his ``asymptotic safety'' scenario Weinberg \cite{wein,wein2} conjectured
that a fundamental quantum field theory of gravity could perhaps be constructed
nonperturbatively by taking the continuum limit at a non-Gaussian fixed point
\cite{smolin}. While originally this idea could be implemented in 
$d=2+\varepsilon$ dimensions only,
the recent results coming from the effective average action strongly support
the hypothesis that this fixed point exists also in 4 dimensions. Within
the Einstein-Hilbert truncation, the existence of a suitable non-Gaussian
fixed point is definitely established by now; the crucial question is whether
it is the projection of a fixed point present in the exact
theory or merely an artifact of the approximation.

Let us assume for a moment that the fixed point indeed exists in the
exact 4-di\-men\-sio\-nal theory and that we define QEG by taking the
$\widehat{k}\rightarrow\infty$ limit there. Then, since
$(\lambda_k,g_k,\cdots)$ approaches $(\lambda_*,g_*,\cdots)$ for 
$k\rightarrow\infty$, the dimensionful couplings behave as
\begin{eqnarray}
\label{3in3}
G_k\approx g_*/k^2\;,\;\;\;\bar{\lambda}_k\approx\lambda_*\,k^2 \;,\;\;\;\cdots
\end{eqnarray}
for large $k$. Obviously $G_k$ vanishes for $k\rightarrow\infty$. At least as
far as this coupling is concerned QEG is asymptotically free similar to 
Yang-Mills theory. The
$1/k^2$-dependence of the running Newton constant will lead to a characteristic
momentum dependence of the cross sections for graviton-graviton scattering and 
graviton
mediated matter-matter scattering. Because of this characteristic momentum
dependence, QEG could be distinguished experimentally from alternative theories
of quantum gravity such as string theory, at least in principle.

Let us generalize the standard definition of the Planck mass,
$m_{\rm Pl}\equiv G^{-1/2}$, and introduce the {\it running Planck mass}
\begin{eqnarray}
\label{3in4}
M_{\rm Pl}(k)\equiv 1/\sqrt{G_k}\;.
\end{eqnarray}
At the laboratory scale $M_{\rm Pl}(k)$ reduces to $m_{\rm Pl}$, most probably,
and its dependence on $k$ is negligible. However, in the fixed point regime
$k\rightarrow\infty$, the asymptotic freedom of $G_k$ implies that
$M_{\rm Pl}(k)$ is proportional to the scale $k$ itself: $M_{\rm Pl}(k)
=k/\sqrt{g_*}$. This shows that the running Planck mass is a rather elusive
``barrier'' which never can be jumped across in any experiment. If we analyze
a system with a probe of increasing momentum $k$ we will always push the
running Planck mass ahead of us and never reach it.

Also the standard constant Planck mass, defined more precisely in terms of the
IR value of $G_k$,
\begin{eqnarray}
\label{3in5}
m_{\rm Pl}\equiv\left[G_{k=0}\right]^{-1/2}\;,
\end{eqnarray}
plays an important role in QEG, similar to that of $\Lambda_{\rm QCD}$ in QCD.
According to the numerical solutions of the $\lambda$-$g$-system \cite{frank},
$m_{\rm Pl}$ marks the lower boundary of the asymptotic scaling region. Near
$k=m_{\rm Pl}$ there is a crossover from the scaling laws (\ref{3in3}) of the
non-Gaussian fixed point to those of the Gaussian fixed 
point.\renewcommand{\baselinestretch}{1}\small\normalsize
\footnote{A similar crossover was already known to occur in Liouville
quantum gravity \cite{liouv}.} 
\renewcommand{\baselinestretch}{1.5}\small\normalsize

According to the UV scaling laws (\ref{3in3}) the dimensionful cosmological
constant diverges for $k\rightarrow\infty$ proportional to $k^2$. This has an
interesting geometrical interpretation. Let us consider the $k$-dependent,
effective field equations implied by the truncation ansatz (\ref{3in2}) with
$\bar{g}_{\mu\nu}=g_{\mu\nu}$ for $d=4$. They happen to coincide with the 
familiar vacuum Einstein equations with the cosmological constant replaced by
the scale-dependent quantity $\bar{\lambda}_k$:
\begin{eqnarray}
\label{3in6}
R_{\mu\nu}-\frac{1}{2}g_{\mu\nu}\,R=-\bar{\lambda}_k\,g_{\mu\nu}\;.
\end{eqnarray}
Since $\bar{\lambda}_k$ is the only quantity which sets a scale, every solution
to (\ref{3in6}) has a typical radius of curvature $r_c(k)\propto 
1/\sqrt{\bar{\lambda}_k}$. (For instance, the maximally symmetric 
$S^4$-solution has the radius $r_c=r=\sqrt{3/\bar{\lambda}_k}$.) The
$k$-dependence of the solutions and in particular of $r_c$ should be 
interpreted as follows. If we want to explore the spacetime structure at a 
fixed length scale $\ell\equiv 1/k$ it is most convenient to use the action
$\Gamma_k[g_{\mu\nu}]$ at $k=1/\ell$ because for this, and only this, 
functional a {\it tree level} analysis is sufficient to describe the essential 
physics at this scale, including all quantum effects. Hence, when we observe
spacetime with a ``microscope'' of resolution $\ell$, we will see an average
radius of curvature given by $r_c(\ell)\equiv r_c(k=1/\ell)$. Once $\ell$ is
smaller than the (standard) Planck length $\ell_{\rm Pl}\equiv m_{\rm Pl}^{-1}$
we are in the fixed point regime (\ref{3in3}) so that $r_c(k)\propto 1/k$, or
\begin{eqnarray}
\label{3in7}
r_c(\ell)\propto\ell
\end{eqnarray}
Thus, when we look at the structure of spacetime with a microscope of 
resolution $\ell$, the average radius 
of curvature which we measure is proportional to the resolution 
itself. If we want to probe finer details and decrease $\ell$ we automatically
decrease $r_c$ and hence {\it in}crease the average curvature. Spacetime seems
to be more strongly curved at small distances than at larger ones. The 
scale-free relation (\ref{3in7}) suggests that at distances below the Planck
length quantum spacetime is a kind of fractal with a self-similar structure.
It has no intrinsic scale.

Before we continue a remark might be in order on which theory
precisely we refer to as ``Quantum Einstein Gravity'' or ``QEG''. While flow
equations can also be used in the effective field theory approach to quantum
gravity \cite{don,Reu96}, in the present context ``QEG'' stands for the
fundamental theory whose continuum limit $\widehat{k}\rightarrow\infty$
is taken at the non-Gaussian fixed point. This theory has 
$\dim({\cal S}_{\rm UV})$ free parameters, and fixing these parameters amounts
to picking a specific trajectory $k\mapsto\Gamma_k$ in the full theory space.
For $k\rightarrow\infty$ this trajectory hits the fixed point action
$\Gamma_*$, regarded as the collection of its infinitely many dimensionless
coordinates on theory space. The fixed point action $\Gamma_*$ 
corresponds to the ``bare'' or ``classical'' action in conventional field 
theory. However, unlike the latter $\Gamma_*$ is not put in by hand but is
rather {\it derived} with the help of the RG equation. The usual canonical or
path integral quantization is always based upon a ``prejudice'' about what the
classical action is. In the asymptotic safety scenario the ``classical''
action is fixed instead by the condition of nonperturbative renormalizability;
it cannot be guessed by simple power-counting, symmetry, or invariance 
arguments, but the effective average action provides a computational framework
to determine it. 

The really crucial property which defines QEG is the requirement of
{\it diffeomorphism invariance}. Before we can write down a flow equation we
must declare what the theory space is on which the renormalization group is
supposed to operate. In the case of QEG it is defined to be the space of
functionals $\Gamma[g_{\mu\nu}]$ depending on a nondegenerate, symmetric 
rank-2 tensor field in a diffeomorphism invariant way. Leaving technical 
details aside, the theory space then fixes the flow equation which in turn 
determines the RG trajectories and the fixed points.

So far our discussion referred to the exact theory on the full theory space.
If we project on the subspace spanned by the Einstein-Hilbert truncation
it is clear that also the fixed point action $\Gamma_*$ must be of the 
Einstein-Hilbert type. However, we emphasize that this is a trivial consequence
of the simple truncation we have chosen, and we have no reason to believe that
the exact $\Gamma_*$ is of the Einstein-Hilbert type, too. In fact, within the
more complicated truncation of the present paper $\Gamma_*$ receives 
corrections which go beyond the Einstein-Hilbert form. Hence ``Quantum Einstein
Gravity'' does {\it not} mean that the Einstein-Hilbert action is the bare
action to be quantized. In this respect our approach is different from 
canonical quantum gravity, along the lines of Ashtekar's program 
\cite{ashtekar}, for instance. (It is intriguing that also in this context
remarkable finiteness properties have been proven recently \cite{finite}.) 

Clearly it is a highly attractive idea that there could be a nonperturbatively
renormalizable field theory of the metric field so that there is no longer
any conceptual need for leaving the framework of quantum field theory in order
to arrive at a consistent microscopic theory of quantum gravity. Therefore
every effort should be made to show that the non-Gaussian fixed point 
found in the Einstein-Hilbert truncation is not just an artifact of this
approximation.

In ref. \cite{LR1} we therefore started an extensive analysis of the 
reliability of the Einstein-Hilbert truncation near the fixed point. There the
strategy was to use the scheme dependence of universal quantities in order to
get a first idea about the precision which can be achieved with this 
truncation. Here ``scheme dependence'' refers to the dependence on the details
of the cutoff procedure, i.e. on the shape of the function ${\cal R}_k(p^2)$.
By definition, universal quantities are exactly scheme independent in the
exact theory, but they might acquire some scheme dependence once we make
approximations. The level of this residual scheme dependence can serve as a
measure for the quality of the approximation. Typical universal quantities are
the critical exponents of fixed points and, as we argued, the product 
$g_*\lambda_*$. The upshot of our analysis was that the Einstein-Hilbert
truncation seems to provide a description that is much more reliable and
precise than originally hoped for, and that it would be very hard to 
understand the approximate scheme independence we found if the fixed point
was just due to a misleading approximation.

These results are certainly very encouraging, but it is clear that the 
ultimate justification of a truncation ansatz consists of adding further
terms to it and verifying that its predictions do not change much. In the
present paper we take a first step in this direction and add one further
invariant constructed from $g_{\mu\nu}$ to the ansatz. 

Which invariant should
we take? In standard renormalized perturbation theory where 
(at least implicitly) the $\widehat{k}\rightarrow\infty$ limit is taken at the
Gaussian fixed point, the relative importance or ``relevance'' of the various
field monomials is measured by their scaling dimensions at the Gaussian fixed
point, i.e. by their canonical dimensions simply. Since at the non-Gaussian
fixed point the anomalous dimensions are large we have no similarly simple
guide line at our disposal, and a priori all invariants are equally plausible.
To get a first idea about what happens away from the Einstein-Hilbert subspace
we shall include the higher-derivative invariant $\int d^dx\,\sqrt{g}R^2$ 
and study the RG flow in the 
``$R^2$-truncation''\renewcommand{\baselinestretch}{1}\small\normalsize
\footnote{Our conventions are $R^\sigma_{\;\,\rho\mu\nu}=-\partial_\nu
\Gamma^\sigma_{\mu\rho}+\cdots$, $R_{\mu\nu}=R^\sigma_{\;\,\mu\sigma\nu}$, 
$R=g^{\mu\nu}R_{\mu\nu}$.} 
\renewcommand{\baselinestretch}{1.5}\small\normalsize
\begin{eqnarray}
\label{3in8}
\Gamma_k[g,\bar{g}]=\int d^dx\,\sqrt{g}\left\{\left(16\pi G_k\right)^{-1}
\left[-R(g)+2\bar{\lambda}_k\right]+\bar{\beta}_k\,R^2(g)\right\}
+\mbox{classical gauge fixing}
\end{eqnarray}
Its truncation subspace is 3-dimensional, with coordinates $G$, 
$\bar{\lambda}$ and the new coupling $\bar{\beta}$.

It is well known that beyond $\int d^dx\,\sqrt{g}R^2$ there exist 
two\renewcommand{\baselinestretch}{1}\small\normalsize\footnote{Except in $d=4$
where one invariant can be eliminated by virtue of the Gauss-Bonnet identity.} 
\renewcommand{\baselinestretch}{1.5}\small\normalsize
more
$({\rm curvature})^2$-invariants: $\int d^dx\,\sqrt{g}R_{\mu\nu}R^{\mu\nu}$ and
$\int d^dx\,\sqrt{g}R_{\mu\nu\rho\sigma}R^{\mu\nu\rho\sigma}$. In a standard
perturbative calculation near the Gaussian fixed point consistency would 
require to include them along with the $R^2$-term because they all have the
same canonical dimension. As for the non-Gaussian fixed point, we have no a
priori information from general principles about the relative importance of the
three terms. Since anyhow the best we can do is to take a ``step into the 
dark'', without knowing whether we walk in the most ``relevant'' direction, we
shall omit the other two invariants here. Including them would go far beyond
the present calculational possibilities, in particular since it would require a
much more complicated projection technique \cite{LR1}.

In this paper we shall derive the (extremely complicated) 3-dimensional RG
equations of the $R^2$-truncation, and we shall use them in order to 
investigate the fixed points of the flow. Our main results will be the 
following: a) The Gaussian fixed point of the Einstein-Hilbert truncation
does {\it not} generalize to a fixed point of the $R^2$-truncation. b) The 
non-Gaussian fixed point does indeed generalize to a fixed point of the
$R^2$-truncation, and the $\lambda$-$g-$projection of this fixed point is
described almost perfectly by the Einstein-Hilbert truncation. Within the 
(weak) residual scheme dependence, the fixed point properties are almost
insensitive to the inclusion of the $R^2$-invariant. This is further strong
evidence against the theoretical possibility that the non-Gaussian fixed point
is a truncation artifact. 

In the second part of the paper we shall address a
very important general problem which is of a more technical nature. It is 
related to a notorious disease of standard Euclidean quantum gravity: the
conformal factor problem. In setting up the truncated RG equation the cutoff
function ${\cal R}_k$ (actually a matrix in field space) is adapted to the 
truncation in such a way that, for $p^2\ll k^2$, the inverse propagator of
every massless mode, $p^2$, is replaced by $p^2+k^2$. A problem arises if there
are modes, such as those of the conformal factor in the Einstein-Hilbert 
truncation, which have a negative kinetic energy, i.e. their inverse propagator
is $-p^2$. In \cite{Reu96} it has been argued that for these modes also the
sign of ${\cal R}_k$ should be reversed so as to obtain the regularized inverse
propagator $-(p^2+k^2)$. While there is little doubt that this procedure is
correct for the Einstein-Hilbert truncation, it leads to the seemingly
paradoxical situation that in the Euclidean path integral the modes of the
conformal factor are enhanced rather than suppressed in the IR \cite{Reu96}.

Contrary to the Einstein-Hilbert truncation, the $R^2$-truncation yields a
functional $\Gamma_k[g_{\mu\nu}]$ which is bounded below and, as we shall see,
gives positive kinetic energy to {\it all} modes, provided one stays close to
the UV fixed point. Hence our investigation of the non-Gaussian fixed point
is not plagued by the conformal factor problem, and the construction of the
cutoff becomes straightforward. This advantage is an independent motivation
for studying the $R^2$-truncation.

In the usual perturbative approach higher-derivative theories of (Lorentzian)
gravity are notoriously problematic as far as causality and unitarity are
concerned \cite{stelle,perturb}. While in $d=4$ the most general 
$({\rm curvature})^2$-theory, when expanded about flat space, is power-counting
renormalizable it suffers from excitations with, classically, negative
linearized energy and, quantum mechanically, a wrong-sign residue of the 
propagator leading to a state space containing negative-norm states 
\cite{stelle}. These ``ghosts'' have masses of the order of the Planck mass.
Correspondingly, if a truncation ansatz is of the $({\rm curvature})^2$-type
the $k$-dependent effective propagator 
$(\delta^2\Gamma_k/\delta g\delta g)^{-1}\equiv(\Gamma^{(2)}_k)^{-1}$, 
evaluated for flat space, has similar ghosts with masses $\propto$ 
$M_{\rm Pl}(k)$.
By itself this does not indicate any real problem because generically flat 
space anyhow is not a solution of the effective equation of motion. Compared
to the perturbative quantization the potential problem of ghost excitations
manifests itself in the Euclidean RG approach in a conceptually different,
more tractable manner. Here the linearization is performed about the
backgrounds needed for the projection procedure, not about flat space.
For a well-defined computation of the RG trajectories on a certain $k$-interval
it is sufficient that the (truncated) $\Gamma_k$ gives positive linearized
action to all modes contributing to the RG running in this interval. In our
calculation this will indeed be the case for $k$ large enough where the
truncation is believed to be reliable \cite{frank}. The much more subtle
issues related to a Lorentzian interpretation of the theory and its causality 
properties can be addressed only
once a complete trajectory, valid down to the IR, and in particular the precise
form of the fixed point action is known. From all what we can tell now the
exact QEG could very well be ``causal'' in an appropriate sense. 

A brief summary of some of the results derived in the present paper appeared in
\cite{LR2}, and an informal introduction to the older work can be found in
\cite{tif}. In the present paper we focus on pure gravity. The gravitational
average action with matter fields included was discussed in \cite{DP97} and
\cite{grano}. The gauge fixing dependence of the original formulation 
\cite{Reu96} was investigated in \cite{sven} and \cite{souma2}. An incomplete
higher-derivative calculation was begun in \cite{odhigher} where the running
of the $R^2$-couplings was neglected, however, and no conclusions about the
fixed point could be drawn.

The remaining sections of this paper are organized as follows. In Section 
\ref{3S2} we review some general properties of the exact RG equation which will
be needed later on. Section \ref{3S3} is devoted to the construction of cutoffs
which are adapted to a specific truncation; in particular the complications
due to the conformal factor problem will be discussed there. In Section 
\ref{$R^2$-trunc} we derive the system of RG equations which results from the
$R^2$-truncation, and in Section \ref{3S5} we analyze its fixed point 
structure. We discuss the fate of the Gaussian fixed point which is present
only in the Einstein-Hilbert truncation, and we reanalyze the non-Gaussian 
fixed point in the more general setting. In Section \ref{3S6} the positivity
properties of the truncated action functional, its Hessian, and the cutoff
operator are investigated; in particular we show that our analysis of the 
non-Gaussian fixed point is not affected by the conformal factor problem.
The conclusions are contained in Section \ref{conclusio}. Many important 
technical results, including the coefficients occuring in the rather 
complicated beta-functions of the $\lambda$-$g$-$\beta-$system, are tabulated
in various appendices.

\section{The exact RG equation}
\renewcommand{\theequation}{2.\arabic{equation}}
\setcounter{equation}{0}
\label{3S2}
In this section we briefly review the construction of the ``type B'' RG
equation for quantum gravity performed in ref. \cite{LR1} to which 
we refer for the details. We start from a scale-dependent modification of the
generating functional for the connected Green's functions, $W_k$. It is defined
by the following Euclidean functional integral:
\begin{eqnarray}
\label{a}
\lefteqn{\exp\left\{W_k[{\rm sources}]\right\}=\int{\cal D}h_{\mu\nu}\,
{\cal D}C^\mu\,{\cal D}\bar{C}_\mu\,\exp\left[-S[\bar{g}+h]\right.}
\nonumber\\
& &\left.-S_{\rm gf}[h;\bar{g}]-S_{\rm gh}
[h,C,\bar{C};\bar{g}]-\Delta_k S[h,C,\bar{C};\bar{g}]-S_{\rm source}\right]\;.
\end{eqnarray}
In (\ref{a}) we use the background gauge fixing
technique which necessitates the decomposition of the full quantum
metric $\gamma_{\mu\nu}$ into a fixed background metric
$\bar{g}_{\mu\nu}$ and a fluctuation variable $h_{\mu\nu}$:
$\gamma_{\mu\nu}(x)=\bar{g}_{\mu\nu}(x)+h_{\mu\nu}(x)$. It allows us to replace
the integration over $\gamma_{\mu\nu}$ by an integration over $h_{\mu\nu}$.
Furthermore, $\bar{C}_\mu$ and $C^\mu$ are the Faddeev-Popov ghosts of the
gravitational field.

The first term of the action, $S[\bar{g}+h]$, is the classical part, which is
assumed to be invariant under general coordinate transformations. For the
time being, we also assume that it is positive definite, $S>0$. The gauge
fixing term is given by
\begin{eqnarray}
\label{02}
S_{\rm gf}[h;\bar{g}]=\frac{1}{2\alpha}\int d^dx\,\sqrt{\bar{g}}\,
\bar{g}^{\mu\nu}\,F_\mu[\bar{g},h]\,F_\nu[\bar{g},h]\;.
\end{eqnarray}
In the present paper we use the linear gauge condition
$F_\mu[\bar{g},h]=\sqrt{2}\kappa\,{\cal F}^{\alpha\beta}_\mu[\bar{g}]\,
h_{\alpha\beta}$ with
${\cal F}^{\alpha\beta}_\mu[\bar{g}]=\delta^\beta_\mu\bar{g}^{\alpha\gamma}
\bar{D}_\gamma-\frac{1}{2}\bar{g}^{\alpha\beta}\bar{D}_\mu$, which amounts to a
background version of the harmonic coordinate condition. Here we introduced 
the constant
$\kappa\equiv(32\pi\bar{G})^{-\frac{1}{2}}$ where $\bar{G}$ is the bare
Newton constant. Moreover, $\bar{D}_\mu$ denotes the covariant derivative
constructed from the background metric $\bar{g}_{\mu\nu}$, while we shall
write $D_\mu$ for the covariant derivative involving the complete metric
$\gamma_{\mu\nu}$. $S_{\rm gh}$ is the Faddeev-Popov ghost action resulting 
from the above gauge fixing. 

Furthermore, $\Delta_k S$ and $S_{\rm source}$ are the cutoff
and the source action, respectively. $\Delta_k S$ provides an appropriate
infrared cutoff for the integration variables and will be discussed in detail
in a moment; $S_{\rm source}$ introduces sources
for the fields $h_{\mu\nu}$, $C^\mu$ and $\bar{C}_\mu$. 

Next we decompose the gravitational field $h_{\mu\nu}$ according to (see e.g.
\cite{York})
\begin{eqnarray}
\label{TT}
h_{\mu\nu}=h^{T}_{\mu\nu}
+\bar{D}_\mu\widehat{\xi}_\nu+\bar{D}_\nu\widehat{\xi}_\mu
+\bar{D}_\mu \bar{D}_\nu\widehat{\sigma}
-\frac{1}{d}\bar{g}_{\mu\nu}\bar{D}^2\widehat{\sigma}
+\frac{1}{d}\bar{g}_{\mu\nu}\phi\;.
\end{eqnarray}
In order to obtain this ``TT-decomposition''
one starts by writing $h_{\mu\nu}$ as a sum of its orthogonal parts:
$h_{\mu\nu}=h_{\mu\nu}^{T}+h_{\mu\nu}^{L}+h^{Tr}_{\mu\nu}$. Here
$h_{\mu\nu}^{T}$, $h_{\mu\nu}^{L}$ and $h^{Tr}_{\mu\nu}$ represent the
transverse traceless, longitudinal traceless and pure trace part, respectively.
Introducing two scalar fields $\phi$ and $\widehat{\sigma}$, and
a transverse vector field $\widehat{\xi}_\mu$, the tensors $h^{Tr}_{\mu\nu}$ 
and $h_{\mu\nu}^{L}$ can be expressed by
$h^{Tr}_{\mu\nu}\equiv\bar{g}_{\mu\nu}\phi/d$ and
$h_{\mu\nu}^{L}=h_{\mu\nu}^{LT}+h_{\mu\nu}^{LL}$ with
$h_{\mu\nu}^{LT}\equiv\bar{D}_\mu\widehat{\xi}_\nu
+\bar{D}_\nu\widehat{\xi}_\mu$ and $h_{\mu\nu}^{LL}\equiv\bar{D}_\mu\bar{D}_\nu
\widehat{\sigma}-\bar{g}_{\mu\nu}\bar{D}^2\widehat{\sigma}/d$. Thereby we end
up with eq. (\ref{TT}). In the following the components of $h_{\mu\nu}$
thus introduced will be referred to as the ``component fields''. They obey
the relations
\begin{eqnarray}
\label{c}
\bar{g}^{\mu\nu}h_{\mu\nu}^{T}=0\;,\;\;\;\bar{D}^\mu h_{\mu\nu}^{T}=0
\;,\;\;\;\bar{D}^\mu\widehat{\xi}_\mu=0\;,\;\;\;\phi=\bar{g}_{\mu\nu}
h^{\mu\nu}\;.
\end{eqnarray}

Obviously the complete field $h_{\mu\nu}$ receives no contribution from those
$\widehat{\xi}_\mu$- and $\widehat{\sigma}$-modes which satisfy the Killing
equation
\begin{eqnarray}
\label{09}
\bar{D}_\mu\widehat{\xi}_\nu+\bar{D}_\nu\widehat{\xi}_\mu=0
\end{eqnarray}
and the scalar equation
\begin{eqnarray}
\label{010}
\bar{D}_\mu \bar{D}_\nu\widehat{\sigma}-\frac{1}{d}\bar{g}_{\mu\nu}\bar{D}^2
\widehat{\sigma}=0\;,
\end{eqnarray}
respectively. Such ``unphysical'' $\widehat{\xi}_\mu$- and
$\widehat{\sigma}$-modes have to be excluded from the functional integral and
all subsequent calculations \cite{LR1}. Having a closer look at the scalar
equation (\ref{010}), one recognizes that there is a one-to-one correspondence
between the nonconstant solutions of (\ref{010}) and the purely longitudinal,
or proper, conformal Killing vectors (PCKV) ${\cal C}_\mu$. They are related
via ${\cal C}_\mu=\bar{D}_\mu\widehat{\sigma}$.

Likewise we decompose the ghost and the antighost into their orthogonal
components:
\begin{eqnarray}
\label{T}
\bar{C}_\mu=\bar{C}_\mu^T+\bar{D}_\mu\widehat{\bar{\eta}}\;,\;\;\;
C^\mu=C^{T\mu}+\bar{D}^\mu\widehat{\eta}\;.
\end{eqnarray}
Here $\bar{C}_\mu^T$ and $C^{T\mu}$ are the transverse components of
$\bar{C}_\mu$ and $C^\mu$: $\bar{D}^\mu\bar{C}_\mu^T=0$, 
$\bar{D}_\mu C^{T\mu}=0$. Furthermore, the scalars $\widehat{\bar{\eta}}$ and 
$\widehat{\eta}$ parametrize the longitudinal part of $\bar{C}_\mu$ and 
$C^\mu$, respectively. The constant $\widehat{\bar{\eta}}$- and 
$\widehat{\eta}$-modes represent unphysical modes which have to be excluded.

For calculational convenience we now introduce new variables
$\{\xi_\mu,\sigma,\bar{\eta},\eta\}$ replacing
$\{\widehat{\xi}_\mu,\widehat{\sigma},\widehat{\bar{\eta}},\widehat{\eta}\}$,
by means of the momentum dependent (nonlocal) redefinitions
\begin{eqnarray}
\label{y}
\xi^\mu &\equiv &\sqrt{-\bar{D}^2-\overline{{\rm Ric}}}\;\,
\widehat{\xi}^\mu\nonumber\\\sigma&\equiv&\sqrt{
(\bar{D}^2)^2+\frac{d}{d-1}\bar{D}_\mu\bar{R}^{\mu\nu}\bar{D}_\nu}\;\,
\widehat{\sigma}\nonumber\\
\bar{\eta}&\equiv &\sqrt{-\bar{D}^2}\;\,
\widehat{\bar{\eta}}\;,\;\;\;\eta\equiv\sqrt{-\bar{D}^2}
\;\,\widehat{\eta}\;.
\end{eqnarray}
Here the operator $\overline{{\rm Ric}}$ maps vectors onto vectors according
to $\left(\overline{{\rm Ric}}\;v\right)^\mu=\bar{R}^{\mu\nu}v_\nu$.
In accordance with the decompositions (\ref{TT}), (\ref{T}) and the 
redefinitions (\ref{y}) we then perform the combined transformation of 
integration variables $h_{\mu\nu}\longrightarrow \{h_{\mu\nu}^T,\xi_\mu,
\sigma,\phi\}$, $\bar{C}_\mu\longrightarrow \{\bar{C}^T_\mu,\bar{\eta}\}$,
$C^\mu\longrightarrow \{C^{T\mu},\eta\}$ in the functional integral (\ref{a}).
The Jacobian induced by this change of variables is such that it boils
down to an umimportant constant if Einstein backgrounds, characterized by
$\bar{R}_{\mu\nu}=C\bar{g}_{\mu\nu}$ with $C$ a constant, are inserted into
(\ref{a}). (For the general case see ref. \cite{LR1}.)

Let us now come to the ``type B'' cutoff term $\Delta_kS$. At the
component field level, it is a sum of inner products,
\begin{eqnarray}
\label{013}
\Delta_k S\left[h,C,\bar{C};\bar{g}\right]
=\frac{1}{2}\sum\limits_{\zeta_1,\zeta_2\in I_1}
\left<\zeta_1,\left({\cal R}_k\right)_{\zeta_1\zeta_2}\,
\zeta_2\right>+\frac{1}{2}
\sum\limits_{\psi_1,\psi_2\in I_2}
\left<\psi_1,\left({\cal R}_k\right)_{\psi_1\psi_2}\psi_2\right>
\end{eqnarray}
with the index sets $I_1\equiv
\{h^T,\xi,\sigma,\phi\}$, $I_2\equiv\{\bar{C}^T,C^T,\bar{\eta},\eta\}$.
At this stage of the discussion it is not necessary to specify the explicit
structure of the cutoff operators ${\cal R}_k$ acting on the component fields.
In order to provide the desired suppression of
low-momentum modes, these operators must vanish
for $p^2/k^2\rightarrow\infty$ (in particular for $k\rightarrow 0$) and must 
behave as ${\cal R}_k\rightarrow {\cal Z}_k k^2$ for $p^2/k^2\rightarrow 
0$. (The meaning of the constant ${\cal Z}_k$ will be explained later.) 
Furthermore, they have to satisfy certain hermiticity conditions \cite{LR1}.

Now we are in a position to construct the effective average action $\Gamma_k$.
It is defined as the difference between the Legendre transform of $W_k$
at fixed $\bar{g}_{\mu\nu}$, denoted $\widetilde{\Gamma}_k[\bar{h},v,\bar{v};
\bar{g}]$, and the 
cutoff action with the classical fields inserted \cite{W93,ym}:
\begin{eqnarray}
\label{nn}
\Gamma_k[g,\bar{g},v,\bar{v}]
\equiv\widetilde{\Gamma}_k\left[g-\bar{g},v,\bar{v};\bar{g}\right]
-\Delta_k S[g-\bar{g},v,\bar{v};\bar{g}]\;.
\end{eqnarray}
Here the classical fields represent the ($k$-dependent) expectation values
of the quantum fluctuations: $\bar{h}_{\mu\nu}\equiv\left<h_{\mu\nu}\right>$,
$\bar{v}_{\mu}\equiv\left<\bar{C}_{\mu}\right>$, $v^{\mu}\equiv\left<C^{\mu}
\right>$. They are obtained in the usual way as functional
derivatives of $W_k$ with respect to the sources. In ({\ref{nn}}) we expressed
$\bar{h}_{\mu\nu}$ in terms of the classical counterpart
$g_{\mu\nu}$ of the quantum metric $\gamma_{\mu\nu}\equiv\bar{g}_{\mu\nu}
+h_{\mu\nu}$ which, by definition, is given by
$g_{\mu\nu}\equiv\bar{g}_{\mu\nu}+\bar{h}_{\mu\nu}$.
The classical analogs of the components $(h^T,\xi,\sigma,\phi,\bar{C}^T,
C^T,\bar{\eta},\eta)$ will be denoted $(\bar{h}^T,\bar{\xi},\bar{\sigma},
\bar{\phi},\bar{v}^T,v^T,\bar{\varrho},\varrho)$.

The exact RG equation for the effective average action describes the
change of $\Gamma_k$ induced by an infinitesimal change of the scale $k$.
Introducing the RG ``time'' $t\equiv\ln\,k$, it can be derived from the
$t$-derivative of the functional integral (\ref{a}). It takes the form
\begin{eqnarray}
\label{oo}
\partial_t\Gamma_k\left[g,\bar{g},v,\bar{v}\right]
&=&\frac{1}{2}{\rm Tr}'\left[\sum\limits_{\zeta_1,\zeta_2\in \bar{I}_1}
\left(\Gamma_k^{(2)}[g,\bar{g},v,\bar{v}]
+{\cal R}_k\right)^{-1}_{\zeta_1\zeta_2}
\partial_t\left({\cal R}_k\right)_{\zeta_2\zeta_1}\right]
\nonumber\\
& &+\frac{1}{2}{\rm Tr}'\left[\sum\limits_{\psi_1,\psi_2\in \bar{I}_2}
\left(\Gamma_k^{(2)}[g,\bar{g},v,\bar{v}]
+{\cal R}_k\right)^{-1}_{\psi_1\psi_2}
\partial_t\left({\cal R}_k\right)_{\psi_2\psi_1}
\right]\;.
\end{eqnarray}
Here $\Gamma_k^{(2)}$ denotes the Hessian of $\Gamma_k$ with respect to the
component fields. Furthermore, we wrote $({\cal R}_k)_{\zeta_1\zeta_2}
\equiv({\cal R}_k)_{\left<\zeta_1\right>\left<\zeta_2\right>}$,
$({\cal R}_k)_{\psi_1\psi_2}\equiv({\cal R}_k)_{\left<\psi_1\right>\left<
\psi_2\right>}$ and introduced the index sets
$\bar{I}_1\equiv\{\bar{h}^T,\bar{\xi},\bar{\sigma},\bar{\phi}\}\;,\;\;\;
\bar{I}_2\equiv\{\bar{v}^T,v^T,\bar{\varrho},\varrho\}$. Furthermore,
the primes at the traces indicate that all unphysical $\bar{\xi}_\mu$- and 
$\bar{\sigma}$-modes, characterized by eqs. (\ref{09}) and (\ref{010}), are to
be excluded from the calculation of the traces.

\section{Truncations and their adapted cutoffs}
\renewcommand{\theequation}{3.\arabic{equation}}
\setcounter{equation}{0}
\label{3S3}
\subsection{Truncating the ghost sector}
\label{3S3A}
In concrete applications of the exact RG equation one encounters the
problem of dealing with an infinite system of coupled differential equations.
Usually it is impossible to find an exact solution so that we are forced to 
rely upon approximations. A powerful nonperturbative approximation scheme is 
the truncation of theory space, which means that only a finite number of 
couplings is considered and the RG flow is projected onto a finite-dimensional
subspace of theory space. In practice one proceeds as follows. One makes an 
ansatz for $\Gamma_k$ that comprises only a few couplings and inserts it on 
both sides of eq. (\ref{oo}). By projecting the RHS of this equation onto the 
space of operators appearing on the LHS one obtains a finite set of coupled 
differential equations for the couplings taken into account.

Given an arbitrary truncation it is not clear a priori whether it is
sensible and leads to at least approximately correct results.
In this respect the modified BRS Ward identities satisfied by the exact 
$\Gamma_k$ \cite{Reu96} are of special importance, since only
those truncations which are (approximately) consistent with them can
be reliable. In \cite{Reu96} it was shown that under certain conditions
truncations of the form
\begin{eqnarray}
\label{truncclass}
\Gamma_k[g,\bar{g},v,\bar{v}]=\bar{\Gamma}_k[g]+\widehat{\Gamma}_k[g,\bar{g}]
+S_{\rm gf}[g-\bar{g};\bar{g}]+S_{\rm gh}[g-\bar{g},v,\bar{v};\bar{g}]
\end{eqnarray}
which neglect the RG running in the ghost sector
are approximate solutions to the Ward identities for the exact 
$\Gamma_k$. Here $\bar{\Gamma}_k[g]$ is defined as
\begin{eqnarray}
\label{024}
\bar{\Gamma}_k[g]\equiv\Gamma_k[g,g,0,0]
\end{eqnarray}
and $\widehat{\Gamma}_k[g,\bar{g}]$ encodes the quantum corrections of the
gauge fixing term. (For the details we refer to \cite{Reu96,LR1}.) Inserting 
the ansatz (\ref{truncclass}) into the exact
evolution equation (\ref{oo}) leads to 
a truncated RG equation which describes the RG flow of $\Gamma_k$ in the
subspace of action functionals spanned by (\ref{truncclass}).
The equation governing the evolution of the purely gravitational action
\begin{eqnarray}
\label{033}
\Gamma_k[g,\bar{g}]\equiv\Gamma_k[g,\bar{g},0,0]=\bar{\Gamma}_k[g]+S_{\rm gf}
[g-\bar{g};\bar{g}]+\widehat{\Gamma}_k[g,\bar{g}]
\end{eqnarray}
takes the form
\begin{eqnarray}
\label{truncflow}
\partial_t\Gamma_k\left[g,\bar{g}\right]
&=&\frac{1}{2}{\rm Tr}'\left[\sum\limits_{\zeta_1,\zeta_2\in\bar{I}_1}
\left(\Gamma_k^{(2)}[g,\bar{g}]
+{\cal R}_k\right)^{-1}_{\zeta_1\zeta_2}
\partial_t\left({\cal R}_k\right)_{\zeta_2\zeta_1}\right]\nonumber\\
& &+\frac{1}{2}{\rm Tr}'\left[\sum\limits_{\psi_1,\psi_2\in\bar{I}_2}
\left(S_{\rm gh}^{(2)}[g,\bar{g}]
+{\cal R}_k\right)^{-1}_{\psi_1\psi_2}
\partial_t\left({\cal R}_k\right)_{\psi_2\psi_1}
\right]\;.
\end{eqnarray}
Here $\Gamma^{(2)}_k$ and $S_{\rm gh}^{(2)}$ are the Hessians of $\Gamma_k[g,
\bar{g}]$ and $S_{\rm gh}[\bar{h},v,\bar{v};\bar{g}]$ with respect to the 
gravitational and the ghost component fields, respectively. They are taken 
at fixed $\bar{g}_{\mu\nu}$.
\subsection{Construction of the cutoff, and the conformal factor problem}
\label{3S3B}
In order to obtain a tractable evolution equation for a given truncation it
is necessary to use a cutoff which is adapted to this truncation but still
has the desired suppression properties for a class of backgrounds which is as 
large as possible.

A convenient cutoff which is adapted to the truncation ansatz can be found in 
the following way \cite{Reu96,DP97}. Given a 
truncation, we assume that for $\bar{g}=g$ the kinetic operators of all modes
with a definite helicity can be brought to the form 
$(\Gamma_k^{(2)})_{ij}=f_{ij}(-\bar{D}^2,k,\ldots)$ where $\{f_{ij}\}$ is a 
set of c-number functions and the indices $i,j$ refer to the different types of
fields. Then we choose the cutoff operator ${\cal R}_k$ in such a way that the
structure 
\begin{eqnarray}
\label{rule}
(\Gamma_k^{(2)}+{\cal R}_k)_{ij}=f_{ij}
\left(-\bar{D}^2+k^2R^{(0)}(-\bar{D}^2/k^2),k,\ldots\right)
\end{eqnarray}
is achieved. Here the function $R^{(0)}(y)$, $y=-\bar{D}^2/k^2$, describes the
details of the mode suppression; it is required to satisfy the boundary 
conditions $R^{(0)}(0)=1$ and $\lim\limits_{y\rightarrow\infty}R^{(0)}(y)
=0$, but is arbitrary otherwise. By virtue of eq. (\ref{rule}), the inverse
propagator of a massless field mode with covariant momentum square $p^2\equiv
-\bar{D}^2$ is proportional to $p^2+k^2R^{(0)}(p^2/k^2)$ which equals $p^2$ for
$p^2\gg k^2$ and $p^2+k^2$ for $p^2\ll k^2$. This means that the small-$p^2$ 
modes, and only those, have acquired a mass $\propto k$ which leads to the 
desired suppression.

In order to see the potential problems of the rule (\ref{rule}) let us be more
specific and assume that the functions $f_{ij}$ are linear in $\bar{D}^2$ and
contain no constant term. Then, after diagonalizing $f_{ij}$ with respect to 
the field indices, $\Gamma_k^{(2)}$ decomposes into a set of (massless, by
assumption) inverse propagators $z_k\,p^2$ with a running wave function
normalization $z_k$. In this diagonal basis ${\cal R}_k$ is diagonal, too. 
It is of the form 
\begin{eqnarray}
\label{Rdiag}
{\cal R}_k(p^2)={\cal Z}_k\,k^2\,R^{(0)}\left(\frac{p^2}{k^2}\right)\;.
\end{eqnarray}
A priori ${\cal Z}_k$ is a free constant, but when we apply the rule 
(\ref{rule}) we are forced to set ${\cal Z}_k=z_k$ for each mode. Only then the
propagator and the cutoff combine in the right way, leading to the modified
inverse propagator $z_k[p^2+k^2R^{(0)}(p^2/k^2)]$.

The choice ${\cal Z}_k=z_k$ is certainly the correct one if $z_k$ is positive.
This is indeed the case in the familiar unitary theories on flat spacetime.
In QEG, however, there are truncations, the Einstein-Hilbert truncation, for
instance, which give a negative kinetic energy to certain modes $\varphi$ of
the metric. In particular, in the Einstein-Hilbert truncation, the conformal
factor $\varphi\equiv\phi$ has $z_k<0$.

The important question is how ${\cal Z}_k$ should be chosen when $z_k$ is 
negative. If we continue to use ${\cal Z}_k=z_k$, the RG equation is still
well-defined because the inverse propagator $-|z_k|[p^2+k^2R^{(0)}(p^2/k^2)]$
never vanishes so that the functional traces on the RHS of (\ref{truncflow}) 
are not 
suffering from any IR problem. In fact, if we write down the perturbative
expansion of the traces, for instance, we see that all propagators are
correctly cut off in the IR, and that loop momenta smaller than $k$ are 
suppressed correctly. This would not have been the case if we had insisted on a
positive ${\cal Z}_k$, setting ${\cal Z}_k=-z_k>0$. In this case the modified
inverse propagator $-|z_k|[p^2-k^2R^{(0)}(p^2/k^2)]$, because of the relative
minus sign between $p^2$ and the ${\cal R}_k$-term, fails to suppress the IR
modes. Even worse, it can introduce a spurious singularity at the value of 
$p^2$ for which $p^2-k^2R^{(0)}(p^2/k^2)=0$.

At first sight the choice ${\cal Z}_k=-z_k>0$ might have appeared to be the
more natural one because only if ${\cal Z}_k>0$ the factor $\exp(-\Delta_k S)
\propto\exp(-\int{\cal R}_k\,\varphi^2)$ is a damped exponential which 
suppresses the low-momentum modes under the path integral. Nevertheless it was
argued in \cite{Reu96} that the ``${\cal Z}_k=z_k$-rule'' is the correct
choice both for
$z_k>0$ and $z_k<0$. The calculations in \cite{Reu96} and all subsequent papers
\cite{LR1,DP97,sven,grano} were based upon this rule. On the one hand, this
rule guarantees that the RG equation is well-defined and consistent. On the
other hand, it is difficult to give a meaning to the Euclidean functional
integral from which this RG equation was derived at the formal level. In the
case ${\cal Z}_k=z_k<0$ the factor $\exp(+\int|{\cal R}_k|\,\varphi^2)$ is a
growing exponential which seems to enhance rather than suppress the
low-momentum modes. However, as suggested by the perturbative argument above,
this conclusion is too naive probably.

Let us now come to the case where the functions
$f_{ij}$ are not linear in $\bar{D}^2$ but, say, of the form
$z^{(1)}_k(\bar{D}^2)^2+z^{(2)}_k\bar{D}^2$. In this case the rule (\ref{rule})
demands that we choose the corresponding operators $({\cal R}_k)_{ij}$ in such
a way that they contain cutoff terms adjusted to both the $(\bar{D}^2)^2$- and
the $\bar{D}^2$-terms of the kinetic operator. For the above example, which is
relevant to the $R^2$-truncation, $({\cal R}_k)_{ij}$ assumes the general form
\begin{eqnarray}
\label{Rquad}
{\cal R}_k(p^2)={\cal Z}^{(1)}_k\left[2p^2\,k^2\,R^{(0)}
\left(\frac{p^2}{k^2}\right)+k^4\,\left(R^{(0)}\left(\frac{p^2}{k^2}\right)
\right)^2\right]
+{\cal Z}^{(2)}_k\,k^2\,R^{(0)}\left(\frac{p^2}{k^2}\right)
\end{eqnarray}
where we omitted the indices referring to the types of fields.
Obviously we have to set ${\cal Z}^{(1)}_k=z^{(1)}_k$ and 
${\cal Z}^{(2)}_k=z^{(2)}_k$
in order to achieve that the propagator and the cutoff
combine as prescribed by the rule (\ref{rule}). This leads to the
modified inverse propagator $z^{(1)}_k[p^2+k^2R^{(0)}
(p^2/k^2)]^2+z^{(2)}_k[p^2+k^2R^{(0)}(p^2/k^2)]$.
For brevity we refer to this prescription, too, as the ${\cal Z}_k=z_k$-rule.

We believe that the ${\cal Z}_k=z_k$-rule is correct also for $z_k<0$, and that
it is the relation between the manifestly well-defined flow equation and the
formal path integral that needs to be understood better. Various attitudes are
possible here. For instance, one could postulate that the fundamental 
definition
of the theory is in terms of the flow equation rather than the path integral.
Since the former is much better defined than the latter (in particular also
with respect to the usual UV and IR problems) one would simply discard the
path integral then. Another way out is to adopt the usual, albeit rather ad 
hoc, prescription of Wick-rotating the conformal factor $(\varphi\rightarrow
{\rm i}\varphi)$ which turns the growing exponential into a decaying one.

A much more attractive and less radical possibility is the following. 
Presumably it will be possible to construct an effective average action for
{\it Lorentzian} quantum gravity by invoking a kind of stationary phase
argument for the mode suppression. Then one deals with oscillating exponentials
$\exp({\rm i S})\exp({\rm i}\Delta_k S)$, and apart from the trivial 
substitutions $\Gamma_k\rightarrow -{\rm i}\Gamma_k$, ${\cal R}_k\rightarrow
-{\rm i}{\cal R}_k$ the flow equation remains the same as in Euclidean gravity.
For ${\cal Z}_k=z_k$ it has all the desired features, and $z_k<0$ poses no
special problem for the path integral. It is interesting that there are also
recent indications \cite{lollambj} coming from the dynamical triangulation
approach to quantum gravity \cite{amb} which suggest that the Lorentzian path
integral might have a better chance of being well-defined than the Euclidean
one.

As the last possibility we mention the best of all situations, namely that
in an exact treatment there are simply no factors $z_k<0$.
If this is actually the case, the conformal factor problem which we encounter
in the Einstein-Hilbert and similar truncations would have the status of an 
unphysical truncation artifact. If so, the ``${\cal Z}_k=z_k$-rule''
could be interpreted as a device which helps in approximating as well as 
possible the exact RG flow by a truncated flow.

It is one of the main results of the present paper that this scenario is 
indeed realized to some extent. We shall see that within the $R^2$-truncation
those terms of $f_{ij}$ which dominate at sufficiently large momenta have 
$z_k>0$ at least for large enough values of $k$ $(k\gg m_{\rm Pl})$.
For too low scales $(k\lesssim m_{\rm Pl})$ some of the $z_k$'s might turn 
negative, but at these scales the $R^2$-truncation becomes unreliable probably 
\cite{frank} so that the negative $z_k$'s might be due to an insufficient
truncation. It is not excluded that in the exact theory the dominating 
$z_k$'s are positive down to 
$k=0$.\renewcommand{\baselinestretch}{1}\small\normalsize\footnote{Recent 
investigations in a scalar toy model \cite{condensate} indeed suggest that
the conformal factor problem could be solved dynamically by strong 
instability-driven renormalization effects.}
\renewcommand{\baselinestretch}{1.5}\small\normalsize

As all those $z_k$'s which determine the sign of the dominating contributions 
to $f_{ij}$ are positive for large values of $k$, the cutoff has the 
standard suppression properties and is not plagued by any conformal factor
problem for $k\rightarrow\infty$. This, then, allows for an unambiguous
investigation of the UV fixed point and its properties. It is quite remarkable
that, as we shall see, all results concerning the fixed point are basically the
same for the $R^2$-truncation (where ${\cal Z}_k=z_k>0$ at least for the
dominating terms) and the Einstein-Hilbert truncation with both positive and 
negative factors ${\cal Z}_k=z_k$. This is
certainly a quite impressive confirmation of the ${\cal Z}_k=z_k$-rule.

\subsection{The cutoff adapted to the $R^2$-truncation}
In the next section we shall see in detail that for the truncation studied in
this paper we can comply with the ${\cal Z}_k=z_k$-rule by using the
following cutoff operators for the component fields:
\begin{eqnarray}
\label{cutoff}
\left({\cal R}_k\right)^{\mu\nu\alpha\beta}_{\bar{h}^T\bar{h}^T}
&=&\frac{1}{4}
\left(\bar{g}^{\mu\alpha}\bar{g}^{\nu\beta}+\bar{g}^{\mu\beta}
\bar{g}^{\nu\alpha}\right)
\left\{{\cal Z}_k^{\bar{h}^T\bar{h}^T}\kappa^2+{\cal Y}_k^{\bar{h}^T\bar{h}^T}
\,\bar{R},k^2R^{(0)}(-D^2/k^2)\right\}\;,\nonumber\\
\left({\cal R}_k\right)^{\mu\nu}_{\bar{\xi}\bar{\xi}}&=&
{\cal Z}_k^{\bar{\xi}\bar{\xi}}\kappa^2\,\bar{g}^{\mu\nu}k^2 R^{(0)}
\left(-\bar{D}^2/k^2\right)\;,\nonumber\\
\left({\cal R}_k\right)_{\bar{\sigma}\bar{\sigma}}&=&{\cal X}_k^{\bar{\sigma}
\bar{\sigma}}\left(-2D^2\,k^2R^{(0)}(-D^2/k^2)
+k^4R^{(0)}(-D^2/k^2)^2\right)\nonumber\\
& &+\frac{1}{2}\left\{{\cal Z}_k^{\bar{\sigma}\bar{\sigma}}\kappa^2
+{\cal Y}_k^{\bar{\sigma}\bar{\sigma}}\,\bar{R},k^2R^{(0)}(-D^2/k^2)\right\}\;,
\nonumber\\
\left({\cal R}_k\right)_{\bar{\phi}\bar{\sigma}}
&=&\left({\cal R}_k\right)_{\bar{\sigma}\bar{\phi}}^\dagger
={\cal X}_k^{\bar{\phi}\bar{\sigma}}\left[\bar{P}_k\,\sqrt{\left(\bar{P}_k
+\frac{d}{d-1}\bar{D}_\mu\bar{R}^{\mu\nu}\bar{D}_\nu(-\bar{D}^2)^{-1}\right)
\bar{P}_k}\right.\nonumber\\
& &\left.+\bar{D}^2\,\sqrt{\left(\bar{D}^2\right)^2+\frac{d}{d-1}
\bar{D}_\mu\bar{R}^{\mu\nu}\bar{D}_\nu}\right]\nonumber\\
& &+\left({\cal Y}_k^{\bar{\phi}\bar{\sigma}}\,\bar{R}+
{\cal Z}_k^{\bar{\phi}\bar{\sigma}}\kappa^2\right)\left[\sqrt{\left(\bar{P}_k
+\frac{d}{d-1}\bar{D}_\mu\bar{R}^{\mu\nu}\bar{D}_\nu(-\bar{D}^2)^{-1}\right)
\bar{P}_k}\right.\nonumber\\
& &-\left.\sqrt{\left(\bar{D}^2\right)^2+\frac{d}{d-1}
\bar{D}_\mu\bar{R}^{\mu\nu}\bar{D}_\nu}\right]\;,\nonumber\\
\left({\cal R}_k\right)_{\bar{\phi}\bar{\phi}}&=&{\cal X}_k^{\bar{\phi}
\bar{\phi}}\left(-2D^2\,k^2R^{(0)}(-D^2/k^2)
+k^4R^{(0)}(-D^2/k^2)^2\right)\nonumber\\
& &+\frac{1}{2}\left\{{\cal Z}_k^{\bar{\phi}\bar{\phi}}\kappa^2
+{\cal Y}_k^{\bar{\phi}\bar{\phi}}\,\bar{R},k^2R^{(0)}(-D^2/k^2)\right\}\;,
\nonumber\\
\left({\cal R}_k\right)^{\mu\nu}_{\bar{v}^Tv^T}&=&-\left({\cal R}_k
\right)^{\mu\nu}_{v^T\bar{v}^T}={\cal Z}_k^{\bar{v}^T v^T}\,
\bar{g}^{\mu\nu}\,k^2 R^{(0)}\left(-\bar{D}^2/k^2\right)\;,\nonumber\\
\left({\cal R}_k\right)_{\bar{\varrho}\varrho}
&=&-\left({\cal R}_k\right)_{\varrho\bar{\varrho}}
={\cal Z}_k^{\bar{\varrho}\varrho}\,k^2 R^{(0)}\left(-\bar{D}^2/k^2\right)\;.
\end{eqnarray}
Here $\bar{P}_k$ is defined as
\begin{eqnarray}
\label{P_k}
\bar{P}_k\equiv -\bar{D}^2+k^2R^{(0)}(-\bar{D}^2/k^2)
\end{eqnarray}
and the curly brackets denote the anticommutator, i.e. $\{A,B\}=AB+BA$ for 
arbitrary operators $A$, $B$. The remaining cutoff operators which appear in 
(\ref{013}) but are not listed in eq. (\ref{cutoff}) are set to zero.

The constants ${\cal X}_k$, ${\cal Y}_k$, and ${\cal Z}_k$ will be adjusted 
later. It should be noted that the terms proportional to the 
${\cal X}_k$'s and ${\cal Y}_k$'s provide the cutoff for those 
contributions to $\Gamma_k^{(2)}$ which come from the higher-derivative terms.
For ${\cal Y}_k^{\bar{h}^T\bar{h}^T}=
{\cal Y}_k^{\bar{\sigma}\bar{\sigma}}={\cal Y}_k^{\bar{\phi}\bar{\sigma}}=
{\cal Y}_k^{\bar{\phi}\bar{\phi}}={\cal X}_k^{\bar{\sigma}\bar{\sigma}}=
{\cal X}_k^{\bar{\phi}\bar{\sigma}}={\cal X}_k^{\bar{\phi}\bar{\phi}}=0$,
eq. (\ref{cutoff}) actually
boils down to the cutoff of type B used in \cite{LR1} in the context
of the Einstein-Hilbert truncation. (This cutoff type has to be distinguished 
from the cutoff of type A used in the original paper \cite{Reu96}, which is
formulated in terms of the complete fields and does not involve the component
fields.) 

Each cutoff contains some ``shape function'' $R^{(0)}$. A particularly
suitable choice is the exponential shape function
\begin{eqnarray}
\label{expcut}
R^{(0)}(y)=y\left[\exp(y)-1\right]^{-1}\;.
\end{eqnarray}
In order to check the scheme independence of universal quantities we employ a
one-parameter generalization of (\ref{expcut}), the class of
exponential shape functions,
\begin{eqnarray}
\label{expshape}
R^{(0)}(y;s)=sy\left[\exp(sy)-1\right]^{-1}\;,
\end{eqnarray}
with the ``shape parameter'' $s>0$ parametrizing the profile of $R^{(0)}$ 
\cite{souma2}. Another admissible choice we are going to use is
the following class of shape functions with compact support: 
\begin{eqnarray}
\label{supp}
R^{(0)}(y;b)=\left\{\begin{array}{ll}1 & \;\;\;y\le b\\
\exp\left[(y-1.5)^{-1}\exp\left[\left(b-y\right)^{-1}\right]\right] & \;\;\;b
<y<1.5\\0 & \;\;\;y\ge 1.5
\end{array}\right.\;.
\end{eqnarray}
Here it is the shape parameter $b\in [0,1.5)$ which parametrizes the profile 
of $R^{(0)}$ \cite{imp}.

\section{The $R^2$-truncation}
\renewcommand{\theequation}{4.\arabic{equation}}
\setcounter{equation}{0}
\label{$R^2$-trunc}
\subsection{The ansatz}
\label{3S4A}
In all previous papers \cite{Reu96,souma1,souma2,sven,LR1,frank} the flow 
equation 
of QEG was used in the Einstein-Hilbert truncation. In this
section we generalize this truncation by taking also an $R^2$-term with
associated running coupling $\bar{\beta}_k$ into account and we derive the RG
flow within this ``$R^2$-truncation''. We assume that, at the UV scale
$\widehat{k}\rightarrow\infty$, gravity in $d$ dimensions is described by the
action
\begin{eqnarray}
\label{init}
\bar{\Gamma}_{\widehat{k}}[g]=S[g]=\int d^dx\,\sqrt{g}
\left\{\frac{1}{16\pi\bar{G}}\left(-R(g)+2\bar{\lambda}\right)
+\bar{\beta}\,R^2(g)\right\}\;.
\end{eqnarray}
It consists of the conventional Einstein-Hilbert action and a
higher-derivative term with bare coupling $\bar{\beta}$.
In order to study the RG flow of $\Gamma_k[g,\bar{g}]$ towards smaller scales
$k<\widehat{k}$ we employ a truncated action functional of the following form:
\begin{eqnarray}
\label{trunc}
\Gamma_k[g,\bar{g}]&=&\int d^dx\,\sqrt{g}\left\{2\kappa^2 Z_{Nk}\left(
-R(g)+2\bar{\lambda}_k\right)+\bar{\beta}_k\,R^2(g)\right\}\nonumber\\
& &+\kappa^2 \frac{Z_{Nk}}{\alpha}\int d^dx\,\sqrt{\bar{g}}\,\bar{g}^{\mu\nu}
\left({\cal F}_\mu^{\alpha\beta}
g_{\alpha\beta}\right)\left({\cal F}_\nu^{\rho\sigma}g_{\rho\sigma}\right)\;.
\end{eqnarray}
The ansatz (\ref{trunc}) is obtained from $S+S_{\rm gf}$ by replacing
\begin{eqnarray}
\label{replace}
\bar{G}\rightarrow G_k\equiv Z_{Nk}^{-1}\,\bar{G}\;,\;\;\bar{\lambda}
\rightarrow\bar{\lambda}_k\;,\;\;\bar{\beta}\rightarrow\bar{\beta}_k\;,\;\;
\alpha\rightarrow Z_{Nk}^{-1}\alpha
\end{eqnarray}
so that its form agrees with that of the gravitational sector in the ansatz
(\ref{truncclass}) with 
\begin{eqnarray}
\label{091}
\widehat{\Gamma}_k[g,\bar{g}]=\kappa^2 \frac{Z_{Nk}-1}{\alpha}\int d^dx\,
\sqrt{\bar{g}}\,\bar{g}^{\mu\nu}\left({\cal F}_\mu^{\alpha\beta}
g_{\alpha\beta}\right)\left({\cal F}_\nu^{\rho\sigma}g_{\rho\sigma}\right)\;.
\end{eqnarray}
In principle, also the gauge fixing parameter $\alpha$ should be treated as
a scale-dependent quantity: $\alpha\rightarrow\alpha_k$. Its evolution is
neglected here for simplicity. However, setting $\alpha=0$ by hand mimics a
dynamical treatment of the gauge fixing parameter since $\alpha=0$ can be
argued to be a RG fixed point \cite{alpha,LR1}.

\subsection{Projecting the flow equation}
\label{3S4B}
The ansatz (\ref{trunc}) comprises three $k$-dependent couplings. They
satisfy the initial conditions $\bar{\lambda}_{\widehat{k}}=\bar{\lambda}$,
$Z_{N\widehat{k}}=1$ which implies $G_{\widehat{k}}=\bar{G}$, and
$\bar{\beta}_{\widehat{k}}=\bar{\beta}$. Here the UV scale $\widehat{k}$ is
taken to be large but finite. In order to determine the evolution of
$\bar{\lambda}_k$, $Z_{Nk}$ and $\bar{\beta}_k$ towards smaller scales we have
to project the flow equation onto the space spanned by the operators
$\int d^dx\,\sqrt{g}$, $\int d^dx\,\sqrt{g}R$ and $\int d^dx\,\sqrt{g}R^2$.
After having inserted the ansatz (\ref{trunc})
into both sides of the flow equation and having performed the 
$g_{\mu\nu}$-derivatives implicit in $\Gamma_k^{(2)}$ we may set 
$g_{\mu\nu}=\bar{g}_{\mu\nu}$.
As a consequence, the gauge fixing term drops out from the LHS which then
reads
\begin{eqnarray}
\label{LHS}
\partial_t\Gamma_k[\bar{g},\bar{g}]=\int d^dx\,\sqrt{\bar{g}}
\left\{2\kappa^2\left[-\bar{R}(\bar{g})\,\partial_t
Z_{Nk}+2\partial_t\left(Z_{Nk}\bar{\lambda}_k\right)\right]
+\bar{R}^2(\bar{g})\,\partial_t\bar{\beta}_k\right\}\;.
\end{eqnarray}
Obviously the LHS is spanned by the operators $\int d^dx\,\sqrt{g}$, 
$\int d^dx\,
\sqrt{g}R$ and $\int d^dx\,\sqrt{g}R^2$. This means that we have
to perform a derivative expansion on the RHS in order to extract precisely
those contributions from the traces which are proportional to these operators.
By equating the result to (\ref{LHS}) and comparing the coefficients we can 
read off the system of coupled differential equations for $\bar{\lambda}_k$, 
$Z_{Nk}$ and $\bar{\beta}_k$. 

In order to make these technically rather involved calculations feasible we may
insert any metric $\bar{g}_{\mu\nu}$ that is general enough to admit a unique
identification of the operators spanning the truncated theory space. We
exploit this freedom by assuming that $\bar{g}_{\mu\nu}$ corresponds to a
maximally symmetric space. Such spaces form a special class of Einstein
spaces and are characterized by
\begin{eqnarray}
\label{maxsym}
\bar{R}_{\mu\nu\rho\sigma}=\frac{\bar{R}}{d(d-1)}\left(\bar{g}_{\mu\rho}
\bar{g}_{\nu\sigma}-\bar{g}_{\mu\sigma}\bar{g}_{\nu\rho}\right)\;,
\;\;\;\bar{R}_{\mu\nu}=\frac{\bar{R}}{d}\bar{g}_{\mu\nu}
\end{eqnarray}
with the curvature scalar $\bar{R}$ considered a constant number rather than a
functional of the metric. It is sufficient to employ spaces with $\bar{R}>0$,
i.e. $d$-spheres $S^d$. They are parametrized by
their radius $r$ which is related 
to the curvature scalar and the volume in the usual way,
\begin{eqnarray}
\label{spheres}
\bar{R}=\frac{d(d-1)}{r^2}\;,\;\;\;
\int d^dx\,\sqrt{\bar{g}}=\frac{\Gamma\left(\frac{d}{2}\right)}{\Gamma(d)}
(4\pi r^2)^{\frac{d}{2}}\;.
\end{eqnarray}

We emphasize that the beta-functions of $\bar{\lambda}_k$, $Z_{Nk}$, and
$\bar{\beta}_k$ do not depend on this choice for $\bar{g}_{\mu\nu}$; it is
simply a technical trick without any physical meaning. In principle the
beta-functions could be computed without any specification of 
$\bar{g}_{\mu\nu}$.

While this projection technique is capable of distinguishing $\int d^dx\,
\sqrt{g}\propto r^{d}$ from both $\int d^dx\,\sqrt{g}R\propto r^{d-2}$ and
$\int d^dx\,\sqrt{g}R^2\propto r^{d-4}$, it cannot disentangle the three
$({\rm curvature})^2$-invariants
$\int d^dx\,\sqrt{g}R^2$, $\int d^dx\,\sqrt{g}R_{\mu\nu}^2$, and 
$\int d^dx\,\sqrt{g}R_{\mu\nu\rho\sigma}^2$ which are all proportional to
$r^{d-4}$. If one wants to project them out individually one has to insert 
non-maximally
symmetric spaces, but then the evaluation of the functional traces on the
RHS of (\ref{truncflow}) is a rather formidable problem with the present
technology. In fact, this is one of the reasons for omitting the other
two $({\rm curvature})^2$-invariants from our truncation ansatz.

In ref. \cite{LR1} we discussed already the expansion of fields defined on
spherical backgrounds. Both the classical and the quantum 
TT-component fields can be expanded in terms of transverse-traceless tensor
harmonics $T^{lm}_{\mu\nu}$, transverse vector harmonics $T^{lm}_\mu$, and 
scalar harmonics $T^{lm}$. They form complete sets of orthogonal 
eigenfunctions with respect to the corresponding covariant Laplacians. 
We summarize the main results of \cite{LR1} in appendix \ref{harm}. In 
particular, 
the expansions of $h^T_{\mu\nu}$, $\phi$, $C^\mu$, $\bar{C}_\mu$ and their 
classical counterparts can be read off from eq. (\ref{067}), while the
remaining component fields are expanded according to
\begin{eqnarray}
\label{092}
\xi_\mu(x)&=&
\sum\limits_{l=2}^{\infty}\sum\limits_{m=1}^{D_l(d,1)}\xi_{lm}\,
T^{lm}_\mu(x)\;,\;\;\;
\sigma(x)=\sum\limits_{l=2}^{\infty}\sum\limits_{m=1}^{D_l(d,0)}\sigma_{lm}\,
T^{lm}(x)\;,\nonumber\\
\eta(x)&=&\sum\limits_{l=1}^{\infty}\sum\limits_{m=1}^{D_l(d,0)}\eta_{lm}\,
T^{lm}(x)\;,\;\;\;
\bar{\eta}(x)=\sum\limits_{l=1}^{\infty}\sum\limits_{m=1}^{D_l(d,0)}
\bar{\eta}_{lm}\,T^{lm}(x)\;.
\end{eqnarray}
Similar expansions hold for the associated classical fields 
(expectation values).

Contrary to eq. (\ref{067}), the summations in eq. (\ref{092}) do {\it not} 
start at $l=1$ for vectors and at $l=0$ for scalars, but at 
$l=2$ for $\xi_\mu$ and $\sigma$, and at $l=1$ for the scalar ghost fields. The
modes omitted here are the Killing vectors $(T^{l=1,m}_\mu)$, the solutions of
the scalar 
equation (\ref{010}) $(T^{l=1,m})$, and the constants $(T^{l=0,m=1})$. As we 
mentioned in section \ref{3S2}, these modes do not correspond to fluctuations
of $h_{\mu\nu}$ or the Faddeev-Popov ghosts.

As in \cite{LR1}, we decompose the quantum field $\phi$ into
a part $\phi_1$ spanned by the same set of eigenfunctions as $\sigma$,
and a part $\phi_0$ containing the contributions from the remaining modes:
\begin{eqnarray}
\label{split}
\phi(x)=\phi_0(x)+\phi_1(x)\,,\;
\phi_0(x)=\sum\limits_{l=0}^{1}\sum\limits_{m=1}^{D_l(d,0)}\phi_{lm}
\,T^{lm}(x)\,,\;
\phi_1(x)=\sum\limits_{l=2}^{\infty}\sum\limits_{m=1}^{D_l(d,0)}
\phi_{lm}\,T^{lm}(x)
\end{eqnarray}
Due to the orthogonality of the spherical harmonics, $\phi_0$ is orthogonal to 
$\phi_1$ and $\sigma$: $\left<\phi_1,\phi_0\right>$ 
$=\left<\sigma,\phi_0\right>=0$. This implies $\left<\phi,\phi\right>
=\left<\phi_0,\phi_0\right>+\left<\phi_1,\phi_1\right>$ and  
$\left<\sigma,\phi\right>=\left<\sigma,\phi_1\right>$. As a consequence, 
splitting $\phi$ according to eq. (\ref{split}) ensures that any nonzero
bilinear cross term of the scalar fields is such that the scalars involved can
be expanded in the same set of eigenfunctions. Of course, the same holds for 
the corresponding classical fields $\bar{\phi}_0$ and $\bar{\phi}_1$.

\subsection{Inserting the ansatz into the RHS of the RG equation}
\label{3S4C}
Let us now start with the evaluation of the RHS of eq. 
(\ref{truncflow}). After having inserted the truncation ansatz (\ref{trunc})
we identify the two metrics $g_{\mu\nu}$ and $\bar{g}_{\mu\nu}$. Therefore it 
is sufficient to calculate the operators $(\Gamma_k^{(2)}[g,\bar{g}]
+{\cal R}_k[\bar{g}])^{-1}$ and $(S^{(2)}_{\rm gh}[g,\bar{g}]
+{\cal R}_k[\bar{g}])^{-1}$ at $g_{\mu\nu}=\bar{g}_{\mu\nu}$.
To this end we first expand the  
ansatz (\ref{trunc}) according to 
\begin{eqnarray}
\label{041}
\Gamma_k[\bar{g}+\bar{h},\bar{g}]=\Gamma_k[\bar{g},\bar{g}]+{\cal O}(\bar{h})
+\Gamma_k^{\rm quad}[\bar{h};\bar{g}]+{\cal O}(\bar{h}^3)
\end{eqnarray}
and concentrate on the part quadratic in $\bar{h}_{\mu\nu}$, i.e. 
$\Gamma_k^{\rm quad}\left[\bar{h};\bar{g}\right]$. This leads to 
\begin{eqnarray}
\label{quadform}
\lefteqn{\Gamma_k^{\rm quad}\left[\bar{h};\bar{g}\right]
=\int d^dx\,\sqrt{\bar{g}}\,\bar{h}_{\mu\nu}\Bigg\{\kappa^2 Z_{Nk}\Bigg[
-\left(\frac{1}{2}\delta^\mu_\rho\delta_\sigma^\nu+\frac{1-2\alpha}{4\alpha}
\bar{g}^{\mu\nu}\bar{g}_{\rho\sigma}\right)\bar{D}^2}\\
& &+\frac{1}{4}\left(2\delta^\mu_\rho\delta^\nu_\sigma-\bar{g}^{\mu\nu}
\bar{g}_{\rho\sigma}\right)\left(\bar{R}-2\bar{\lambda}_k\right)
+\bar{g}^{\mu\nu}\bar{R}_{\rho\sigma}-\delta_\sigma^\mu\bar{R}^\nu_{\;
\rho}-\bar{R}^{\nu\;\;\mu}_{\;\rho\;\;\;\sigma}\nonumber\\
& &+\frac{1-\alpha}{\alpha}
\left(\bar{g}^{\mu\nu}\bar{D}_\rho\bar{D}_\sigma-\delta^\mu_\sigma
\bar{D}^\nu\bar{D}_\rho\right)\Bigg]\nonumber\\
%
%
& &+\frac{1}{2}\bar{\beta}_k\Bigg[
\frac{1}{2}\left(\frac{1}{2}\bar{g}^{\mu\nu}\bar{g}_{\rho\sigma}
-\delta^\mu_\rho\delta^\nu_\sigma\right)\bar{R}^2
+2\bar{g}^{\mu\nu}\bar{R}\left(-\bar{R}_{\rho\sigma}+\bar{D}_{\rho}
\bar{D}_\sigma-\bar{g}_{\rho\sigma}\bar{D}^2\right)\nonumber\\
& &+2R\left(\delta_\sigma^\nu\bar{R}^\mu_{\;\,\rho}
-\bar{R}^{\mu\;\;\;\;\nu}_{\;\;\rho\sigma}
-3 \delta^\nu_\sigma\bar{D}^{\mu}\bar{D}_{\rho}
+2\bar{g}_{\rho\sigma}\bar{D}^\mu\bar{D}^\nu
+2\delta^\mu_\rho\delta^\nu_\sigma \bar{D}^2
-\delta^\nu_\sigma\bar{D}_\rho\bar{D}^\mu\right)
+2\delta^\nu_\sigma \bar{D}_\rho\bar{R}\bar{D}^\mu\nonumber\\
& &
-3\delta^\mu_\rho\delta^\nu_\sigma\bar{D}^{\lambda}\bar{R}\bar{D}_\lambda 
+4\delta^\nu_\sigma \bar{D}^\mu\bar{R}\bar{D}_\rho
-4\bar{g}_{\rho\sigma}\bar{D}^{\mu}\bar{R}\bar{D}^\nu
+\bar{g}^{\mu\nu}\bar{g}_{\rho\sigma}\bar{D}^{\lambda}\bar{R}\bar{D}_\lambda 
+2\bar{R}^{\mu\nu}\bar{R}_{\rho\sigma}\nonumber\\
& &-4\bar{R}^{\mu\nu}
\left(\bar{D}_\rho\bar{D}_\sigma-\bar{g}_{\rho\sigma}\bar{D}^2\right)
+2\bar{D}^\mu\bar{D}^\nu\bar{D}_\rho\bar{D}_\sigma
-4\bar{g}^{\mu\nu}\bar{D}^2\bar{D}_\rho\bar{D}_\sigma
+2\bar{g}^{\mu\nu}\bar{g}_{\rho\sigma}\left(\bar{D}^2\right)^2
\Bigg]\Bigg\}\bar{h}^{\rho\sigma}\;.\nonumber
\end{eqnarray}
At this stage $\bar{g}_{\mu\nu}$ is still arbitrary. In order to 
(partially) diagonalize this quadratic form we insert the family of $S^d$
background metrics into eq. (\ref{quadform}) and decompose $\bar{h}_{\mu\nu}$
according to eq. (\ref{TT}). Then we apply eq. (\ref{split}) to the classical
field $\bar{\phi}$ to decompose it as well. This yields
\begin{eqnarray}
\label{gravquad}
\lefteqn{\Gamma_k^{\rm quad}\left[\bar{h};\bar{g}\right]
=\int d^dx\,\sqrt{\bar{g}}\,\frac{1}{2}\Bigg\{
\bar{h}^T_{\mu\nu}\bigg[
Z_{Nk}\kappa^2\left(-\bar{D}^2+A_T(d)\,\bar{R}-2\bar{\lambda}_k\right)}
\nonumber\\
& &+\bar{\beta}_k\left(\bar{R}\,\bar{D}^2+G_T(d)\,\bar{R}^2\right)
\bigg]\bar{h}^{T\mu\nu}\nonumber\\
& &+\bar{\xi}_\mu\left[\frac{2}{\alpha}
Z_{Nk}\kappa^2\left(-\bar{D}^2+A_V(d,\alpha)\,\bar{R}-2\alpha
\bar{\lambda}_k\right)+G_V(d)\,\bar{\beta}_k\,\bar{R}^2\right]
\bar{\xi}^\mu\nonumber\\
& &+\bar{\sigma}\bigg[
C_{S2}(d,\alpha)\,Z_{Nk}\kappa^2\left(-\bar{D}^2+A_{S2}(d,\alpha)\,\bar{R}
+B_{S2}(d,\alpha)\,\bar{\lambda}_k\right)\nonumber\\
& &+\bar{\beta}_k\left(H_S(d)\,\left(\bar{D}^2\right)^2
-G_{S1}(d)\,\left(2\bar{R}\,\bar{D}^2+\bar{R}^2\right)\right)
\bigg]\bar{\sigma}\nonumber\\
& &+2\bar{\phi}_1\bigg[C_{S2}(d,\alpha)\,C_{S3}(d,\alpha)\,Z_{Nk}\kappa^2
+\bar{\beta}_k\left(-H_S(d)\,\bar{D}^2+2G_{S1}(d)\,\bar{R}\right)\bigg]
\nonumber\\
& &\times\sqrt{-\bar{D}^2}\sqrt{-\bar{D}^2-\frac{\bar{R}}{d-1}}\,\bar{\sigma}
\nonumber\\
& &+\sum\limits_{\bar{\phi}\in\{\bar{\phi}_0,\bar{\phi}_1\}}
\bar{\phi}\bigg[C_{S2}(d,\alpha)\,C_{S1}(d,\alpha)\,Z_{Nk}\kappa^2
\left(-\bar{D}^2+A_{S1}(d,\alpha)\,\bar{R}
+B_{S1}(d,\alpha)\,\bar{\lambda}_k\right)\nonumber\\
& &+\bar{\beta}_k\left(H_S(d)\,\left(\bar{D}^2\right)^2
-G_{S2}(d)\,\bar{R}\,\bar{D}^2+G_{S3}(d)\,\bar{R}^2\right)
\bigg]\bar{\phi}\Bigg\}\;.
\end{eqnarray}
Here the various $A$'s, $B$'s, $C$'s and $G$'s and $H_S$ are functions of the
dimensionality $d$ and the gauge fixing parameter $\alpha$. The explicit
expressions for these coefficients are given in appendix \ref{S_kcoeff}.

This partial diagonalization is performed in order to simplify the inversion
of the operator $\Gamma_k^{(2)}[g,\bar{g}]+{\cal R}_k[\bar{g}]$. In fact,
this is the main reason for using the TT-decomposition (\ref{TT}) and
specifying a concrete background. Note that in the pure Einstein-Hilbert 
truncation it is only the term in (\ref{quadform}) which is proportional to 
$1-\alpha$ that gives rise to mixings between the traceless
part of $\bar{h}_{\mu\nu}$ and $\bar{\phi}$ and therefore necessitates the 
complete decomposition (\ref{TT}). For $\alpha=1$, a complete diagonalization 
can be achieved by merely splitting off the trace part  
\cite{Reu96,LR1}. This has to be contrasted with the $R^2$-truncation where the
higher-derivative term introduces additional mixings between the traceless part
and $\bar{\phi}$. These cross terms do {\it not} vanish for $\alpha=1$. Hence
the complete decomposition of $\bar{h}_{\mu\nu}$ is necessary for a partial
diagonalization even in the case $\alpha=1$.

At the component field level the cross terms boil down to a purely scalar
$\bar{\sigma}$-$\bar{\phi}$ mixing term that vanishes for the spherical
harmonics $T^{l=0,m=1}$ and $T^{l=1,m}$.
Since these modes contribute to $\bar{\phi}$, but not to
$\bar{\sigma}$, we cannot directly invert the associated matrix differential 
operator $\left((\Gamma_k^{(2)})_{ij}\right)_{i,j\in\{\bar{h}^T,\bar{\xi},
\bar{\sigma},\bar{\phi}\}}$. As a way out, we split $\bar{\phi}$ according to 
eq. (\ref{split}) into $\bar{\phi}_0$ and $\bar{\phi}_1$. This has the effect 
that only mixings between the scalars $\bar{\sigma}$ and $\bar{\phi}_1$ 
survive, which can be expanded in the same set of eigenfunctions $T^{lm}$ 
starting at $l=2$.
Hence the resulting matrix differential operator 
$\left((\Gamma_k^{(2)})_{ij}\right)_{i,j\in\{\bar{h}^T,\bar{\xi},\bar{\phi}_0,
\bar{\sigma},\bar{\phi}_1\}}$ is invertible. However, it should be noted that
this additional split of $\bar{\phi}$ leads to a slightly modified flow 
equation since it affects the matrix structure of this operator. In fact, the 
summation in the gravitational sector of eq. (\ref{truncflow}) now runs over 
the set of fields $\{\bar{h}^T,\bar{\xi},\bar{\phi}_0,\bar{\sigma},
\bar{\phi}_1\}$, with $({\cal R}_k)_{\bar{\phi}_0\bar{\phi}_0}\equiv
({\cal R}_k)_{\bar{\phi}_1\bar{\phi}_1}\equiv({\cal R}_k)_{\bar{\phi}
\bar{\phi}}$ and $({\cal R}_k)_{\bar{\sigma}\bar{\phi}_1}\equiv
({\cal R}_k)_{\bar{\sigma}\bar{\phi}}$.

As a next step we calculate the contributions from the ghost fields appearing
on the RHS of eq. (\ref{truncflow}). For this purpose we insert the family of
spherical background spaces $S^d$ into $S_{\rm gh}$ and set $g_{\mu\nu}
=\bar{g}_{\mu\nu}$. Then we use eq. (\ref{T}) to decompose the ghost fields.
This yields
\begin{eqnarray}
\label{ghquad}
S_{\rm gh}\left[0,v,\bar{v};g\right]=\sqrt{2}\int 
d^dx\,\sqrt{g}\left\{\bar{v}_\mu^T\left[-D^2-\frac{R}{d}\right]
v^{T\mu}+\bar{\varrho}\left[-D^2-2\frac{R}{d}\right]\varrho\right\}\;.
\end{eqnarray}

From now on the bars are omitted from the metric, the curvature and the
operators $D^2$ and $P_k$. 

Before we can continue with our evaluation we have to specify the precise form
of the cutoff operators. Adapting them to $\Gamma_k^{(2)}$ and 
$S_{\rm gh}^{(2)}$ of eqs. (\ref{gravquad}), (\ref{ghquad}) by applying the 
rule (\ref{rule}) leads precisely to the structure (\ref{cutoff}) with the 
following choices for the ${\cal X}_k$'s, ${\cal Y}_k$'s and ${\cal Z}_k$'s: 
\begin{eqnarray}
\label{49}
& &{\cal X}_k^{\bar{\phi}_1\bar{\sigma}}={\cal X}_k^{\bar{\sigma}\bar{\sigma}}=
{\cal X}_k^{\bar{\phi}_0\bar{\phi}_0}={\cal X}_k^{\bar{\phi}_1\bar{\phi}_1}=
H_S(d)\,\bar{\beta}_k\;,\;\;
{\cal Y}_k^{\bar{h}^T\bar{h}^T}=-\bar{\beta}_k\;,\nonumber\\
& &{\cal Y}_k^{\bar{\phi}_1\bar{\sigma}}=
{\cal Y}_k^{\bar{\sigma}\bar{\sigma}}=2G_{S1}(d)\,\bar{\beta}_k\;,
\;\;
{\cal Y}_k^{\bar{\phi}_0\bar{\phi}_0}
={\cal Y}_k^{\bar{\phi}_1\bar{\phi}_1}=G_{S2}(d)\,\bar{\beta}_k\;,\;\;
{\cal Z}_k^{\bar{h}^T\bar{h}^T}=Z_{Nk}\;,
\nonumber\\
& &{\cal Z}_k^{\bar{\xi}
\bar{\xi}}=\frac{2}{\alpha}Z_{Nk}\;,\;\;{\cal Z}_k^{\bar{\phi}_1\bar{\sigma}}
=C_{S2}(d,\alpha)C_{S3}(d,\alpha)Z_{Nk}\;,\;\;
{\cal Z}_k^{\bar{\sigma}\bar{\sigma}}=C_{S2}(d,\alpha)Z_{Nk}\;,\nonumber\\
& &{\cal Z}_k^{\bar{\phi}_0\bar{\phi}_0}
={\cal Z}_k^{\bar{\phi}_1\bar{\phi}_1}=C_{S2}(d,\alpha)C_{S1}(d,\alpha)
Z_{Nk} \;,\;\;{\cal Z}_k^{\bar{v}^Tv^T}={\cal Z}_k^{\bar{\varrho}\varrho}
=\sqrt{2}\;.
\end{eqnarray}
Thus, for $g_{\mu\nu}=\bar{g}_{\mu\nu}$, the nonvanishing entries of the
matrix differential operators $\Gamma_k^{(2)}+{\cal R}_k$ and
$S^{(2)}_{\rm gh}+{\cal R}_k$ take the form
\begin{eqnarray}
\label{entries}
\left(\Gamma^{(2)}_k[g,g]+{\cal R}_k\right)_{\bar{h}^T\bar{h}^T}
&=&Z_{Nk}\kappa^2\left(P_k+A_T(d)\,R-2\bar{\lambda}_k\right)
+\bar{\beta}_k\left(-R\,P_k+G_T(d)\,R^2\right)\;,\nonumber\\
\left(\Gamma^{(2)}_k[g,g]+{\cal R}_k\right)_{\bar{\xi}\bar{\xi}}
&=&\frac{2}{\alpha}
Z_{Nk}\kappa^2\left(P_k+A_V(d,\alpha)\,R-2\alpha
\bar{\lambda}_k\right)+G_V(d)\,\bar{\beta}_k\,R^2\;,\nonumber\\
\left(\Gamma^{(2)}_k[g,g]+{\cal R}_k\right)_{\bar{\sigma}\bar{\sigma}}
&=&C_{S2}(d,\alpha)\,Z_{Nk}\kappa^2\left(P_k+A_{S2}(d,\alpha)\,R
+B_{S2}(d,\alpha)\,\bar{\lambda}_k\right)\nonumber\\
& &+\bar{\beta}_k\left(H_S(d)\,P_k^2
+G_{S1}(d)\,\left(2R\,P_k-R^2\right)\right)\;,\nonumber\\
\left(\Gamma^{(2)}_k[g,g]+{\cal R}_k\right)_{\bar{\phi}_1\bar{\sigma}}
&=&\left(\Gamma^{(2)}_k[g,g]+{\cal R}_k\right)_{\bar{\sigma}\bar{\phi}_1}
\nonumber\\
&=&\bigg[C_{S2}(d,\alpha)\,C_{S3}(d,\alpha)\,Z_{Nk}\kappa^2
+\bar{\beta}_k\left(H_S(d)\,P_k+2G_{S1}\,R\right)\bigg]
\nonumber\\
& &\times\sqrt{P_k}\sqrt{P_k-\frac{R}{d-1}}\;,\nonumber\\
\left(\Gamma^{(2)}_k[g,g]+{\cal R}_k\right)_{\bar{\phi}_0\bar{\phi}_0}
&=&\left(\Gamma^{(2)}_k[g,g]+{\cal R}_k\right)_{\bar{\phi}_1\bar{\phi}_1}
\nonumber\\
&=&C_{S2}(d,\alpha)\,C_{S1}(d,\alpha)\,Z_{Nk}\kappa^2
\left(P_k+A_{S1}(d,\alpha)\,R
+B_{S1}(d,\alpha)\,\bar{\lambda}_k\right)\nonumber\\
& &+\bar{\beta}_k\left(H_S(d)\,P_k^2
+G_{S2}(d)\,R\,P_k+G_{S3}(d)\,R^2\right)\;,\nonumber\\
\left(S_{\rm gh}^{(2)}[g,g]+{\cal R}_k\right)_{\bar{v}^Tv^T}
&=&-\left(S_{\rm gh}^{(2)}[g,g]+{\cal R}_k\right)_{v^T\bar{v}^T}
=\sqrt{2}\left[P_k-\frac{R}{d}\right]\;,\nonumber\\
\left(S_{\rm gh}^{(2)}[g,g]+{\cal R}_k\right)_{\bar{\varrho}\varrho}
&=&-\left(S_{\rm gh}^{(2)}[g,g]+{\cal R}_k\right)_{\varrho\bar{\varrho}}
=\sqrt{2}\left[P_k-2\frac{R}{d}\right]\;.
\end{eqnarray}
For notational simplicity we set 
$\left(S_{\rm gh}^{(2)}[0,v,\bar{v};g]\right)_{\psi_1\psi_2}\equiv
\left(S_{\rm gh}^{(2)}[g,g]\right)_{\psi_1\psi_2}$ with $\psi_1,\psi_2\in
\bar{I}_2$.

Now we are in a position to write down the RHS of the flow equation with
$g_{\mu\nu}=\bar{g}_{\mu\nu}$. We shall denote it ${\cal S}_k(R)$ in the 
following. Obviously we need the inverse operators
$(\Gamma_k^{(2)}+{\cal R}_k)^{-1}$ and $(S^{(2)}_{\rm gh}+{\cal R}_k)^{-1}$. 
This inversion is carried out in appendix \ref{inv}. Inserting the inverse 
operators into ${\cal S}_k(R)$ leads to the somewhat complicated result
\begin{eqnarray}
\label{squareflow}
\lefteqn{{\cal S}_k(R)=
{\rm Tr}_{(2ST^2)}\bigg[\left(P_k+A_T(d)\,R-2\bar{\lambda}_k
+(Z_{Nk}\kappa^2)^{-1}\bar{\beta}_k\left(-R\,P_k+G_T(d)\,R^2\right)
\right)^{-1}}\nonumber\\
& &\times\Big({\cal N}-(Z_{Nk}\kappa^2)^{-1}\bar{\beta}_k\,R\,{\cal T}_1
\Big)\bigg]\nonumber\\
& &+{\rm Tr}_{(1T)}'\left[\left(P_k+A_V(d,\alpha)R-2\alpha
\bar{\lambda}_k+G_V(d)\,(Z_{Nk}\kappa^2)^{-1}\bar{\beta}_k\,R^2\right)^{-1}
{\cal N}\right]\nonumber\\
& &+{\rm Tr}_{(0)}''\Bigg[
\bigg\{\Big(C_{S2}(d,\alpha)\left(P_k+A_{S2}(d,\alpha)\,R
+B_{S2}(d,\alpha)\,\bar{\lambda}_k\right)\nonumber\\
& &+(Z_{Nk}\kappa^2)^{-1}\bar{\beta}_k\left(H_S(d)\,P_k^2
+G_{S1}(d)\,\left(2R\,P_k-R^2\right)\right)\Big)\nonumber\\
& &\times\Big(C_{S2}(d,\alpha)\,C_{S1}(d,\alpha)
\left(P_k+A_{S1}(d,\alpha)\,R
+B_{S1}(d,\alpha)\,\bar{\lambda}_k\right)\nonumber\\
& &+(Z_{Nk}\kappa^2)^{-1}\bar{\beta}_k\left(H_S(d)\,P_k^2
+G_{S2}(d)\,R\,P_k+G_{S3}(d)\,R^2\right)\Big)\nonumber\\
& &-\Big(C_{S2}(d,\alpha)\,C_{S3}(d,\alpha)
+(Z_{Nk}\kappa^2)^{-1}\bar{\beta}_k\left(H_S(d)\,P_k+2G_{S1}\,R\right)\Big)^2\,
P_k\Big(P_k-\frac{R}{d-1}\Big)\bigg\}^{-1}\nonumber\\
& &\times\bigg\{\Big(C_{S2}(d,\alpha)\left(P_k+A_{S2}(d,\alpha)\,R
+B_{S2}(d,\alpha)\,\bar{\lambda}_k\right)\nonumber\\
& &+(Z_{Nk}\kappa^2)^{-1}\bar{\beta}_k\left(H_S(d)\,P_k^2
+G_{S1}(d)\,\left(2R\,P_k-R^2\right)\right)\Big)\nonumber\\
& &\times\Big(C_{S2}(d,\alpha)\,C_{S1}(d,\alpha)
\,{\cal N}
+(Z_{Nk}\kappa^2)^{-1}\bar{\beta}_k\left(H_S(d)\,{\cal T}_2
+G_{S2}(d)\,R\,{\cal T}_1\right)\Big)\nonumber\\
& &+\Big(C_{S2}(d,\alpha)\,C_{S1}(d,\alpha)
\left(P_k+A_{S1}(d,\alpha)\,R
+B_{S1}(d,\alpha)\,\bar{\lambda}_k\right)\nonumber\\
& &+(Z_{Nk}\kappa^2)^{-1}\bar{\beta}_k\left(H_S(d)\,P_k^2
+G_{S2}(d)\,R\,P_k+G_{S3}(d)\,R^2\right)\Big)\nonumber\\
& &\times\Big(C_{S2}(d,\alpha)\,{\cal N}
+(Z_{Nk}\kappa^2)^{-1}\bar{\beta}_k\left(H_S(d)\,{\cal T}_2
+2G_{S1}(d)\,R\,{\cal T}_1\right)\Big)\nonumber\\
& &+2\Big(C_{S2}(d,\alpha)\,C_{S3}(d,\alpha)
+(Z_{Nk}\kappa^2)^{-1}\bar{\beta}_k\left(H_S(d)\,P_k+2G_{S1}\,R\right)\Big)\,
\sqrt{P_k}\sqrt{P_k-\frac{R}{d-1}}\nonumber\\
& &\times \frac{1}{2Z_{Nk}\kappa^2}\,\partial_t\Bigg(
H_S(d)\,\bar{\beta}_k\Bigg[P_k\,\sqrt{\left(P_k
+\frac{d}{d-1}\,D_\mu\,R^{\mu\nu}\,D_\nu(-D^2)^{-1}\right)
P_k}\nonumber\\
& &+D^2\,\sqrt{\left(D^2\right)^2+\frac{d}{d-1}
\,D_\mu\,R^{\mu\nu}\,D_\nu}\Bigg]\nonumber\\
& &+\Big(2G_{S1}(d)\,\bar{\beta}_k\,R+
C_{S2}(d,\alpha)\,C_{S3}(d,\alpha)\,Z_{Nk}\kappa^2\Big)\nonumber\\
& &\times
\Bigg[\sqrt{\left(P_k+\frac{d}{d-1}\,D_\mu\,R^{\mu\nu}\,D_\nu(-D^2)^{-1}\right)
P_k}-\sqrt{\left(D^2\right)^2+\frac{d}{d-1}
\,D_\mu\,R^{\mu\nu}\,D_\nu}\Bigg]\Bigg)\bigg\}\Bigg]\nonumber\\
& &-2{\rm Tr}_{(1T)}\left[\left(P_k-\frac{R}{d}\right)^{-1}{\cal N}_0\right]
-2{\rm Tr'}_{(0)}\left[\left(P_k-2\frac{R}{d}\right)^{-1}{\cal N}_0\right]
\nonumber\\
& &+\frac{1}{2Z_{Nk}\kappa^2}\sum\limits_{l=0}^{1}\Bigg[
D_l(d,0)\bigg\{C_{S2}(d,\alpha)\,C_{S1}(d,\alpha)\,\Big(\Lambda_l(d,0)+
k^2R^{(0)}(\Lambda_l(d,0)/k^2)\nonumber\\
& &+A_{S1}(d,\alpha)\,R
+B_{S1}(d,\alpha)\,\bar{\lambda}_k\Big)\nonumber\\
& &+(Z_{Nk}\kappa^2)^{-1}\bar{\beta}_k\Big(H_S(d)\,\left(\Lambda_l(d,0)+
k^2R^{(0)}(\Lambda_l(d,0)/k^2)\right)^2\nonumber\\
& &+G_{S2}(d)\,R\left(\Lambda_l(d,0)+
k^2R^{(0)}(\Lambda_l(d,0)/k^2)\right)+G_{S3}(d)\,R^2\Big)\bigg\}^{-1}
\nonumber\\
& &\times\partial_t\bigg\{
C_{S2}(d,\alpha)\,C_{S1}(d,\alpha)\,Z_{Nk}\kappa^2
\,k^2R^{(0)}(\Lambda_l(d,0)/k^2)\nonumber\\
& &+\bar{\beta}_k\Big(H_S(d)\,
\left(2\Lambda_l(d,0)\,k^2R^{(0)}(\Lambda_l(d,0)/k^2)
+k^4R^{(0)}(\Lambda_l(d,0)/k^2)^2\right)\nonumber\\
& &+G_{S2}(d)\,R\,k^2R^{(0)}(\Lambda_l(d,0)/k^2)\Big)\bigg\}
\Bigg]
\end{eqnarray}
The new quantities ${\cal N}$, ${\cal N}_0$, ${\cal T}_1$ and ${\cal T}_2$  
introduced in eq. (\ref{squareflow}) are defined as 
\begin{eqnarray}
\label{ntop}
{\cal N}&\equiv&\left(2 Z_{Nk}\right)^{-1}
\partial_t \left[Z_{Nk}\,k^2R^{(0)}(-D^2/k^2)\right]\nonumber\\
&=&\left[1-\frac{1}{2}\eta_N(k)\right]k^2R^{(0)}(-D^2/k^2)
+D^2R^{(0)'}(-D^2/k^2)\;,\nonumber\\
{\cal N}_0&\equiv&2^{-1}
\partial_t \left[k^2R^{(0)}(-D^2/k^2)\right]
=k^2R^{(0)}(-D^2/k^2)+D^2R^{(0)'}(-D^2/k^2)\;,\nonumber\\
{\cal T}_1&\equiv&\left(2\bar{\beta}_k\right)^{-1}
\partial_t \left[\bar{\beta}_k\,k^2R^{(0)}(-D^2/k^2)\right]\nonumber\\
&=&\left[1-\frac{1}{2}\eta_{\beta}(k)\right]k^2R^{(0)}(-D^2/k^2)
+D^2R^{(0)'}(-D^2/k^2)\;,\nonumber\\
{\cal T}_2&\equiv&\left(2\bar{\beta}_k\right)^{-1}
\partial_t \left[\bar{\beta}_k\left(-2D^2\,k^2R^{(0)}(-D^2/k^2)
+k^4R^{(0)}(-D^2/k^2)^2\right)\right]\nonumber\\
&=&2 P_k\left[k^2R^{(0)}(-D^2/k^2)
+D^2R^{(0)'}(-D^2/k^2)\right]\nonumber\\
& &-\frac{1}{2}\eta_{\beta}(k)\left(-D^2\,k^2R^{(0)}(-D^2/k^2)
+k^4R^{(0)}(-D^2/k^2)^2\right)\;.
\end{eqnarray}
Here
\begin{eqnarray}
\label{anomdim1}
\eta_N(k)\equiv -\partial_t \ln Z_{Nk}
\end{eqnarray}
and
\begin{eqnarray}
\label{anomdim2}
\eta_{\beta}(k)\equiv -\partial_t \ln \bar{\beta}_k
\end{eqnarray}
are the anomalous dimensions of the operators $\int d^dx\,\sqrt{g}R$ and 
$\int d^dx\,\sqrt{g}R^2$, respectively. Furthermore, the prime at $R^{(0)}$ 
denotes the derivative with respect to the argument.

In eq. (\ref{squareflow}) we refined our notation concerning the primes at the 
traces. From now on one prime indicates that the mode corresponding to the
lowest eigenvalue has to be excluded, while two primes indicate the subtraction
of the contributions from the lowest two eigenvalues. The
subscripts at the traces describe on which kind of field the 
operators under the traces act. We use the subscripts $(0)$, $(1T)$ and 
$(2ST^2)$ for spin-0 fields, transverse spin-1 fields, 
and symmetric transverse traceless spin-2 fields, respectively. 

\subsection{The system of flow equations for $\lambda_k$, $g_k$ and $\beta_k$}
\label{3S4D}
Next we derive the flow equations for the couplings. In order to make the
rather complicated calculations feasible we are forced to work from now on
in the technically convenient gauge $\alpha=1$. 
Here we merely summarize the main steps, 
the details of the calculation can be found in appendix \ref{trace}.

By expanding ${\cal S}_k(R)$ where $R\propto r^{-2}$ with respect to $r$ and 
evaluating the traces 
by means of heat kernel techniques we extract those pieces from the RHS
of the flow equation, eq. (\ref{squareflow}), which are proportional to the 
appropriate powers of the
radius, i.e. $r^d\propto\int d^dx\,\sqrt{g}$, $r^{d-2}\propto
\int d^dx\,\sqrt{g}R$, and $r^{d-4}\propto\int d^dx\,\sqrt{g}R^2$.
Then we equate the result to the LHS, eq. (\ref{LHS}), and compare the 
coefficients of the various powers of $r$. This leads to a system of coupled
differential equations for $\bar{\lambda}_k$, $Z_{Nk}$ and $\bar{\beta}_k$.

In order to present it in a transparent manner
we introduce the dimensionless running cosmological constant
\begin{eqnarray}
\label{lambda}
\lambda_k\equiv k^{-2}\,\bar{\lambda}_k\;,
\end{eqnarray}
the dimensionless running Newton constant
\begin{eqnarray}
\label{g}
g_k\equiv k^{d-2}\,G_k\equiv k^{d-2}\,Z_{Nk}^{-1}\,\bar{G}\;,
\end{eqnarray}
and the dimensionless running $R^2$-coupling
\begin{eqnarray}
\label{beta}
\beta_k\equiv k^{4-d}\,\bar{\beta}_k\;.
\end{eqnarray}
$G_k\equiv\bar{G}/Z_{Nk}$ denotes the dimensionful running Newton constant. 

In terms of the couplings $\lambda_k$, $g_k$, and $\beta_k$, our final result
for the 3-dimensional flow equation reads
\begin{eqnarray}
\label{del}
\partial_t\lambda_k&=&\mbox{\boldmath$\beta$}_\lambda(\lambda_k,g_k,\beta_k;d)
\nonumber\\
&\equiv & A_1(\lambda_k,g_k,\beta_k;d)+\eta_N(k)\,A_2(\lambda_k,g_k,
\beta_k;d)+\eta_{\beta}(k)\,A_3(\lambda_k,g_k,\beta_k;d)\;,
\end{eqnarray}
\begin{eqnarray}
\label{deg}
\partial_t g_k=\mbox{\boldmath$\beta$}_g(\lambda_k,g_k,\beta_k;d)
\equiv\left[d-2+\eta_N(k)\right]g_k\;,
\end{eqnarray}
\begin{eqnarray}
\label{deb}
\partial_t\beta_k=\mbox{\boldmath$
\beta$}_\beta(\lambda_k,g_k,\beta_k;d)
\equiv\left[4-d-\eta_{\beta}(k)\right]\beta_k\;.
\end{eqnarray}
The anomalous dimensions are explicitly given by
\begin{eqnarray}
\label{eta}
\lefteqn{\eta_N(k)\equiv\eta_N(\lambda_k,g_k,\beta_k;d)=}\\
& &g_k\frac{B_1(\lambda_k,g_k,\beta_k;d)\left[\beta_k
+C_3(\lambda_k,g_k,\beta_k;d)
\right]-C_1(\lambda_k,g_k,\beta_k;d)B_3(\lambda_k,g_k,\beta_k;d)}
{\left[1-g_k B_2(\lambda_k,g_k,\beta_k;d)\right]\left[\beta_k
+C_3(\lambda_k,g_k,\beta_k;d)\right]+g_k C_2(\lambda_k,g_k,\beta_k;d) 
B_3(\lambda_k,g_k,\beta_k;d)}\nonumber
\end{eqnarray}
and
\begin{eqnarray}
\label{rho}
\lefteqn{\eta_{\beta}(k)\equiv\eta_{\beta}(\lambda_k,g_k,\beta_k;d)=}\\
& &-\frac{C_1(\lambda_k,g_k,\beta_k;d)\left[1-g_k 
B_2(\lambda_k,g_k,\beta_k;d)\right]+g_kB_1(\lambda_k,g_k,\beta_k;d)
C_2(\lambda_k,g_k,\beta_k;d)}{\left[1-g_k B_2(\lambda_k,g_k,\beta_k;d)\right]
\left[\beta_k+C_3(\lambda_k,g_k,\beta_k;d)\right]+g_k 
C_2(\lambda_k,g_k,\beta_k;d) B_3(\lambda_k,g_k,\beta_k;d)}\;.\nonumber
\end{eqnarray}
The three $\mbox{\boldmath$\beta$}$-functions 
$\mbox{\boldmath$\beta$}_\lambda$, $\mbox{\boldmath$\beta$}_g$ and 
$\mbox{\boldmath$\beta$}_\beta$ contain the quantities $A_i$, $B_i$, $C_i$,
$i=1,2,3$, which are extremely complicated functions of the couplings and the 
dimensionality $d$. The explicit expressions for these coefficient functions
can be found in appendix \ref{coefffnct}. They contain the new threshold 
functions $\Psi$ and $\widetilde{\Psi}$ of eqs. (\ref{psi}) and 
(\ref{psitilde})
which functionally depend on $R^{(0)}$. They generalize the familiar threshold
functions $\Phi$ and $\widetilde{\Phi}$ which occur in the Einstein-Hilbert
truncation. The three $\mbox{\boldmath$\beta$}$-functions are one of the main
results of this paper.

\section{The fixed points}
\label{3S5}
\renewcommand{\theequation}{5.\arabic{equation}}
\setcounter{equation}{0}
\subsection{Fixed points, critical exponents, and nonperturbative 
renormalizability}
\label{3S5A}
Because of its complexity it is clearly impossible to solve the system of flow
equations for $\lambda_k$, $g_k$ and $\beta_k$, eqs. (\ref{del}),
(\ref{deg}) and (\ref{deb}), exactly.
Even a numerical solution would be a formidable task. However, it is possible 
to gain important information about the general structure of the RG flow 
by looking at its fixed point structure.

Given a set of $\mbox{\boldmath $\beta$}$-functions corresponding to an 
arbitrary set of dimensionless essential couplings ${\rm g}_i(k)$, it is 
often possible to predict their scale dependence for very small
and/or very large scales $k$ by investigating their fixed points. They
are those points in the space spanned by the ${\rm g}_i$ where
all $\mbox{\boldmath $\beta$}$-functions vanish. Fixed points are 
characterized by their
stability properties. A given  eigendirection of the linearized flow is said to
be UV or IR attractive (or stable) if, for $k\rightarrow\infty$ or 
$k\rightarrow 0$, respectively, the trajectories are attracted towards the 
fixed point along this particular direction. The UV (IR) critical hypersurface 
${\cal S}_{\rm UV}$ (${\cal S}_{\rm IR}$) in the space of all couplings is 
defined to consist of all trajectories that run into a given fixed point for 
$k\rightarrow\infty$ ($k\rightarrow 0$).

In quantum field theory, fixed points play an important role in the modern 
approach to renormalization theory \cite{wilson}. At a UV fixed point the 
infinite
cutoff limit can be taken in a controlled way, the theory can be renormalized
nonperturbatively there. As for gravity, Weinberg
\cite{wein} argued that a theory described by a RG trajectory lying on a 
{\it finite-dimensional} UV critical hypersurface of some fixed point is 
presumably free from unphysical singularities. It is predictive since it 
depends only on a {\it finite} number of free (essential) parameters. In 
Weinberg's words, such a theory is {\it asymptotically safe}. Asymptotic 
safety has to be regarded as a generalized, nonperturbative version of 
renormalizability. It covers the class of perturbatively renormalizable 
theories, whose infinite cutoff limit is taken at the Gaussian fixed point 
${\rm g}_{*i}=0$, as well as those perturbatively nonrenormalizable theories 
which are described by a RG trajectory on a finite-dimensional UV critical 
hypersurface of a non-Gaussian fixed point ${\rm g}_{*i}\neq 0$ and are
nonperturbatively renormalizable therefore \cite{wein}.

Let us now consider the component form of the exact RG equation, i.e the 
system of differential equations 
\begin{eqnarray}
\label{fp0}
k\,\partial_k{\rm g}_i(k)=\mbox{\boldmath$\beta$}_i({\rm g})
\end{eqnarray} 
for a set of dimensionless essential couplings ${\rm g}(k)\equiv
\big({\rm g}_1(k),\ldots,{\rm g}_n(k)\big)$. 
In an exact treatment the number $n$ is infinite; in a
specific truncation it might be finite. We assume that ${\rm g}_*$ is a fixed 
point of eq. (\ref{fp0}), i.e. $\mbox{\boldmath$\beta$}_i({\rm g}_*)=0$ for 
all $i=1,\ldots,n$. We linearize the RG flow about ${\rm g}_*$ which leads to
\begin{eqnarray}
\label{gfp4}
k\,\partial_k\,{\rm g}_i(k)=\sum\limits_{j=1}^n B_{ij}\,\left({\rm g}_j(k)
-{\rm g}_{*j}\right)
\end{eqnarray}
where $B_{ij}\equiv\partial_j\mbox{\boldmath$\beta$}_i({\rm g}_*)$ are the 
entries of the stability matrix ${\bf B}=(B_{ij})$. Diagonalizing ${\bf B}$ 
according to $S^{-1}{\bf B}S=-{\rm diag}(\theta_1,\dots,\theta_n)$, 
$S=(V^1,\ldots,V^n)$, where $V^I$ is the right-eigenvector of ${\bf B}$ with 
eigenvalue $-\theta_I$ we have
\begin{eqnarray}
\label{gfp7}
\sum\limits_{j=1}^n B_{ij}\,V^I_j=-\theta_I\,V^I_i\;,\;\; I=1,\ldots ,n\;.
\end{eqnarray}
The general solution to eq. (\ref{gfp4}) may be written as 
\begin{eqnarray}
\label{gfp5}
{\rm g}_i(k)={\rm g}_{*i}+\sum\limits_{I=1}^n C_I\,V^I_i\,
\left(\frac{k_0}{k}\right)^{\theta_I}\;.
\end{eqnarray}
Here
\begin{eqnarray}
\label{fp1}
C_I\equiv\sum\limits_{j=1}^n(S^{-1})_{Ij}\,{\rm g}_j(k_0)
\end{eqnarray}
are arbitrary real parameters and $k_0$ is a reference scale. 

Obviously a fixed point $g_*$ is UV attractive for a given trajectory (i.e. 
attractive for $k\rightarrow\infty$) only if all its $C_I$ corresponding to 
negative $\theta_I<0$ are set to zero. Therefore the dimensionality 
$\Delta_{\rm UV}\equiv{\rm dim}({\cal S}_{\rm UV})$ of the UV critical 
hypersurface belonging to this particular fixed point equals the
number of positive $\theta_I$'s. Conversely, for a trajectory where all $C_I$ 
corresponding to positive $\theta_I$ are set to zero, ${\rm g}_*$ is an IR 
attractive fixed point (approached in the limit $k\rightarrow 0$). As a 
consequence, the IR critical hypersurface ${\cal S}_{\rm IR}$ of a fixed point
has a dimensionality $\Delta_{\rm IR}\equiv{\rm dim}({\cal S}_{\rm IR})$ which
equals the number of negative $\theta_I$'s. 

In a slight abuse of language we 
shall refer to the $\theta_I$'s as the {\it critical exponents}. 

Strictly speaking, the solution (\ref{gfp5}) and its above interpretation
is valid only in such cases where all eigenvalues $-\theta_I$ are real, 
which is
not  guaranteed since the matrix $\bf B$ is not symmetric in general. If 
complex eigenvalues occur one has to consider complex $C_I$'s and to take the 
real part of eq. (\ref{gfp5}), see below. Then the real parts of the critical 
exponents determine which directions in coupling constant space 
are attractive or repulsive.

At this point it is necessary to discuss the impact a change of the cutoff
scheme has on the scaling behavior. Since the path integral for $\Gamma_k$ 
depends on the cutoff scheme, i.e. on the $\Delta_k S$
chosen, it is clear that generically the $k$-dependent couplings and their 
fixed point values are scheme dependent. Hence a variation of the 
cutoff scheme, i.e. of ${\cal R}_k$, induces a change in the 
corresponding ${\bf B}$-matrix. So one might naively expect that also its 
eigenvalues, the critical exponents, are scheme dependent. In fact, this is 
not the case. According to the general theory of critical phenomena and a 
recent reanalysis in the framework of the exact RG equations \cite{kana} any 
variation of the cutoff scheme can be generated by a specific coordinate 
transformation in the space of couplings, with the cutoff held fixed. Such 
transformations leave the eigenvalues of the ${\bf B}$-matrix invariant,
so that the critical behavior near the corresponding fixed point 
is universal. The positions of fixed points are scheme dependent but their
(non)existence and the qualitative structure of the RG flow are 
${\cal R}_k$-independent features. Quantities like the $\theta_I$'s which are
${\cal R}_k$-independent are called {\it universal}. Their residual scheme
dependence present in an approximate treatment (truncation, etc.) can be used
in order to judge the quality of the approximation.
A truncation can be considered reliable only if it predicts
the same fixed point structure for all admissible choices of ${\cal R}_k$.

In the context of the $R^2$-truncation the space of couplings is
parametrized by ${\rm g}_1=\lambda$, ${\rm g}_2=g$ and ${\rm g}_3=\beta$. The
$\mbox{\boldmath
$\beta$}$-functions occurring in the three flow equations
\begin{eqnarray}
\label{fp2}
\partial_t\lambda_k=\mbox{\boldmath$\beta$}_\lambda(\lambda_k,g_k,\beta_k)\;,
\;\;\;\partial_t g_k=\mbox{\boldmath$\beta$}_g(\lambda_k,g_k,\beta_k)\;,
\;\;\;\partial_t\beta_k=\mbox{\boldmath$\beta$}_\beta(\lambda_k,g_k,\beta_k)
\end{eqnarray}
are given in eqs. (\ref{del}), (\ref{deg}) and (\ref{deb}), respectively.

In ref. \cite{LR1} the fixed point structure of the pure Einstein-Hilbert
truncation was investigated. In this case the 
$\mbox{\boldmath$\beta$}$-functions were found to have both a trivial 
zero at $\lambda_*=g_*=0$, referred to as the {\bf Gaussian fixed point}, and 
a {\bf non-Gaussian fixed point} at $\lambda_*\neq 0$, $g_*\neq 0$. As we will
see in subsection \ref{3S5B}, the Gaussian fixed point is {\it not} present 
any more in the generalized truncation.
This has to be contrasted with the non-Gaussian fixed point which is 
found with the $R^2$-truncation, too. In subsection \ref{3S5C} we study its 
cutoff
dependence and the cutoff dependence of the associated critical exponents
employing the above $\mbox{\boldmath$\beta$}$-functions with the families of
shape functions (\ref{expshape}) or (\ref{supp}) inserted. 

\subsection{The fate of the Gaussian fixed point}
\label{3S5B}
In this subsection we study the fate of the Gaussian fixed point 
found in the context of the pure Einstein-Hilbert truncation. In \cite{LR1}
we investigated the 2-dimensional RG flow near this fixed point
$(\lambda_*,g_*)=(0,0)$ and discussed its stability properties. It is an
important question how the situation changes by enlarging the parameter
space. 

Quite remarkably, we find that in the 3-dimensional $\lambda$-$g$-$\beta-$space
of the $R^2$-truncation there is no Gaussian fixed point, i.e. 
$(\lambda,g,\beta)=(0,0,0)$ is not a simultaneous zero of all three 
$\mbox{\boldmath$\beta$}$-functions. While $\mbox{\boldmath$\beta$}_\lambda$ 
and $\mbox{\boldmath$\beta$}_g$ vanish at the origin
$(\lambda,g,\beta)=(0,0,0)$, 
setting $\lambda_k=g_k=0$ in $\mbox{\boldmath$\beta$}_\beta$ leads to
\begin{eqnarray}
\label{unigamma}
\mbox{\boldmath$\beta$}_\beta(0,0,\beta_k;d)=
\mbox{\boldmath$\beta$}_\beta(0,0,0;d)=\gamma_d\;\;\;\forall\beta_k\;.
\end{eqnarray}
The nonzero constant $\gamma_d$ is given by
\begin{eqnarray}
\label{gamma_d}
\gamma_d=(4\pi)^{-\frac{d}{2}}\left\{h_{31}(d)\,\Phi^1_{d/2-2}(0)
+h_{32}(d)\,\Phi^2_{d/2-1}(0)+h_{33}(d)\,\Phi^3_{d/2}(0)\right\}\;.
\end{eqnarray}
Here the $h_i$'s are defined as in subsection \ref{betacoeff}.
In $d=4$ dimensions
\begin{eqnarray}
\label{unicon}
\gamma_4=\frac{419}{1080}\,(4\pi)^{-2}
\end{eqnarray}
is a universal quantity
since $\Phi^1_0(0)=\Phi^2_1(0)=2\Phi^3_2(0)=1$ independently of the cutoff,
see appendix \ref{threshprop}.

Although there is no fixed point at the origin of the parameter space
it is nevertheless very interesting to study the RG flow in the vicinity of
$(\lambda,g,\beta)=(0,0,0)$. For simplicity we restrict our considerations to
the case $d>2$. Expanding $(\mbox{\boldmath$\beta$}_\lambda,
\mbox{\boldmath$\beta$}_g,\mbox{\boldmath$\beta$}_\beta$) about the origin we
obtain instead of (\ref{gfp4}) the {\it inhomogeneous} system
\begin{eqnarray}
\label{lingauss}
k\,\partial_k\,{\rm g}_i(k)=\gamma_d\,\delta_{i,3}+\sum\limits_{j=1}^3
M_{ij}\,{\rm g}_j(k)\;.
\end{eqnarray}
The linearized renormalization group flow is governed by the Jacobi-Matrix 
${\bf M}=(M_{ij})$,
$M_{ij}\equiv\partial_j\mbox{\boldmath $\beta$}_i(0,0,0;d)$, which takes the
form
\begin{eqnarray}
\label{gaussianstab}
{\bf M}=\left(\begin{array}{ccc}
-2 & \nu_d\,d & 0\\ 0 & d-2 & 0\\ \varsigma_d & \tau_d & 4-d\end{array}\right)
\;.
\end{eqnarray}
Its entries follow from the expanded 
$\mbox{\boldmath$\beta$}$-functions (\ref{expandbeta}) of appendix 
\ref{approxngfp}. Here $\nu_d$, $\varsigma_d$ and 
$\tau_d$ are $d$-dependent parameters defined as
\begin{eqnarray}
\label{nu_d}
\nu_d&\equiv&(d-3)(4\pi)^{1-\frac{d}{2}}\,\Phi^1_{d/2}(0)\;,
\end{eqnarray}
\begin{eqnarray}
\label{varsigma_d}
\varsigma_d&\equiv&(4\pi)^{-\frac{d}{2}}\left\{h_{34}(d)\,\Phi^2_{d/2-2}(0)
+h_{35}(d)\,\Phi^3_{d/2-1}(0)+h_{36}(d)\,\Phi^4_{d/2}(0)\right\}\;,
\end{eqnarray}
\begin{eqnarray}
\label{tau_d}
\lefteqn{\tau_d\equiv-(4\pi)^{1-d}\Bigg\{
\left[h_{37}(d)\,\Phi^1_{d/2-1}(0)+h_{38}(d)\,\Phi^2_{d/2}(0)\right]}
\nonumber\\
& &\times\left[\frac{1}{4}h_{34}(d)\,\widetilde{\Phi}^1_{d/2-2}(0)
+\frac{1}{8}h_{35}(d)\,\widetilde{\Phi}^2_{d/2-1}(0)
+\frac{1}{12}h_{36}(d)\,\widetilde{\Phi}^3_{d/2}(0)\right]\nonumber\\
& &+\left[h_{31}(d)\,\Phi^1_{d/2-2}(0)
+h_{32}(d)\,\Phi^2_{d/2-1}(0)+h_{33}(d)\,\Phi^3_{d/2}(0)\right]\\
& &\times\left[h_{39}(d)\,\widetilde{\Phi}^0_{d/2-2}(0)
+h_{40}(d)\,\widetilde{\Phi}^1_{d/2-1}(0)
+h_{41}(d)\,\widetilde{\Phi}^2_{d/2}(0)
+h_{42}(d)\,\widetilde{\Phi}^3_{d/2+1}(0)+\frac{3}{2}\right]\Bigg\}\;.\nonumber
\end{eqnarray}

At this point it should be noted that the sub-matrix
$(M_{ij})_{i,j\in\{1,2\}}$ coincides precisely with the stability matrix
of the Gaussian fixed point which was calculated in \cite{LR1} in 
the Einstein-Hilbert truncation.

Diagonalizing the matrix (\ref{gaussianstab}) yields the (obviously universal)
eigenvalues $\vartheta_1=-2$, $\vartheta_2=d-2$ and $\vartheta_3=4-d$ which 
are associated with the eigenvectors 
\begin{eqnarray}
\label{EV}
V^1=\Big(1,0,\varsigma_d/(d-6)\Big)^{\bf T}\;,\;\; 
V^2=\Big(\nu_d,1,(\varsigma_d\nu_d+\tau_d)/(2(d-3))\Big)^{\bf T}\;,\;\; 
V^3=(0,0,1)^{\bf T}\;
\end{eqnarray}
Eq. (\ref{EV}) is valid only for $d\neq 3, 6$. In $d=3$ 
we obtain $V^1=(1,0,-\varsigma_3/3)$, $V^2=V^3=(0,0,1)$, and
in the $6$-dimensional case the eigenvectors are $V^1=V^3=(0,0,1)$,
$V^2=(\nu_6,1,(\varsigma_6\nu_6+\tau_6)/6)$. Thus in both cases the space
spanned by the eigenvectors in only 2-dimensional, i.e. they do not form a 
complete system. For all values of $d$, including $d=3$ and $d=6$, the 
solutions for $\lambda_k$ and 
$g_k$ obtained from the linearized system (\ref{lingauss}) 
assume the following form:
\begin{eqnarray}
\label{lambdag}
\lambda_k&=&\left(\lambda_{k_0}-\nu_d\,g_{k_0}\right)\left(\frac{k_0}{k}
\right)^2+\nu_d\,g_{k_0}\,\left(\frac{k}{k_0}\right)^{d-2}\;,\nonumber\\
g_k&=&g_{k_0}\,\left(\frac{k}{k_0}\right)^{d-2}\;.
\end{eqnarray}

Since the expanded $\mbox{\boldmath$\beta$}$-function 
$\mbox{\boldmath$\beta$}_g$ of eq. (\ref{expandbeta}) does not depend on 
$\lambda_k$ and $\beta_k$
up to terms of third order in the couplings we can easily  
calculate also the next-to-leading approximation for $g_k$ near the origin. In
terms of the dimensionful quantity $G_k$ this improved solution reads
\begin{eqnarray}
\label{G_kflow}
G_k=G_{k_0}\left[1-\omega_d\,G_{k_0}\left(k_0^{d-2}-k^{d-2}\right)\right]^{-1}
\end{eqnarray}
with
\begin{eqnarray}
\label{omega_d}
\omega_d\equiv-\frac{1}{d-2}\,B_1(0,0,0;d)=(4\pi)^{1-\frac{d}{2}}
\left\{h_{43}(d)\,\Phi^1_{d/2-1}(0)+h_{44}(d)\,\Phi^2_{d/2}(0)\right\}
\end{eqnarray}
a $d$-dependent parameter. It agrees with the $\omega_d$ defined
in \cite{LR1} in the context of the pure Einstein-Hilbert truncation. For 
$k\ll |\omega_d G_{k_0}|^{-1/(d-2)}$ and with the reference scale $k_0=0$ 
(which is admissible only for specific initial conditions of the cosmological
constant) eq. (\ref{G_kflow}) yields
\begin{eqnarray}
\label{G_kflow2}
G_k=G_0\left[1-\omega_d G_0 k^{d-2}+{\cal O}\left(G_0^2k^{2(d-2)}\right)\right]
\;.
\end{eqnarray}
For the dimensionful cosmological constant we obtain from eq. (\ref{lambdag})
\begin{eqnarray}
\label{gfp11}
\bar{\lambda}_k=\bar{\lambda}_{k_0}+\nu_d G_{k_0}\left(k^d-k_0^d\right)\;.
\end{eqnarray}
Eqs. (\ref{G_kflow2}) and (\ref{gfp11}) agree completely with the corresponding
results from the Einstein-Hilbert truncation.

Let us now discuss the solution for $\beta_k$. In order to derive it we start
by picking $i=3$ in (\ref{lingauss}) and rewrite the corresponding equation as
\begin{eqnarray}
\label{rewrite}
\partial_k\left[k^{d-4}\,\beta_k\right]=k^{d-5}
\left[\gamma_d+\varsigma_d\,\lambda_k+\tau_d\,g_k\right]\;.
\end{eqnarray}  
Then we insert the solutions for $g_k$ and $\lambda_k$ of eq. (\ref{lambdag}) 
into (\ref{rewrite}). The resulting differential equation may easily be
solved. For $d\neq 3,4,6$, the solution reads 
\begin{eqnarray}
\label{betan4}
\beta_k&=&\frac{\gamma_d}{d-4}+\frac{\left(\lambda_{k_0}-\nu_d\,g_{k_0}\right)
\varsigma_d}{d-6}\left(\frac{k_0}{k}\right)^2
+\frac{\nu_d\varsigma_d+\tau_d}{2(d-3)}\,g_{k_0}
\,\left(\frac{k}{k_0}\right)^{d-2}\nonumber\\
& &+\left[\beta_{k_0}-\frac{\gamma_d}{d-4}
-\frac{\left(\lambda_{k_0}-\nu_d\,g_{k_0}\right)
\varsigma_d}{d-6}-\frac{\nu_d\varsigma_d+\tau_d}{2(d-3)}\,g_{k_0}\right]
\,\left(\frac{k_0}{k}\right)^{d-4}\;.
\end{eqnarray}
The solutions in $d=3$, $d=4$, and $d=6$ can be obtained from (\ref{betan4}) by
a careful evaluation of the limits $d\rightarrow 3$, $d\rightarrow 4$, and
$d\rightarrow 6$, 
respectively. In the most interesting case of $d=4$ dimensions this leads to
the following solution:
\begin{eqnarray}
\label{beta4}
\beta_k&=&\beta_{k_0}+\frac{\left(\lambda_{k_0}-\nu_4\,g_{k_0}\right)
\varsigma_4}{2}-\frac{\nu_4\varsigma_4+\tau_4}{2}\,g_{k_0}
+\frac{419}{1080}(4\pi)^{-2}\,\ln\left(\frac{k}{k_0}\right)\nonumber\\
& &-\frac{\left(\lambda_{k_0}-\nu_4\,g_{k_0}\right)
\varsigma_4}{2}\left(\frac{k_0}{k}\right)^2
+\frac{\nu_4\varsigma_4+\tau_4}{2}\,g_{k_0}
\,\left(\frac{k}{k_0}\right)^{2}.
\end{eqnarray}
The parameters appearing in eq. (\ref{beta4}) are 
\begin{eqnarray}
\label{gaussianpara4}
\nu_4&\equiv&\frac{1}{4\pi}\,\Phi^1_{2}(0)\nonumber\\
\varsigma_4&\equiv&(4\pi)^{-2}\left\{-\frac{559}{432}
+\frac{71}{36}\,\Phi^3_1(0)+\frac{347}{24}\,\Phi^4_{2}(0)\right\}\nonumber\\
\tau_4&\equiv&(4\pi)^{-3}\Bigg\{
\left[\frac{13}{3}\,\Phi^1_1(0)+\frac{79}{3}\,\Phi^2_2(0)\right]
\left[-\frac{559}{1728}
+\frac{71}{288}\,\widetilde{\Phi}^2_{1}(0)
+\frac{347}{288}\,\widetilde{\Phi}^3_{2}(0)\right]\nonumber\\
& &+\frac{419}{1080}
\left[-\frac{299}{180}+\frac{13}{3}\,\widetilde{\Phi}^1_1(0)
+\frac{40}{3}\,\widetilde{\Phi}^2_{2}(0)\right]\Bigg\}
\end{eqnarray}
Employing the exponential shape function with $s=1$ and inserting the 
corresponding values of the $\Phi^p_n(0)$- and $
\widetilde{\Phi}^p_n(0)$-integrals given in appendix \ref{threshprop} we obtain
\begin{eqnarray}
\label{numpara}
& &\nu_4=\zeta(3)/(2\pi)\approx 0.19\;,\;\;\varsigma_4=(4\pi)^{-2}
\left\{-\frac{559}{432}+\frac{278}{4}\ln(2)-\frac{347}{24}\ln(3)\right\}
\approx 0.20\;,\nonumber\\
& &\tau_4=\frac{2817356+25(474+13\pi^2)(-559+4590 \ln(2)-2082\ln(3))}
{49766400 \pi^3}
\approx 0.0051\;.
\end{eqnarray}

Let us now analyze the RG flow near the origin of the parameter space.
Strictly speaking our analysis even extends to all points of the 
$\lambda$-$g$-$\beta-$space which satisfy $|\lambda|,|g|\ll 1$ and $|\beta|\ll
1/|g|$. This is because in any of the three $\mbox{\boldmath$\beta$}$-functions
all terms of second and higher orders in $\beta_k$ appear as products 
$g_k^n\beta_k^m$ with $n\ge m$. 

Since $\vartheta_1=-2<0$ and, for $d>2$, $\vartheta_2=d-2>0$, $\lambda_k$ of 
eq. (\ref{lambdag}) starts growing as soon as $k$ falls below $k_0$ and the
linearization breaks down, unless the couplings run along a trajectory which
satisfies
\begin{eqnarray}
\label{IRstable}
\lambda_k=\nu_d\,
g_k\;\;\;\Longleftrightarrow\;\;\;\bar{\lambda}_k=\nu_d\,G_k\,k^d
\end{eqnarray}
for sufficiently small values of $k$ \cite{frank}, with $G_k$ given by eq. 
(\ref{G_kflow}). In this case both $\lambda_k$ and $g_k$ approach
$\lambda=g=0$ in the limit $k\rightarrow 0$ as long as $|\beta_k|\ll 1/|g_k|$ 
is satisfied as well. Since $\Phi^1_{d/2}(0)$ depends on the shape function 
$R^{(0)}$, $\nu_d$ is not a universal quantity. Therefore the slope of the 
distinguished trajectories characterized by (\ref{IRstable}) is not fixed in a
universal manner. 

Eq. (\ref{IRstable}) is exactly the condition for the ``separatrix'' found 
in \cite{frank} in the context of the $4$-dimensional Einstein-Hilbert 
truncation. In the terminology of \cite{frank}, the separatrix is the ``type 
IIa trajectory'' that interpolates between the Gaussian and the non-Gaussian
fixed point of the Einstein-Hilbert truncation, thereby separating the
region of trajectories with $\lambda_{k\rightarrow 0}\rightarrow -\infty$
(type Ia) from those which hit the boundary of parameter space 
($\lambda=1/2$) for some finite value of $k$ (type IIIa).
In FIG. \ref{plotx}(a) we depict the separatrix, a type Ia and
a type IIIa trajectory in the vicinity of $(\lambda,g)=(0,0)$. 
The plot should be thought of as a projection from $\lambda$-$g$-$\beta-$space
onto its $\beta=0-$plane.

For the separatrix, the equation (\ref{G_kflow2}) for the running of $G_k$ is
valid down to $k=0$ because $g_k$ stays near the origin.
For $d\neq 2$ the parameters $\Phi^1_{d/2-1}(0)$ and $\Phi^2_{d/2}(0)$ 
appearing in $\omega_d$ are scheme dependent, and $\omega_d$ is  
nonuniversal. In the most interesting case of $d=4$,
\begin{eqnarray}
\label{omega4}
\omega_4=\frac{1}{24\pi}\left[13\Phi^1_1(0)+79\Phi^2_2(0)\right]\;.
\end{eqnarray} 
Since $\Phi^1_1(0)$ and $\Phi^2_2(0)$ are positive for any admissible shape
function\renewcommand{\baselinestretch}{1}\small\normalsize
\footnote{Using the exponential 
shape function $R^{(0)}$ with $s=1$, for instance,  we have $\Phi^1_1(0)
=\pi^2/6$, $\Phi^2_2(0)=1$ so that $\omega_4\approx 1.33$. 
Furthermore, we have $\Phi^1_2(0)=2\zeta(3)$ where $\zeta$ denotes the
Riemann zeta function, and thus $\nu_4\approx 0.19$.} 
\renewcommand{\baselinestretch}{1.5}\small\normalsize
we can infer from eq. (\ref{omega4}) that $\omega_4$ is positive. 
Thus, if we define QEG with vanishing renormalized 
cosmological constant to be the theory described by a 
trajectory in $\lambda$-$g$-$\beta-$space, whose $\lambda$- and $g$-coordinates
follow the separatrix in the limit $k\rightarrow 0$, eq. (\ref{G_kflow}) 
implies that QEG is antiscreening in the IR, i.e. $G_k$ decreases as $k$ 
increases.\vspace{5mm}

Up to now we investigated the $\beta=0$-projection of the flow on 
$\lambda$-$g$-$\beta-$space.
Next we discuss the linearized RG flow of the $\beta$-component.

\subsubsection{\boldmath$d=4$}
In $d=4$ the solution for $\beta_k$, eq. (\ref{beta4}), 
diverges in the limit $k\rightarrow 0$. If $\lambda_k\neq\nu_d\,g_k$ as $k
\rightarrow 0$, the leading divergence is quadratic in $k$. However, in this
case the linearization cannot be trusted down to arbitrarily small values of
$k$ anyhow since trajectories with $\lambda_k\neq\nu_d\,g_k$ ultimately run 
away from $(\lambda,g)=(0,0)$ for $k\rightarrow 0$.

For the distinguished trajectories which satisfy the separatrix condition
(\ref{IRstable}) for $k\rightarrow 0$, the coefficient of the term 
$\propto(k_0/k)^2$ in (\ref{beta4}) vanishes and only a logarithmic running
with a {\it universal} coefficient remains:
\begin{eqnarray}
\label{beta4.2}
\beta_k=\beta_{k_0}-\frac{\nu_4\varsigma_4+\tau_4}{2}\,g_{k_0}
+\frac{419}{1080}(4\pi)^{-2}\,\ln\left(\frac{k}{k_0}\right)
+{\cal O}(k^2)\;.
\end{eqnarray}

Since higher orders in $\beta_k$ appear 
exclusively as products $g_k^n\beta_k^m$ with $n\ge m$, the 
vanishing of $g_k^n(\ln(k/k_0))^m\propto(k/k_0)^{2n}(\ln(k/k_0))^m$ 
in the limit $k\rightarrow 0$ then implies that terms of order
$\beta_k^2$ remain negligible as $k\rightarrow 0$. As a
consequence, the linearization does not break down for $k\rightarrow 0$
although $\beta_k$ diverges in this limit. 

According to eqs. (\ref{lambdag}), (\ref{IRstable}), $\lambda_k$ and $g_k$ 
quickly approach $\lambda=g=0$ so that the
corresponding trajectories run almost along the $\beta$-axis for 
$k\rightarrow 0$, and the RG flow becomes essentially one-dimensional.
This logarithmic running of $\beta_k$ was expected on the basis
of conventional perturbation theory \cite{perturb}. We observe that $|\beta_k|$
{\it de}creases logarithmically with {\it in}creasing $k$.
This is what is usually referred to as the ``asymptotic freedom'' of the
$({\rm curvature})^2$-coupling. We emphasize, however, that according to our
results this logarithmic running occurs only close to $g=\lambda=0$ and does
not represent the true short-distance behavior of the theory. In fact, we shall
find that $\beta_k$ runs towards a fixed point value $\beta_*$ for 
$k\rightarrow\infty$.

\renewcommand{\baselinestretch}{1}
\small\normalsize
\begin{figure}[ht]
\begin{minipage}{7.9cm}
        \epsfxsize=7.9cm
        \epsfysize=5.2cm
        \centerline{\epsffile{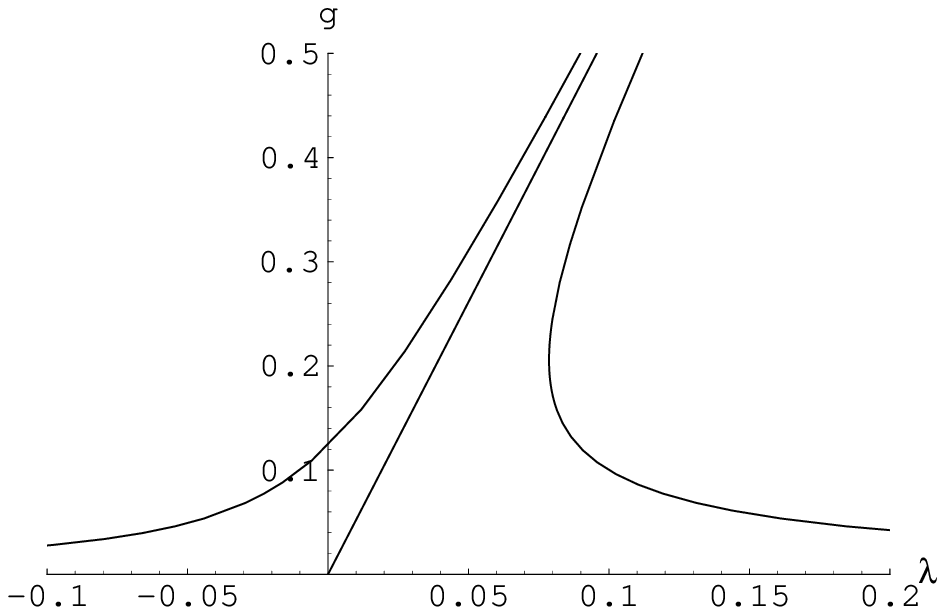}}
\centerline{(a)}
\end{minipage}
\hfill
\begin{minipage}{7.9cm}
        \epsfxsize=7.9cm
        \epsfysize=5.2cm
        \centerline{\epsffile{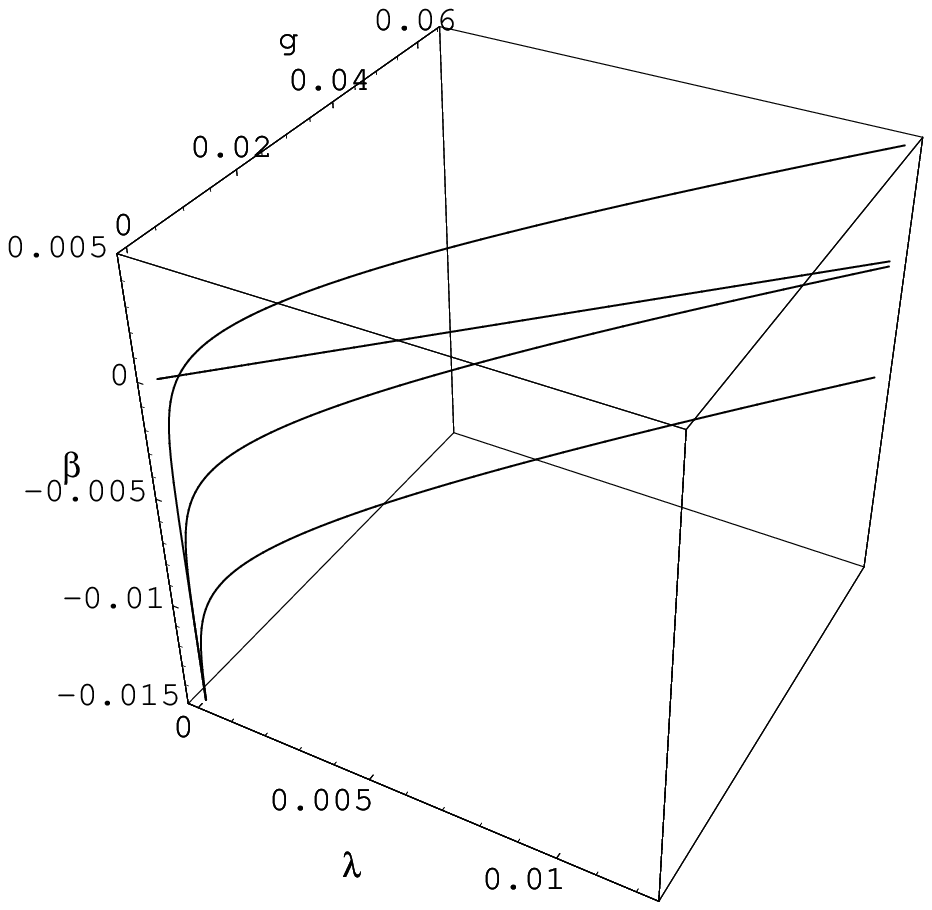}}
\centerline{(b)}
\end{minipage}
\caption{The case $d=4$: (a) Type Ia, type IIa, and type IIIa trajectories
(from left to right) obtained from the $\lambda$-$g-$projection 
(\ref{lambdag}). They coincide precisely with the corresponding trajectories of
the Einstein-Hilbert 
truncation. The type IIa trajectory is the separatrix with $\lambda_{k=0}=0$, 
which separates the region of trajectories with $\lambda_{k\rightarrow0}
\rightarrow-\infty$ (type Ia) from those running towards more positive 
$\lambda$'s (type IIIa).
(b) Three typical trajectories of the linearized 
$\lambda$-$g$-$\beta-$equation. They correspond to
different values of $\beta_{k_0}$, but all of them satisfy (\ref{IRstable}). 
We also depict their projection onto the $\lambda$-$g-$plane. It coincides 
with the separatrix of the pure Einstein-Hilbert truncation.}  
\label{plotx}
\end{figure}
\renewcommand{\baselinestretch}{1.5}
\small\normalsize

FIG. \ref{plotx}(b) shows three typical trajectories of the $R^2$-truncation
close to $(\lambda,g,\beta)=(0,0,0)$; all of them satisfy the separatrix 
condition 
(\ref{IRstable}). Their $\beta$-component diverges logarithmically towards 
$-\infty$ as $k$ goes to zero, which is due to the positive coefficient 
in front of the $\ln(k/k_0)$-term in (\ref{beta4.2}).
In this figure we also depict the common projection of the trajectories onto 
the $\lambda$-$g-$plane. 
It coincides precisely with the separatrix of the 
Einstein-Hilbert truncation \cite{frank}, i.e. the curve in FIG. 
\ref{plotx}(a) that hits
$(\lambda,g)=(0,0)$. Conversely, all trajectories of the $R^2$-truncation
satisfying (\ref{IRstable}) represent specific ``lifts'' of the separatrix
with nonvanishing $\beta$-components; they are distinguished by their 
$\beta_{k_0}$- and $g_{k_0}$-values.

\subsubsection{\boldmath$d\neq 4$}
Let us now discuss the $\beta$-evolution for $2<d\neq 4$. Again, the 
linearization breaks down 
for trajectories which do not satisfy (\ref{IRstable}) for $k\rightarrow 0$
since $|\lambda_{k\rightarrow 0}|\rightarrow\infty$ in this case. Therefore 
we restrict our considerations to the trajectories with 
$\lambda_k=\nu_d\,g_k$ for sufficiently small $k$. In this case the
second, quadratically divergent term of eq. (\ref{betan4}) drops out, and
the only powers of $k$ which occur in $\beta_k$ are $k^{d-2}$ and $k^{4-d}$. 

In $\mbox{\boldmath$2<d<4$}$, both $\vartheta_2=d-2$ and $\vartheta_3=4-d$ are
positive which 
implies that the RG trajectories considered are attracted towards 
$(\lambda,g,\beta)=(0,0,\gamma_d/(d-4))$ as $k$ is sent to zero. 

For $\mbox{\boldmath$d>4$}$, $\vartheta_3$ is negative and thus $\beta_k$ 
contains a divergent term $\propto k^{4-d}$. As a consequence, the 
coefficient of this term must vanish, if a trajectory is to hit the point
$(\lambda,g,\beta)=(0,0,\gamma_d/(d-4))$ in the limit $k\rightarrow0$. 
The distinguished trajectory which runs into this point as $k\rightarrow 0$
satisfies, for sufficiently small values of $k$,
\begin{eqnarray}
\label{IRstable2}
\beta_k&=&\frac{\nu_d\varsigma_d+\tau_d}{2(d-3)}\,g_k
=\frac{\nu_d\varsigma_d+\tau_d}{2(d-3)\nu_d}\,\lambda_k\nonumber\\
\Longleftrightarrow\;\;\bar{\beta}_k&=&
\frac{\nu_d\varsigma_d+\tau_d}{2(d-3)}\,G_k\,k^{2(d-3)}
=\frac{\nu_d\varsigma_d+\tau_d}{2(d-3)\nu_d}\,\bar{\lambda}_k\,k^{d-6}\;.
\end{eqnarray}
For all other trajectories the 
$\beta$-component diverges for $k\rightarrow 0$. However, higher orders of
$\beta_k$ are again suppressed by powers of $g_k$ and may therefore be 
neglected. (Note that $\lim_{k\rightarrow 0}g_k^n\beta_k^m\propto
\lim_{k\rightarrow 0}k^{n(d-2)-m(d-4)}=0$ for $n\ge m$.) As a consequence, the
linearization can be 
trusted down to arbitrarily small scales $k$ even in this case. The shape of
the corresponding trajectories resembles the one found in $d=4$. While 
$|\beta_k|\rightarrow\infty$, the $\lambda$- and $g$-components approach 
$\lambda=g=0$ in the limit $k\rightarrow 0$. Thus, for sufficiently small $k$,
the trajectories are almost straight lines which virtually coincide with the
$\beta$-axis.    

Having a closer look at the $\mbox{\boldmath$\beta$}$-functions one recognizes
that the IR scaling behavior in $d\neq4$ dimensions is actually 
governed by a {\bf ``quasi-Gaussian'' fixed point} at 
\begin{eqnarray}
\label{QGFP}
(\lambda_*,g_*,\beta_*)=\Big(0,0,\gamma_d/(d-4)\Big)\;.
\end{eqnarray}
The quasi-Gaussian fixed point is not present in $d=4$.
Linearizing the RG flow about this fixed point yields essentially the same 
results as our expansion about $(0,0,0)$ above. The linearized
$\mbox{\boldmath$\beta$}$-functions with stability matrix ${\bf B}$, and the 
linear solutions associated to this fixed point, may be obtained from eqs. 
(\ref{lingauss}), (\ref{gaussianstab}), (\ref{lambdag}), (\ref{betan4})
and (\ref{IRstable2}) simply by replacing $\tau_d$ with
\begin{eqnarray} 
\label{widehattau}
\widehat{\tau}_d\equiv\tau_d-\frac{2(4\pi)^{1-d}}{d-4}\,
\left(h_{39}(d)+h_{45}(d)\,\Phi^2_{d/2}(0)\right)\;.
\end{eqnarray}
In particular, ${\bf B}={\bf M}(\tau_d\rightarrow\widehat{\tau}_d)$.
The constants $\theta_I=-\vartheta_I$ assume the meaning of critical exponents
now, and their signs determine the dimensionality $\Delta_{\rm IR}$ of the
(truncated) IR critical hypersurface ${\cal S}_{\rm IR}$ of the quasi-Gaussian
fixed point. 

In $2<d<4$ we have 
one positive critical exponent $\theta_1>0$ and two negative critical exponents
$\theta_2,\theta_3<0$. Therefore, within the truncation, $\Delta_{\rm IR}=2$, 
as suggested by the corresponding solutions
discussed above. 

In $d>4$, $\theta_1$ and $\theta_3$ are positive and 
$\theta_2$ is negative. Hence, in this case $\Delta_{\rm IR}=1$, i.e. 
${\cal S}_{\rm IR}$ consists of a single trajectory. For sufficiently small
values of $k$ this IR critical trajectory is given by eq. (\ref{IRstable2})
with $\tau_d$ replaced with $\widehat{\tau}_d$.
Since the parameters $\nu_d$, $\varsigma_d$ and $\widehat{\tau}_d$ contain 
$R^{(0)}$-dependent integrals $\Phi_n^p(0)$, $\widetilde{\Phi}^p_n(0)$,
they are not universal. Therefore the slopes in both directions of 
the distinguished trajectory (\ref{IRstable2}) are not fixed in a universal 
manner. This is in accordance with the general expectation that the 
eigenvalues of ${\bf B}$ should be universal, but not its eigenvectors.

\renewcommand{\baselinestretch}{1}
\small\normalsize
\begin{figure}[ht]
\begin{minipage}{7.9cm}
        \epsfxsize=7.9cm
        \epsfysize=5.2cm
        \centerline{\epsffile{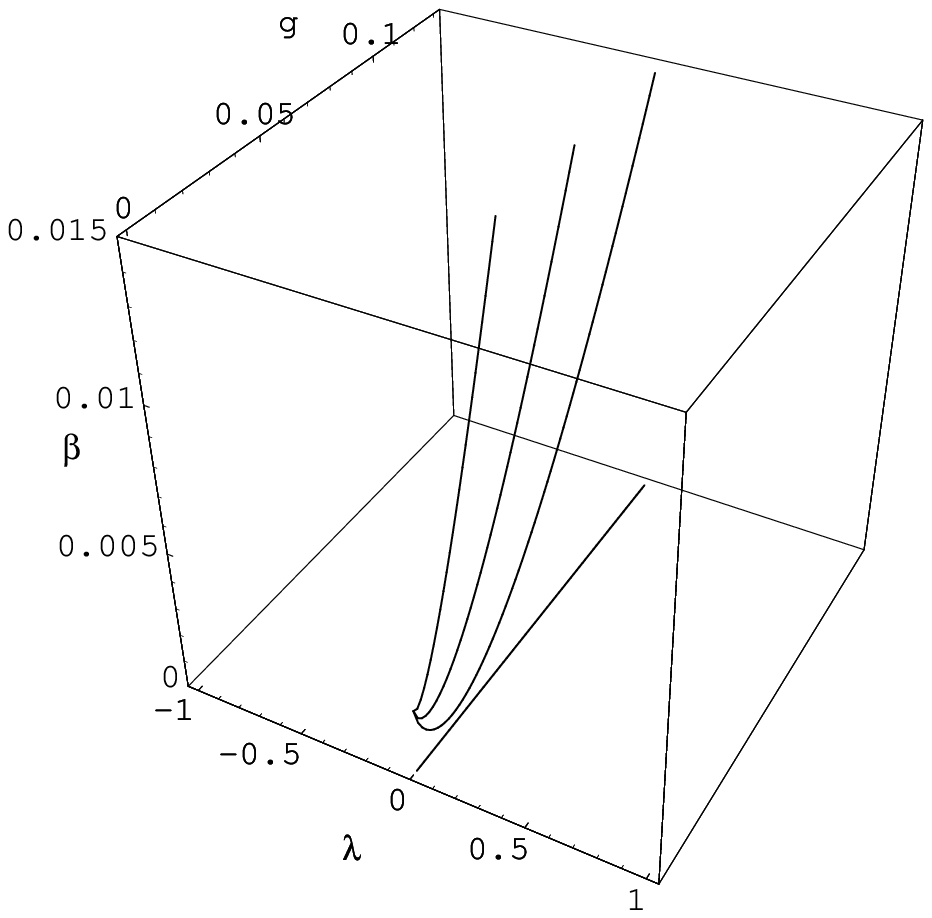}}
\centerline{(a)}
\end{minipage}
\hfill
\begin{minipage}{7.9cm}
        \epsfxsize=7.9cm
        \epsfysize=5.2cm
        \centerline{\epsffile{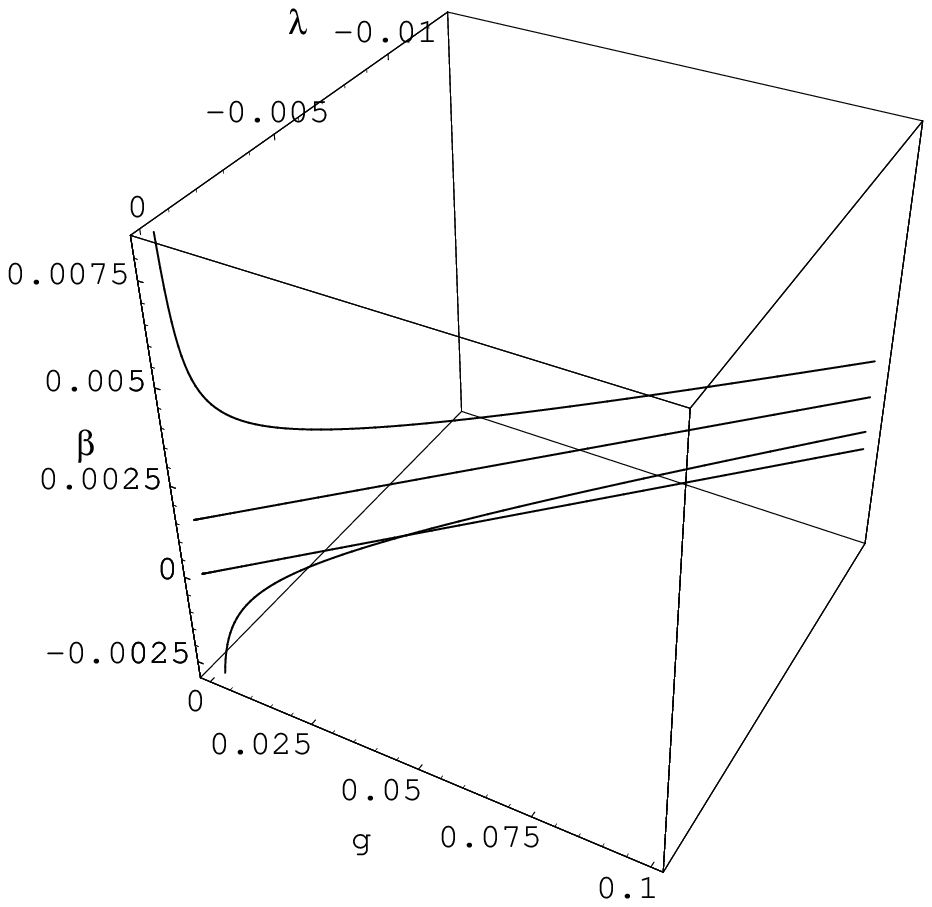}}
\centerline{(b)}
\end{minipage}
\caption{(a) The case $d=3$: three typical trajectories of the linearized flow.
They correspond to different values of
$\beta_{k_0}$, but all of them satisfy (\ref{IRstable}). As a consequence, all
3 trajectories hit the quasi-Gaussian fixed point for $k\rightarrow 0$. 
(b) The case $d=5$: three typical trajectories  
of the linearized flow corresponding to different values of
$\beta_{k_0}$. All of them satisfy (\ref{IRstable}), but in contrast to the 
other two curves, the one in the middle satisfies also (\ref{IRstable2}). As a 
consequence, it hits the quasi-Gaussian fixed point for $k\rightarrow 0$. In
both (a) and (b) we also depict the projection of the curves onto the 
$\beta=0-$plane.}  
\label{ploty}
\end{figure}
\renewcommand{\baselinestretch}{1.5}
\small\normalsize

We illustrate our results for $d\neq4$ in FIG. \ref{ploty}.  In FIG. 
\ref{ploty}(a) we consider $d=3$ and in FIG. 
\ref{ploty}(b) the $5$-dimensional case. Each figure shows three typical 
trajectories in the vicinity of the quasi-Gaussian fixed point. All of them
satisfy (\ref{IRstable}), so that in both $d=3$ and $d=5$ the projections of 
the 3 trajectories onto the $\beta=0-$plane coincide with the 
separatrix. 
 
In $d=3$, FIG. \ref{ploty}(a), all 3 trajectories
hit the fixed point, independently of their $\beta_{k_0}$-value. As we already
pointed out above, the quasi-Gaussian fixed point in $d<4$ is IR attractive 
for {\it all} trajectories satisfying (\ref{IRstable}). In $d>4$ this is no
longer the case. This is confirmed by FIG. \ref{ploty}(b) for $d=5$. Here only
one of the trajectories hits
the quasi-Gaussian fixed point for $k\rightarrow 0$, and this is precisely the
one which satisfies the additional condition (\ref{IRstable2}). The other two 
trajectories shown in FIG. \ref{ploty}(b) correspond to $\beta_{k_0}$-values 
which are different from the one in (\ref{IRstable2}) and thus their 
$\beta$-component diverges in the limit $k\rightarrow 0$. Depending on the 
$\beta_{k_0}$-value, $\beta_k$ runs towards $+\infty$ or $-\infty$.

\subsection{The non-Gaussian fixed point}
\label{3S5C}
Now we turn to the nontrivial simultaneous zeros of the set of
$\mbox{\boldmath$\beta$}$-functions $\{\mbox{\boldmath$\beta$}_\lambda,
\mbox{\boldmath$\beta$}_g,\mbox{\boldmath$\beta$}_\beta\}$ given by eqs. 
(\ref{del}), (\ref{deg}), (\ref{deb}). Such 
non-Gaussian fixed points with $\lambda_*$, $g_*$, $\beta_*$ all different from
zero have the anomalous dimensions 
\begin{eqnarray}
\label{ngfp0}
\eta_{N*}=2-d\;,\;\;\;\;\eta_{\beta*}=d-4
\end{eqnarray}
which follow immediately from eqs. (\ref{deg}) and (\ref{deb}).

\subsubsection{Results obtained from the pure Einstein-Hilbert truncation}
\label{3S5.3.1}
In $d=4$ dimensions, and for the cutoff of the type A introduced in \cite{LR1},
the non-Gaussian fixed point of the pure Einstein-Hilbert truncation was first
discussed in \cite{souma1,bh2}, and in ref. \cite{souma2} the $\alpha$- and 
$R^{(0)}$-dependence of its projection $(0,g_*)$ onto the $g$-direction has 
been investigated. However, since for $\alpha\neq 1$ the cutoff of type A 
was defined in \cite{sven} by an ad hoc modification of the standard one-loop 
determinants it is not clear whether it can be derived from an action 
$\Delta_k S$, except for the case $\alpha=1$ \cite{Reu96}. Since a 
specification of $\Delta_k S$ is indispensable for the actual construction of 
$\Gamma_k$, the status of the results derived in \cite{souma2} is somewhat 
unclear. In ref. \cite{LR1} we performed a comprehensive analysis of the fixed 
point properties using different cutoffs of type B, for which a $\Delta_k S$ 
is known to exist. In particular, we investigated the cutoff scheme dependence
of various universal quantities of interest, both by looking at their 
dependence on the shape function $R^{(0)}$ and by comparing the ``type A'' 
and ``type B'' results. 

In this respect universal quantities are of special importance because,
by definition, they are strictly cutoff scheme independent in the exact 
theory. Any truncation leads to a scheme dependence of these quantities whose
magnitude is a measure for the reliability of the 
truncation \cite{kana}. Typical examples of universal quantities are the 
critical exponents $\theta_I$. The existence or nonexistence of a fixed point 
is also a universal, scheme independent feature, but its precise location in 
parameter space is scheme dependent. Nevertheless it can be argued that, 
in $d=4$, the product $g_*\lambda_*$ is universal \cite{nin,LR1} while $g_*$ 
and $\lambda_*$ separately are not.

For later comparison with the $R^2$-truncation, let us briefly list some of 
the results we obtained in \cite{LR1} with the pure Einstein-Hilbert 
truncation:

{\boldmath$(1_{\rm E.H.})$} \underline{Universal Existence}: 
Both for type A and type B cutoffs the non-Gaussian fixed point exists for all
shape functions $R^{(0)}$ we considered. This result is highly nontrivial 
since in higher dimensions $(d\gtrsim 5)$ the fixed point exists for some but 
does not exist for other cutoffs \cite{frank}.

{\boldmath$(2_{\rm E.H.})$} \underline{Positive Newton Constant}:
While the position of the fixed point is scheme dependent, all cutoffs yield
{\it positive} values of $g_*$ and $\lambda_*$. A negative $g_*$ might be
problematic for stability reasons, but there is no mechanism in the flow
equation which would exclude it on general grounds.

{\boldmath$(3_{\rm E.H.})$} \underline{Stability}:
For any cutoff employed, the non-Gaussian fixed point is found to be UV
attractive in both directions of the $\lambda$-$g-$plane. Linearizing the
flow equation according to eq. (\ref{gfp4}) we obtain a pair of complex
conjugate critical exponents $\theta_1=\theta_2^*$ with positive real part 
$\theta'$ and imaginary parts $\pm\theta''$. Due to the positivity of 
$\theta'$, all trajectories in its basin of attraction hit the fixed point as 
$k$ is sent to infinity. Because of 
the nonvanishing imaginary part $\theta''$ the trajectories spiral into the 
fixed point for $k\rightarrow\infty$. 

Solving the full, nonlinear flow equations \cite{frank} shows that the 
asymptotic scaling region where the linearization is valid 
extends from $k``=\mbox{''}\infty$ down to about $k\approx m_{\rm Pl}$ with the
Planck mass defined as $m_{\rm Pl}\equiv G_0^{-1/2}$. It is the regime above 
the Planck scale where the asymptotic freedom of $G_k$ sets in.

{\boldmath$(4_{\rm E.H.})$} \underline{Scheme- and Gauge Dependence}:
The critical exponents  are reasonably constant within about a factor of 2. 
For the gauges $\alpha=1$ and $\alpha=0$, for instance, they assume values in 
the ranges $1.4\lesssim
\theta'\lesssim 1.8$, $2.3\lesssim\theta''\lesssim 4$ and $1.7\lesssim
\theta'\lesssim 2.1$, $2.5\lesssim\theta''\lesssim 5$, respectively. The
universality properties of the product $g_*\lambda_*$ are much more
impressive though. Despite the rather strong scheme dependence of $g_*$ and 
$\lambda_*$ separately, their product exhibits almost no visible 
$R^{(0)}$-dependence. Its value is $g_*\lambda_*\approx 0.12$ for $\alpha=1$ 
and $g_*\lambda_*\approx 0.14$ for $\alpha=0$.
The differences between the ``physical'' (fixed point) value of the gauge
parameter, $\alpha=0$, and the technically more convenient $\alpha=1$ are at 
the level of about 10 to 20 per-cent.\\[12pt]
\indent The above results suggest that the UV attractive non-Gaussian fixed 
point occuring in the Einstein-Hilbert truncation is very unlikely to be an
artifact of this truncation but should rather be the projection of a fixed
point in the exact theory. We interpreted them as nontrivial 
indications supporting the conjecture that 4-dimensional QEG
 is ``asymptotically safe'' in Weinberg's sense. 

\subsubsection{Results obtained from the $R^2$-truncation}
\label{3S5.3.2}
The actual justification of a truncation is that when one adds 
further terms to it its physical predictions do not change significantly any 
more. In order to test the stability of the Einstein-Hilbert
truncation against the inclusion of other invariants we shall now reanalyze
the non-Gaussian fixed point in the 
generalized truncation (\ref{trunc}) including the $R^2$-term. 
Starting from the $\mbox{\boldmath$\beta$}$-functions of the $R^2$-truncation,
eqs. (\ref{del}), (\ref{deg}) and (\ref{deb}), we determine the location of 
the fixed point in $\lambda$-$g$-$\beta-$space and the linearized
flow in its vicinity. Then we investigate the residual cutoff scheme dependence
of the associated universal quantities, and we compare
our results to those obtained from the pure Einstein-Hilbert truncation.

Note that, contrary to the pure Einstein-Hilbert truncation, only a cutoff of 
type B is used in the context of the generalized truncation. Therefore
we omit the specification of the cutoff type when we refer to results 
obtained from the $R^2$-truncation.\\[12pt]
{\bf Location of the fixed point \boldmath$(d=4)$}\\[6pt]
In a first attempt at finding the non-Gaussian fixed point in the 
$R^2$-truncation we neglect the cosmological constant and the coupling of the 
$R^2$-invariant. We approximate $\lambda_k\equiv\lambda_*=0$, 
$\beta_k\equiv\beta_*=0$,
thereby projecting the renormalization group flow onto the one-dimensional 
space parametrized by $g$. In this case the non-Gaussian fixed point is 
obtained as the nontrivial solution of $\mbox{\boldmath$\beta$}_g
(0,g_*,0;d)=0$. It is determined in appendix \ref{approxngfp} with the result 
given by eq. (\ref{gstar1}). For any $d$, this solution coincides precisely 
with the analogous approximate solution (H2) of ref. \cite{LR1} with 
$\alpha=1$, obtained in the pure Einstein-Hilbert truncation.
In order to get a numerical value for the fixed point we have to specify 
$R^{(0)}$. Inserting the exponential shape function with $s=1$ into eq. 
(\ref{gstar1}) and setting $d=4$ leads to $g_*\approx 0.590$.

Assuming that for the combined $\lambda$-$g$-$\beta$ system the numbers 
$\lambda_*$, $g_*$ and $\beta_*$ are of the same order of magnitude as $g_*$ 
above we expand the $\mbox{\boldmath$\beta$}$-functions about 
$(\lambda_k,g_k,\beta_k)=(0,0,0)$ and neglect terms of higher orders in the 
couplings. Again in appendix \ref{approxngfp} we determine the non-Gaussian 
fixed point from the corresponding system of differential equations. Inserting
the shape function (\ref{expcut}) and setting $d=4$, we find 
$(\lambda_*,g_*,\beta_*)\approx(0.287,0.751,0.002)$. Quite remarkably, for any
cutoff $\lambda_*$ and $g_*$ agree perfectly with the 
corresponding values obtained in \cite{LR1} by the same approximation applied
to the pure Einstein-Hilbert truncation.

In order to determine the {\it exact} position of the non-Gaussian fixed point 
$(\lambda_*,g_*,\beta_*)$ we have to resort to numerical methods. Given a 
starting value for the fixed point, for instance one of the approximate 
solutions above, the program we use determines a numerical solution which is 
exact up to an arbitrary degree of accuracy. Under the same conditions as 
above, i.e. for $s=1$ and $d=4$, we obtain
\begin{eqnarray}
\label{R^2fixedpoint}
(\lambda_*,g_*,\beta_*)=(0.330,0.292,0.005)
\end{eqnarray}

In the pure Einstein-Hilbert truncation the corresponding coordinates of the
fixed point are $(\lambda_*,g_*)=(0.348,0.272)$ \cite{LR1}. Obviously the 
values of $\lambda_*$ and $g_*$ are almost the same in both cases. While
$\lambda_*$ and $g_*$ are of the same order of magnitude, we find that 
$\beta_*$ is significantly smaller than $\lambda_*$ and $g_*$.

In order to test whether these properties of the fixed point coordinates are
universal we study their scheme dependence by looking at the $s$- or 
$b$-dependence introduced
via the one-parameter families of shape functions (\ref{expshape}) or 
(\ref{supp}), respectively. Here $s$ parametrizes the family of exponential 
shape functions (\ref{expshape}), while the shape parameter $b$ allows us
to change the profile of the shape functions with compact support (\ref{supp}).

As for the family of exponential shape functions, we 
are forced to restrict our considerations to shape parameters $s\ge 1$. This 
is because for $s<1$ the numerical integrations are plagued by convergence 
problems. They are due to the fact that in $d=4$ some of the 
threshold functions appearing in $\mbox{\boldmath$\beta$}_\lambda$, 
$\mbox{\boldmath$\beta$}_g$ and $\mbox{\boldmath$\beta$}_\beta$ diverge in 
the limit $s\rightarrow 0$, see also \cite{souma2}. As for the family of 
shape functions with compact support, we have to restrict ourselves to
$b\le 1.2$ for similar reasons. Here $R^{(0)}(y;b)$ approaches a sharp cutoff 
as $b\rightarrow 1.5$, which introduces discontinuities into the
integrands of the threshold functions $\Phi^p_n$ and $\widetilde{\Phi}_n^p$.
Already for $b\gtrsim 1.2$ the $\mbox{\boldmath$\beta$}$-functions
start to ``feel'' the sharp cutoff limit, which leads to convergence problems. 

\renewcommand{\baselinestretch}{1}
\small\normalsize
\begin{figure}[ht]
\begin{minipage}{7.9cm}
        \epsfxsize=7.9cm
        \epsfysize=5.2cm
        \centerline{\epsffile{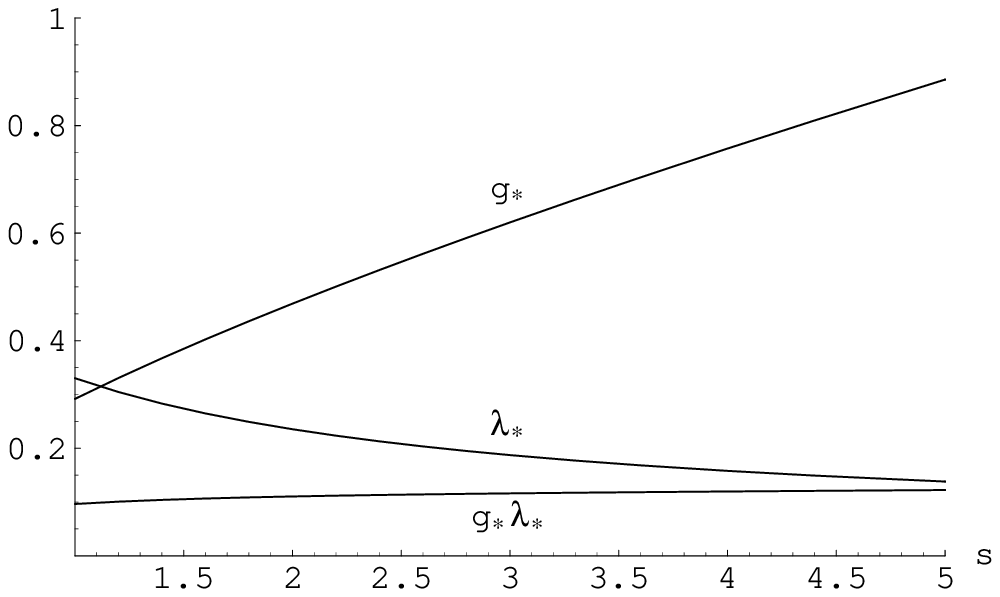}}
\centerline{(a)}
\end{minipage}
\hfill
\begin{minipage}{7.9cm}
        \epsfxsize=7.9cm
        \epsfysize=5.2cm
        \centerline{\epsffile{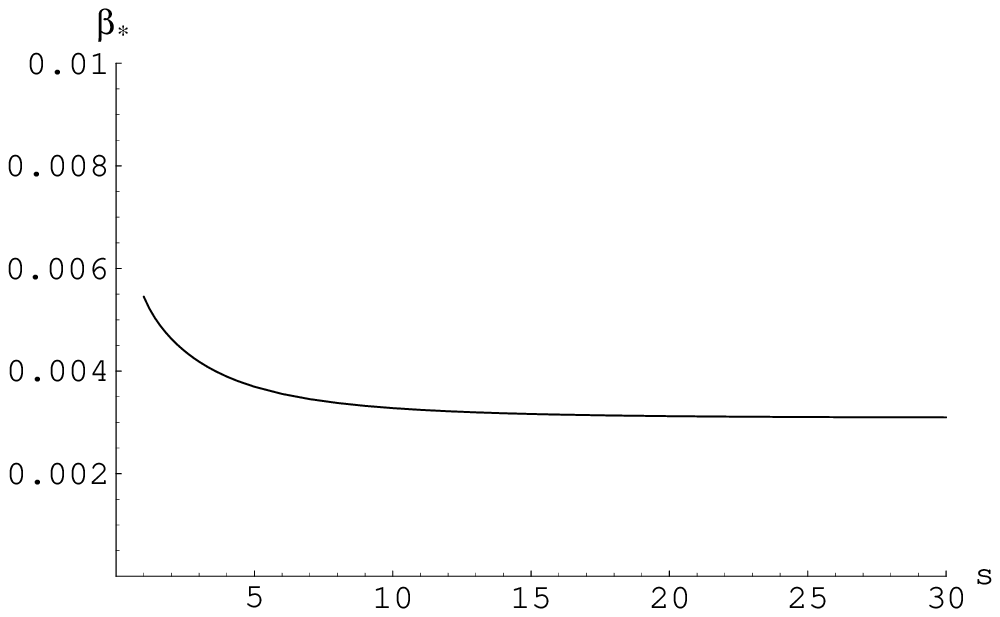}}
\centerline{(b)}
\end{minipage}
\caption{(a) $g_*$, $\lambda_*$, and $g_*\lambda_*$ as functions of $s$ 
for $1\le s\le 5$, and (b) $\beta_*$ as a function of $s$ for $1\le s\le 30$,
using the family of exponential shape functions.}  
\label{plot1}
\end{figure}
\renewcommand{\baselinestretch}{1.5}
\small\normalsize
\renewcommand{\baselinestretch}{1}
\small\normalsize
\begin{figure}[ht]
\begin{minipage}{7.9cm}
        \epsfxsize=7.9cm
        \epsfysize=5.2cm
        \centerline{\epsffile{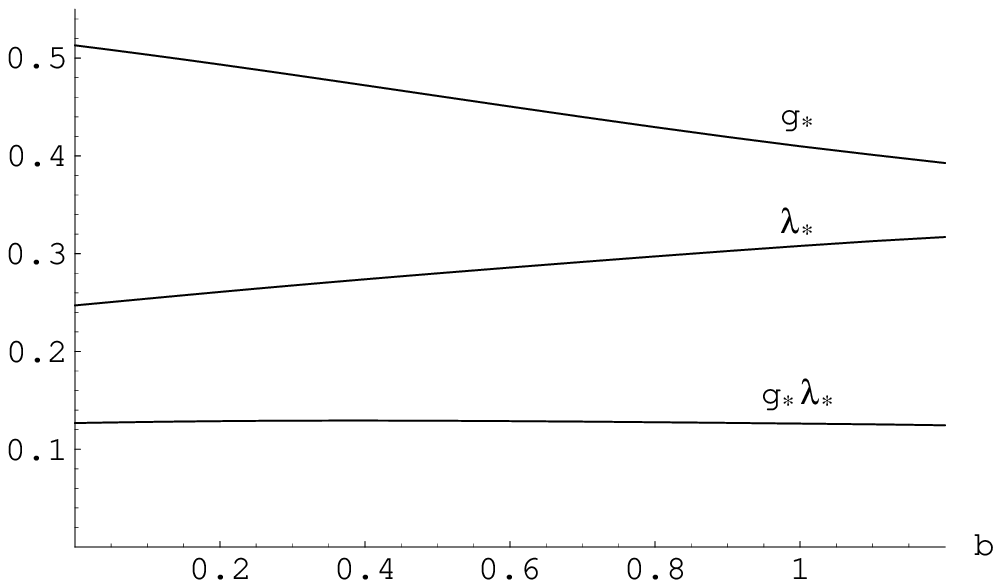}}
\centerline{(a)}
\end{minipage}
\hfill
\begin{minipage}{7.9cm}
        \epsfxsize=7.9cm
        \epsfysize=5.2cm
        \centerline{\epsffile{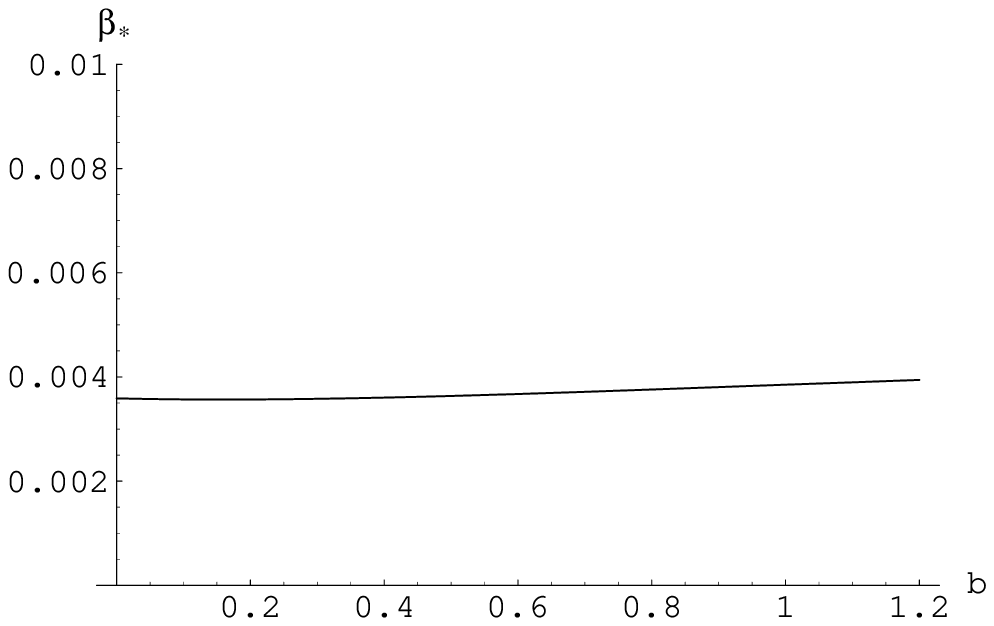}}
\centerline{(b)}
\end{minipage}
\caption{(a) $g_*$, $\lambda_*$ and $g_*\lambda_*$, and 
(b) $\beta_*$ as functions of $b$ for $0\le b\le 1.2$,
using the family of shape functions with compact support.}  
\label{plot4}
\end{figure}
\renewcommand{\baselinestretch}{1.5}
\small\normalsize

As in the case of the pure Einstein-Hilbert truncation \cite{LR1} our results
establish the existence of the non-Gaussian fixed point in a wide range of 
$s$- and $b$-values. As expected, the position of the fixed point turns out to
depend on $s$ or $b$, i.e. on the cutoff scheme, but the crucial point is that
it exists for any of the cutoffs employed. FIGS. \ref{plot1} and \ref{plot4}
show its coordinates $(\lambda_*,g_*,\beta_*)$ as well as the product 
$g_*\lambda_*$ for the shape functions (\ref{expshape}) and (\ref{supp}), 
respectively. In Fig \ref{plot1}(a) we plotted the various quantities in the
range $1\le s\le 5$ where the largest changes in $\lambda_*$ and $g_*$ occur,
but we calculated them for $1\le s\le 30$. For every shape parameter $s$ or 
$b$, the values of $\lambda_*$ and $g_*$ are almost the same as those 
obtained with the Einstein-Hilbert 
truncation \cite{LR1}. As a consequence, the product $g_*\lambda_*$ is again
almost constant and its value differs only slighlty from the one in \cite{LR1} 
for the same gauge $\alpha=1$. Both FIGS. \ref{plot1}(a) and 
\ref{plot4}(a) suggest the universal value $g_*\lambda_*\approx 0.14$ while 
we obtained $g_*\lambda_*\approx 0.12$ from the pure Einstein-Hilbert 
truncation.
Thus we may expect that our $g_*\lambda_*$-value is precise at the 10 to 20
per-cent level. Presumably this degree of precision is the best we can achieve
in the present calculation because we saw already that the error due to using
$\alpha=1$ instead of the ``correct'' $\alpha=0$ leads to an uncertainty of the
same size.

Furthermore, our results show that $\beta_*$ is always 
significantly smaller than $g_*$ and $\lambda_*$ for both families of shape
functions, which is quite remarkable. Within the limited precision of our 
calculation this means that in the three-dimensional 
$\lambda$-$g$-$\beta-$space the fixed point practically lies in the 
$\lambda$-$g-$plane with $\beta=0$, i.e. on the parameter space of the pure 
Einstein-Hilbert truncation.

It is also interesting to note that the scheme dependence of $\beta_*$ is 
unexpectedly small. As for the family of exponential shape functions 
(\ref{expshape}), the function $\beta_*(s)$ depicted in FIG. \ref{plot1}(b)
develops a plateau-like shape for not too small values of $s$. Employing the
family of shape functions with compact support, the scheme dependence of 
$\beta_*$ is even weaker. The function $\beta_*(b)$ plotted in FIG.
\ref{plot4}(b) is almost constant in the range $0\le b\le 1.2$. Moreover,
the positions of the two plateaus are nearly identical. While FIG. 
\ref{plot1}(b) suggests the value $\beta_*\approx0.0031$, we obtain 
$\beta_*\approx0.0036$ from FIG. \ref{plot4}(b). This indicates that in $d=4$
dimensions also $\beta_*$ might be a universal quantity.\\[12pt]
{\bf The linearized flow \boldmath$(d=4)$}\\[6pt]
Let us now analyze the critical behavior near the non-Gaussian fixed point.
Quite remarkably, the non-Gaussian fixed point of the $R^2$-truncation proves
to be UV attractive in any of the three directions of 
$\lambda$-$g$-$\beta-$space, for all cutoffs used. The linearized flow in
its vicinity is always governed by a pair of complex conjugate critical
exponents $\theta_1=\theta'+{\rm i}\theta''=\theta_2^*$ with $\theta'>0$ and
a single real, positive critical exponent $\theta_3>0$. (We define
$\theta_1$ as the critical exponent with the positive imaginary part so that
$\theta''>0$.) The general solution to the linearized flow equations is
obtained by taking the real part of eq. (\ref{gfp5}). Introducing the RG time
$t\equiv\ln(k/k_0)$ it may be written as
\begin{eqnarray}
\label{linflow2}
\left(\lambda_k,g_k,\beta_k\right)^{\bf T}
&=&\left(\lambda_*,g_*,\beta_*\right)^{\bf T}
+2\Bigg\{\left[{\rm Re}\,C\,\cos\left(\theta''\,t\right)
+{\rm Im}\,C\,\sin\left(\theta''\,t\right)\right]
{\rm Re}\,V\nonumber\\
& &+\left[{\rm Re}\,C\,\sin\left(\theta''\,t\right)-{\rm Im}\,C
\,\cos\left(\theta''\,t\right)\right]{\rm Im}\,V\Bigg\}\,e^{-\theta' t}
+C_3 V^3\,e^{-\theta_3 t}
\end{eqnarray}
with arbitrary complex $C\equiv C_1=(C_2)^*$ and arbitrary real $C_3$. 
Furthermore, $V\equiv V^1=(V^2)^*$ and $V^3$ are  the right-eigenvectors of 
the stability
matrix $(B_{ij})_{i,j\in\{\lambda,g,\beta\}}$ with eigenvalues $-\theta_1=
-\theta_2^*$ and $-\theta_3$, respectively. Obviously the conditions for UV
stability are $\theta'>0$ and $\theta_3>0$. They are indeed
satisfied for all cutoffs. As a consequence, all RG trajectories which reach
its basin of attraction hit the fixed point as $t$ is sent to infinity.
The trajectories
(\ref{linflow2}) comprise three independent normal modes with amplitudes 
proportional to ${\rm Re}\,C$, ${\rm Im}\,C$ and $C_3$, respectively.
The first two are of the spiral type, the third one is a straight line.

\begin{figure}[ht]
\begin{minipage}{7.9cm}
        \epsfxsize=7.9cm
        \epsfysize=5.2cm
        \centerline{\epsffile{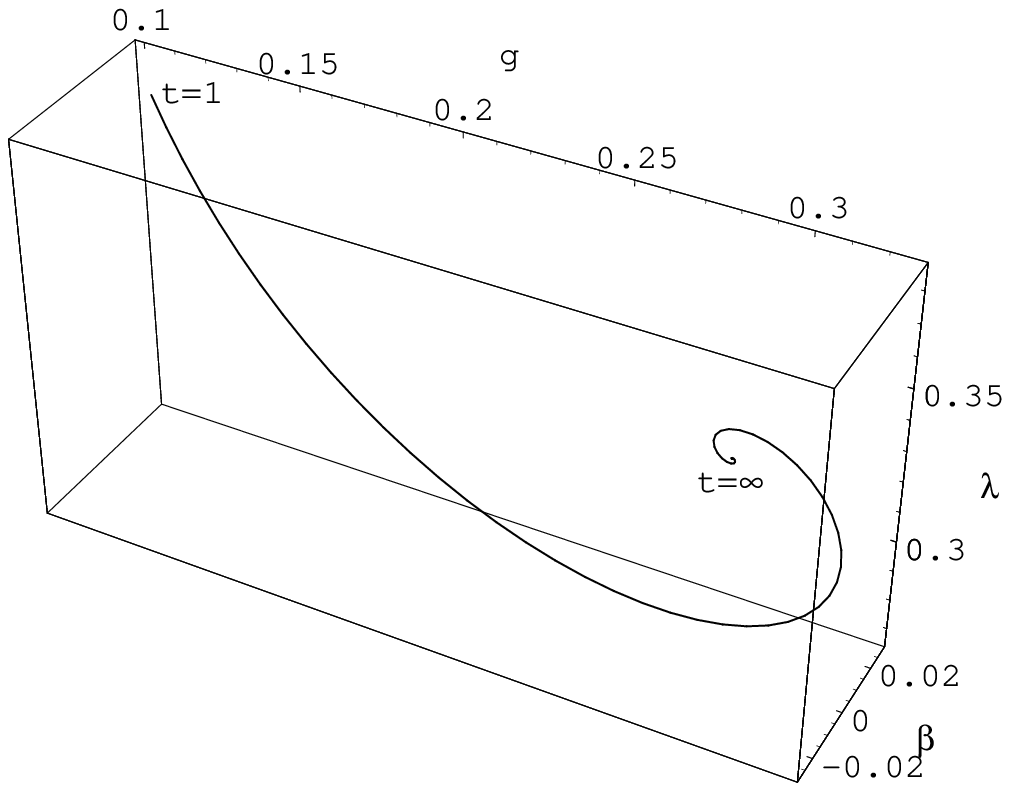}}
\centerline{(a)}
\end{minipage}
\hfill
\begin{minipage}{7.9cm}
        \epsfxsize=7.9cm
        \epsfysize=5.2cm
        \centerline{\epsffile{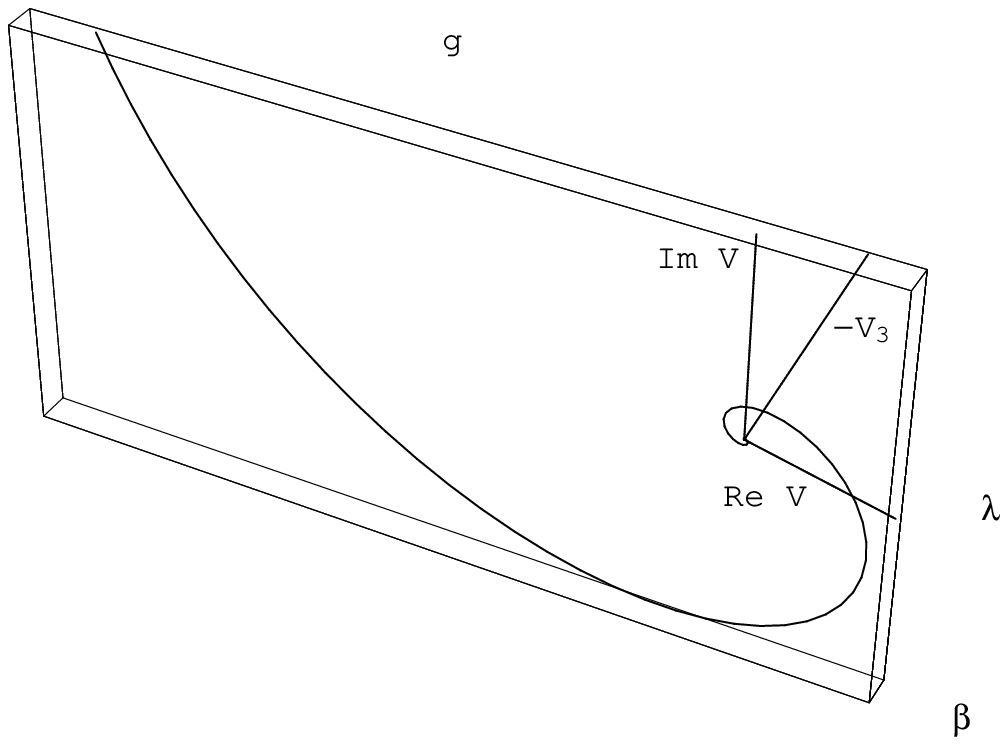}}
\centerline{(b)}
\end{minipage}
\caption{Trajectory of the linearized flow equation obtained from the
$R^2$-truncation for $1\le t=\ln(k/k_0)<\infty$. In (b) we depict
the eigendirections and the ``box'' to which the trajectory is confined.}  
\label{plot2}
\end{figure}

Let us illustrate these features by means of an example. For the
exponential shape function (\ref{expshape}) with $s=1$, for instance,
we have $(\lambda_*,g_*,\beta_*)=(0.330,0.292,0.005)$. The
corresponding stability matrix $\bf B$ takes the form
\begin{eqnarray}
\label{ngfp1}
{\bf B}=-\left(\begin{array}{rrr}8.83 & 2.61 & 401.75\\
6.18 & 4.46 & 89.24\\
0.29 & 0.32 & 19.82\end{array}\right)\;.
\end{eqnarray}
It leads to the pair of complex critical exponents 
$\theta_1=\theta_2^*$ with $\theta'=2.15$, $\theta''=3.79$,
and to the real critical exponent $\theta_3=28.8$.
For the associated right-eigenvectors we find
\begin{eqnarray}
\label{eigenvectors} 
{\rm Re}\,V&=&(-0.164,0.753,-0.008)^{\bf T}\;,\nonumber\\
{\rm Im}\,V&=&(0.64,0,-0.01)^{\bf T}\;,\nonumber\\
V^3&=&-(0.92,0.39,0.04)^{\bf T}\;.
\end{eqnarray}
(The vectors are normalized such that
$\left\|V\right\|=\left\|V^3\right\|=1$.) In FIG. \ref{plot2} we show a
typical trajectory which has all three normal modes
excited with equal strength $({\rm Re}\,C={\rm Im}\,C=1/\sqrt{2}$, $C_3=1)$.
All its way down from $k``=\mbox{''}\infty$ to about $k=m_{\rm Pl}$ it is
confined to a very thin box surrounding the $\beta=0-$plane, i.e the parameter
space of the Einstein-Hilbert truncation.

In fact, the linearized flow is
characterized by the following quite remarkable properties, independently
of the cutoff. They all indicate that, close to the non-Gaussian fixed point, 
the RG flow is rather well approximated by the pure Einstein-Hilbert 
truncation. 

\renewcommand{\baselinestretch}{1}
\small\normalsize
\renewcommand{\baselinestretch}{1.5}
\small\normalsize

{\bf (a)}
The $\beta$-components of ${\rm Re}\,V$ and ${\rm Im}\,V$ are very
tiny. Hence these two vectors span a plane which virtually coincides with the
$\lambda$-$g-$subspace at $\beta=0$, i.e. with the parameter space of the
Einstein-Hilbert truncation. As a consequence, the  ${\rm Re}\,C$- and
${\rm Im}\,C$- normal modes are essentially the same trajectories as the
``old'' normal modes already found without the $R^2$-term. Also the
corresponding $\theta'$- and $\theta''$-values coincide within the scheme
dependence, see below.

{\bf (b)}
For all cutoffs employed, the new eigenvalue $\theta_3$ introduced by the 
$R^2$-term is significantly larger than $\theta'$, see below. When a 
trajectory approaches the fixed
point from below $(t\rightarrow\infty)$, the ``old'' normal modes $\propto
{\rm Re}\,C,{\rm Im}\,C$ are proportional to $\exp(-\theta' t)$, but the new
one is proportional to $\exp(-\theta_3 t)$, so that it decays much more
quickly. For every trajectory running into the fixed point, i.e. for every
set of constants $({\rm Re}\,C,{\rm Im}\,C,C_3)$, we find therefore that, once
$t$ is sufficiently large, the trajectory lies entirely in the
${\rm Re}\,V$-${\rm Im}\,V-$subspace, i.e. the $\beta=0-$plane practically.

Due to the large value of $\theta_3$, the new scaling field is very
``relevant''. However, when we start at the fixed point $(t``=\mbox{''}\infty)$
and lower $t$ it is only at the low energy scale $k\approx m_{\rm Pl}$
$(t\approx 0)$ that $\exp(-\theta_3 t)$ reaches unity, and only then, i.e.
far away from the fixed point, the new scaling field starts growing
rapidly.

{\bf (c)}
Since the matrix ${\bf B}$ is not symmetric its eigenvectors have
no reason to be orthogonal. In fact, we find that $V^3$ lies almost in the
${\rm Re}\,V$-${\rm Im}\,V-$plane. For the angles between the eigenvectors
given above we obtain $\sphericalangle ({\rm Re}\,V,{\rm Im}\,V)=102.3^\circ$, 
$\sphericalangle({\rm Re}\,V,V^3)=100.7^\circ$, 
$\sphericalangle({\rm Im}\,V,V^3)=156.7^\circ$. Their sum is $359.7^\circ$
which confirms that ${\rm Re}\,V$, ${\rm Im}\,V$ and $V^3$ are almost 
coplanar. This implies that when we lower $t$ and move away from the fixed
point so that the $V^3$- scaling field starts growing, it is again
predominantly the $\int d^dx\,\sqrt{g}$- and $\int d^dx\,\sqrt{g}R$-invariants
which get excited, but not $\int d^dx\,\sqrt{g}R^2$ in the first place.

\begin{figure}[ht]
\begin{minipage}{7.9cm}
        \epsfxsize=7.9cm
        \epsfysize=5.2cm
        \centerline{\epsffile{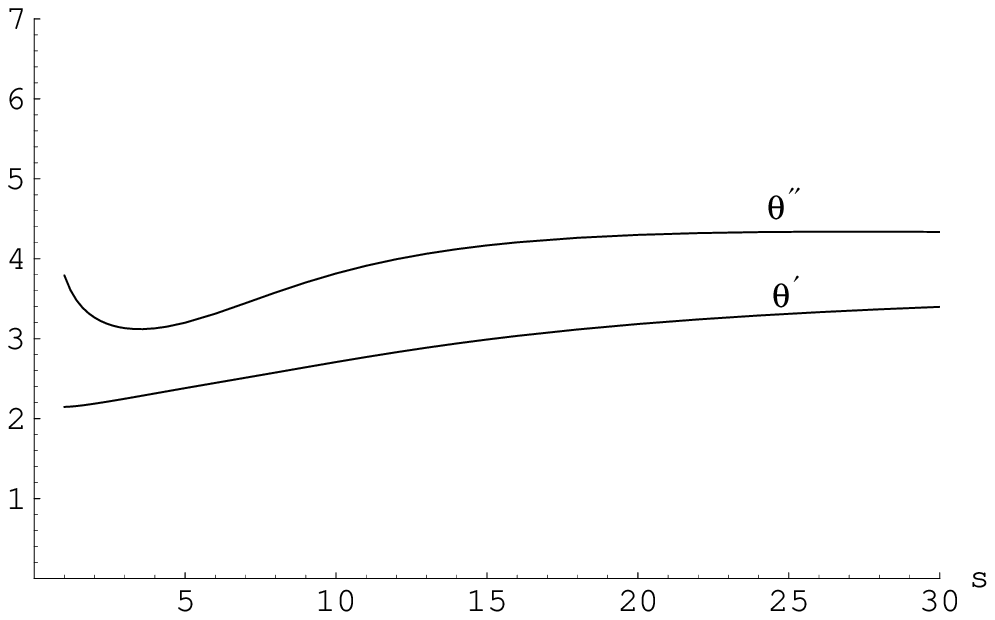}}
\centerline{(a)}
\end{minipage}
\hfill
\begin{minipage}{7.9cm}
        \epsfxsize=7.9cm
        \epsfysize=5.2cm
        \centerline{\epsffile{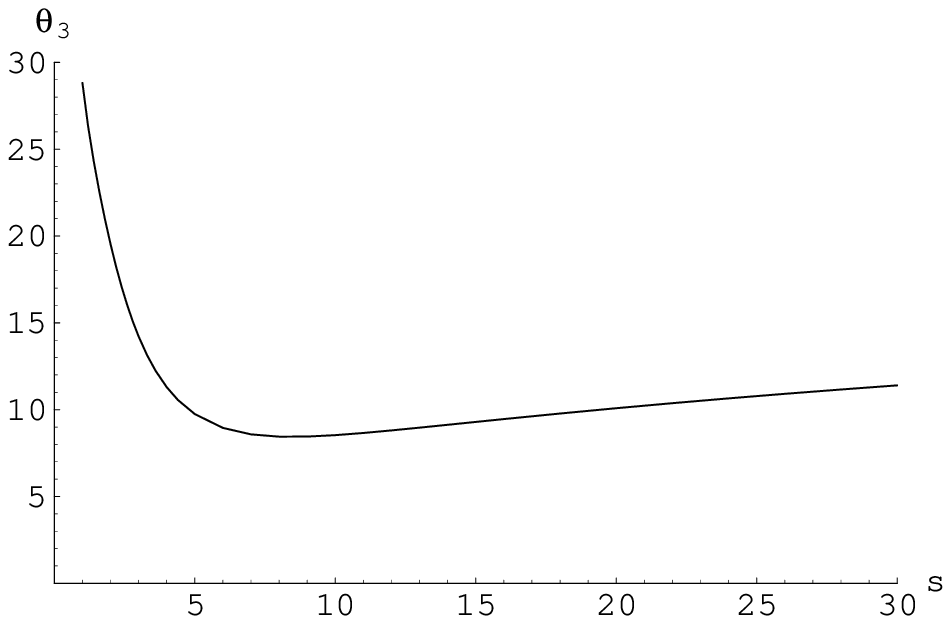}}
\centerline{(b)}
\end{minipage}
\caption{(a) $\theta'={\rm Re}\,\theta_1$ and $\theta''={\rm Im}\,\theta_1$, 
and (b) $\theta_3$ as functions of $s$, using the family of exponential shape 
functions.}  
\label{plot3}
\end{figure}

\begin{figure}[ht]
\begin{minipage}{7.9cm}
        \epsfxsize=7.9cm
        \epsfysize=5.2cm
        \centerline{\epsffile{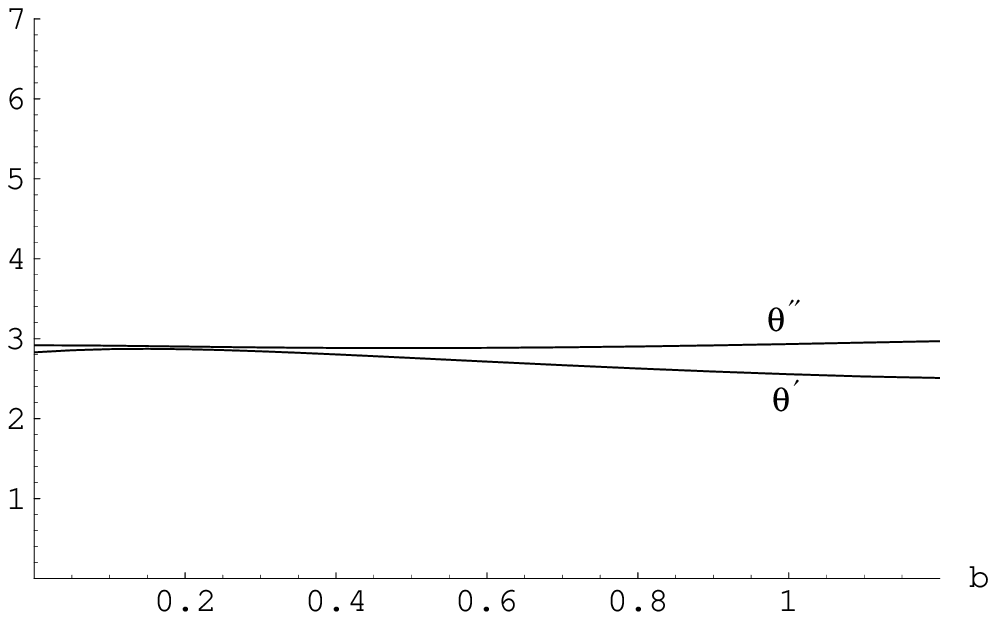}}
\centerline{(a)}
\end{minipage}
\hfill
\begin{minipage}{7.9cm}
        \epsfxsize=7.9cm
        \epsfysize=5.2cm
        \centerline{\epsffile{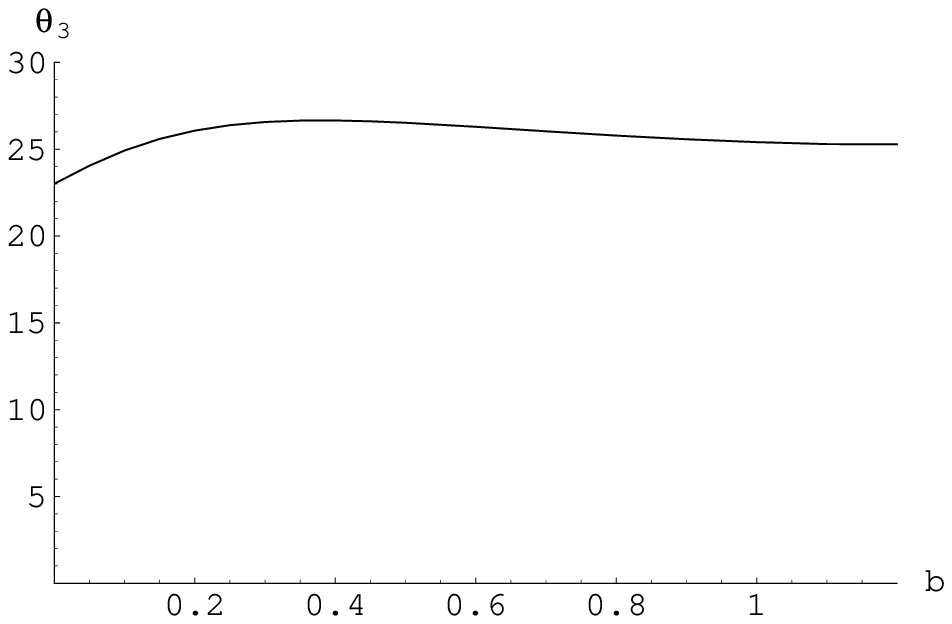}}
\centerline{(b)}
\end{minipage}
\caption{(a) $\theta'={\rm Re}\,\theta_1$ and $\theta''={\rm Im}\,\theta_1$, 
and (b) $\theta_3$ as functions of $b$, using the family of shape functions
with compact support.}  
\label{plot5}
\end{figure}

Summarizing the three points above we can say that very close to the fixed
point the RG flow seems to be essentially two-dimensional, and that this
two-dimensional flow is well approximated by the RG equations of the
Einstein-Hilbert truncation.\\[12pt] 
{\bf Scheme dependence of the critical exponents \boldmath($d=4$)}\\[6pt]
As we pointed out already the critical exponents are universal
in an exact treatment, but in a truncated parameter space a scheme dependence 
is expected to occur as an artifact of the truncation. We may use it to judge 
the quality of our truncation.
Also in this respect the $R^2$-truncation yields satisfactory results, which
we display in FIGS. \ref{plot3} and \ref{plot5}. FIGS. \ref{plot3}(a) and
\ref{plot5}(a) show the real and the imaginary part $\theta'$ and $\theta''$ of
the complex conjugate pair $\theta_1=\theta_2^*$ while $\theta_3$ is depicted
in FIGS.\ref{plot3}(b) and \ref{plot5}(b). The plots in FIG. \ref{plot3}
are based on the family of exponential shape functions (\ref{expshape}) and
those in FIG. \ref{plot5} are obtained by employing the family of shape
functions with compact support (\ref{supp}). They display the $s$- and the
$b$-dependence of the critical exponents, respectively.

\renewcommand{\baselinestretch}{1}
\small\normalsize
\renewcommand{\baselinestretch}{1.5}
\small\normalsize
\renewcommand{\baselinestretch}{1}
\small\normalsize
\renewcommand{\baselinestretch}{1.5}
\small\normalsize

As for the complex conjugate pair of critical exponents, the scheme dependence
is of the same order of magnitude as in the case of the Einstein-Hilbert
truncation \cite{LR1}. While the scheme dependence of $\theta''$ is
weaker than that found in \cite{LR1} we see that it is slightly larger for
$\theta'$. For the exponential shape functions with $1\le s\le 30$,
$\theta'$ and $\theta''$ assume values in the ranges $2.1\lesssim\theta'(s)
\lesssim 3.4$ and $3.1\lesssim\theta''(s)\lesssim 4.3$, respectively.
Employing the shape functions with compact support leads to a weaker 
dependence on the shape parameter $b$. However, the corresponding values
$\theta'(b)$ and $\theta''(b)$ are in good agreement with those obtained with
the exponential cutoffs. In fact, they all lie in the $\theta'(s)$- and
$\theta''(s)-$intervals given above. The average values of $\theta'$ and
$\theta''$ are slightly larger than those obtained from the pure
Einstein-Hilbert truncation. The difference between the corresponding average
values is approximately 1 for both $\theta'$ and $\theta''$.

Let us now come to the new critical exponent $\theta_3$ which was not present 
in the Einstein-Hilbert truncation. Using the exponential shape functions
(\ref{expshape}) it suffers from relatively strong variations as the shape
parameter $s$ is changed. It assumes values in the range $8.4\lesssim
\theta_3(s)\lesssim 28.8$. As compared to the exponential cutoffs, the cutoffs
with compact support lead to a much weaker scheme dependence. For $b\in[0,1.2]$
we have $23.0\lesssim\theta_3(b)\lesssim 26.7$. However, the results obtained
with the two families of shape functions agree within the scheme dependence.
Moreover, $\theta_3$ is always systematically larger than $\theta'$ (and 
$\theta''$) with both families of cutoffs.
As a consequence, the hierarchy of critical exponents which was mentioned in 
(b) above and which squeezes the trajectories into a thin box is a universal
feature.

Obviously the critical exponents, in particular $\theta_3$, exhibit a
much stronger scheme dependence than $g_*\lambda_*$. This is most
probably due to neglecting further relevant operators in the truncation so
that the ${\bf B}$-matrix we are diagonalizing is still too small.\\[12pt]
{\bf In {\boldmath$2+\varepsilon$} dimensions}\\[6pt]
The above results and their mutual consistency strongly suggest that
4-dimensional Quantum Einstein Gravity indeed possesses a RG
fixed point with precisely the properties needed for its nonperturbative
renormalizability or ``asymptotic safety''. However, with the present approach
it is clearly not possible to determine the dimensionality $\Delta_{\rm UV}$
of the UV critical hypersurface, which coincides with the number of
invariants relevant at the non-Gaussian fixed point. According 
to the canonical dimensional analysis,
the $({\rm cur}$\-va\-${\rm ture})^n$-invariants in 4 dimensions are
{\it classically} marginal for $n=2$ and irrelevant for $n>2$. The results
for $\theta_3$ indicate that there are large {\it nonclassical} contributions 
so that there might be relevant operators perhaps even beyond $n=2$. However, 
as it is hardly conceivable that the quantum effects change the signs of
arbitrarily large (negative) classical scaling dimensions, $\Delta_{\rm UV}$
should be finite \cite{wein}. 

A first confirmation of this picture comes from our
$R^2$-calculation in $d=2+\varepsilon$ where the dimensional count is shifted
by two units. In this case we find indeed that the third scaling field is
{\it irrelevant} for any cutoff employed, $\theta_3<0$. 

For our analysis of the $R^2$-truncation in $d=2+\varepsilon$ dimensions
with $0<\varepsilon\ll 1$ we had to resort to numerical methods. Using the
$\varepsilon$-expansion we calculated the fixed point coordinates and the
critical exponents for selected values of the shape parameter $s$. For all
quantities only the leading nontrivial order of the $\varepsilon$-expansion
was retained. In Table 1 we present the corresponding numerical results. 

\vspace{0.5cm} 
\begin{center}
\begin{tabular}
{|c|c|c|c|c|c|c|}
\hline\multicolumn{7}{|c|}{Table 1: Fixed point coordinates and critical 
exponents}\\
\hline $s$ & $\lambda_*\,(+{\cal O}(\varepsilon^2)$) & 
$g_*\,(+{\cal O}(\varepsilon^2))$ & $\beta_*\,(+{\cal O}(\varepsilon))$ & 
$\theta_1\,(+{\cal O}(\varepsilon))$ & $\theta_2\,(+{\cal O}(\varepsilon^2))$
& $\theta_3\,(+{\cal O}(\varepsilon))$\\
\hline 1 & $-0.131\varepsilon$ & $0.087\varepsilon$ & $-0.083$ & $2$ & 
$0.963\varepsilon$ & $-1.968$\\
\hline 5 & $-0.055\varepsilon$ & $0.092\varepsilon$ & $-0.312$ & $2$ &
$0.955\varepsilon$ & $-1.955$\\
\hline 10 & $-0.035\varepsilon$ & $0.095\varepsilon$ & $-0.592$ & $2$ & 
$0.955\varepsilon$ & $-1.956$\\
\hline\end{tabular}
\end{center}
\vspace{0.5cm}

For all cutoffs used we
obtain three {\it real} critical exponents, the first two are positive and the
third is negative. Thus, the corresponding $V^3$-direction is UV repulsive.
This suggests that the dimensionality of ${\cal S}_{\rm UV}$ could be as
small as $\Delta_{\rm UV}=2$, but this is not a proof, of course. If so,
the quantum theory would be characterized by only two free parameters, the
renormalized Newton constant $G_0$ and the renormalized cosmological constant
$\bar{\lambda}_0$, for instance.

Let us now compare the results to those from the
Einstein-Hilbert truncation \cite{Reu96,LR1}. The $\lambda$- and
$g$-coordinates of the fixed point and the critical exponents $\theta_1$
and $\theta_2$ are found to be similar to those in \cite{Reu96,LR1}. However, 
in the Einstein-Hilbert truncation the leading-order results $g_*=3/38\,
\varepsilon+{\cal O}(\varepsilon^2)\approx 0.079\,\varepsilon
+{\cal O}(\varepsilon^2)$ and $\theta_2=\varepsilon+{\cal O}
(\varepsilon^2)$ are scheme independent, which is not quite true for the 
results above. Both truncations agree on $\theta_1=2
+{\cal O}(\varepsilon)$.\\[12pt]
{\bf Summary:}\\[6pt]
Our main results concerning the non-Gaussian fixed point in the 
$R^2$-truncation are:

{\boldmath$(1_{R^2})$} \underline{Position of the FP}:
The fixed point is found to exists for all cutoffs used. This result is
highly nontrivial since the example of the Gaussian fixed point clearly shows
that a fixed point of the Einstein-Hilbert truncation does not necessarily 
generalize to a fixed point of the $R^2$-truncation. 
For every shape parameter the fixed point practically lies on the
$\lambda$-$g-$plane, and its position almost exactly coincides
with that from the Einstein-Hilbert truncation. 

{\boldmath$(2_{R^2})$} \underline{Eigenvalues and -vectors}:
The fixed point is UV attractive in any of the three directions of the
$\lambda$-$g$-$\beta-$space for all cutoffs employed. The linearized flow
in its vicinity is always governed by a pair of complex conjugate critical
exponents $\theta_1=\theta'+{\rm i}\theta''=\theta_2^*$ with $\theta'>0$ and
a single real, positive critical exponent $\theta_3>0$. It is essentially
two-dimensional, and this two-dimensional flow is well descibed by the
RG equations of the Einstein-Hilbert truncation.

{\boldmath$(3_{R^2})$} \underline{Scheme dependence}:
The scheme dependence of the critical exponents and of the product
$g_*\lambda_*$ is of the same order of magnitude as in the case
of the Einstein-Hilbert truncation. While the scheme dependence of
$\theta''$ is weaker than in the case of the Einstein-Hilbert truncation we
find that it is slightly larger for $\theta'$. The exponent $\theta_3$
shows a relatively strong dependence on the cutoff. The product 
$g_*\lambda_*$ again exhibits an impressively weak scheme dependence.

{\boldmath$(4_{R^2})$} \underline{Dimensionality of ${\cal S}_{\rm UV}$}:
The dimensionality $\Delta_{\rm UV}$ of the UV critical hypersurface cannot be
determined within the present approach. However, the results from our
$R^2$-calculation in $2+\varepsilon$ dimensions suggest that
$\Delta_{\rm UV}$ should be finite also in 4 dimensions.\vspace{4mm}

On the basis of the above results we believe that the non-Gaussian fixed
point occuring in the Einstein-Hilbert truncation is very unlikely to be an
artifact of this truncation but rather should be the projection of a fixed
point in the exact theory. We demonstrated explicitly that the fixed point
and all its qualitative properties are stable against the inclusion of a
further invariant in the truncation. These results strongly support the
hypothesis that 4-dimensional QEG is indeed
nonperturbatively renormalizable.

\section{Positivity of action, Hessian, and cutoff}
\renewcommand{\theequation}{6.\arabic{equation}}
\setcounter{equation}{0}
\label{3S6}
\subsection{Positivity of the action}
\label{3S6A}
It is a well known problem that in $d>2$ dimensions the Euclidean 
Einstein-Hilbert action 
\begin{eqnarray}
\label{EH}
S_{\rm EH}[g]=\frac{1}{16\pi\bar{G}}\int d^dx\,\sqrt{g}\left\{-R(g)
+2\bar{\lambda}\right\}
\end{eqnarray}
is not bounded below. In fact, decomposing the metric as $
g_{\mu\nu}=\exp(2\chi)\bar{g}_{\mu\nu}$ where $\bar{g}_{\mu\nu}$ is a fixed 
reference metric we obtain
\begin{eqnarray}
\label{m1}
S_{\rm EH}[g]=\frac{1}{16\pi \bar{G}}\int d^dx\,
\sqrt{\bar{g}}\,e^{(d-2)\chi}\,\left[-\bar{R}+2\bar{\lambda}\,e^{2\chi}
-(d-1)(d-2)\,\bar{g}^{\mu\nu}\,
\left(\bar{D}_\mu\chi\right)\left(\bar{D}_\nu\chi\right)\right]\,.
\end{eqnarray}
This shows that $S_{\rm EH}$ can become arbitrarily negative if the conformal
factor $\chi(x)$ varies rapidly enough so that $(\bar{D}_\mu\chi)^2$ is large.
Therefore it seems difficult to define a path integral $Z
=\int {\cal D}g_{\mu\nu}\exp(-S_{\rm EH})$ for Euclidean quantum gravity.

The situation improves by including the term $\int d^dx\sqrt{g}R^2$ with a 
positive coefficient since the resulting action is {\it bounded below} 
\cite{nonloc}. While the Einstein-Hilbert term $\int d^dx\sqrt{g}R$ leads to 
a negative contribution to the kinetic term of the conformal factor, which 
dominates at 
small momenta, the $R^2$-term gives rise to a positive contribution dominating
at large momenta. As a consequence, both the truncated action functional
$\Gamma_k[g,\bar{g}]$ of eq. (\ref{trunc}) and the bare action $S[g]=
\Gamma_{\widehat{k}\rightarrow\infty}[g,g]$ possess an absolute minimum. 
Moreover, rewriting the 
truncation ansatz (\ref{trunc}) with $\bar{g}_{\mu\nu}=g_{\mu\nu}$ as
\begin{eqnarray}
\label{S>0}
\Gamma_k[g,g]=\int d^dx\,\sqrt{g}\left\{\bar{\beta}_k\left(R-\frac{Z_{Nk}
\kappa^2}{\bar{\beta}_k}\right)^2+Z_{Nk}\kappa^2\left(4\bar{\lambda}_k
-\frac{Z_{Nk}\kappa^2}
{\bar{\beta}_k}\right)\right\}
\end{eqnarray}
one can easily determine a sufficient condition for a manifestly {\it positive}
action $\Gamma_k[g,g]>0$. In terms of the dimensionless couplings it reads:
$g_k>0$, $\beta_k>0$, and 
\begin{eqnarray}
\label{posit}
128\pi\,g_k\lambda_k\beta_k>1\,.
\end{eqnarray}

\subsection{Positivity of the Hessian}
At the level of the flow equation, $\Gamma_k$ appears on the RHS in terms of
its Hessian $\Gamma_k^{(2)}$ to which the cutoff operator ${\cal R}_k$ is 
adapted by the rule (\ref{rule}). Thus, only if $\Gamma_k^{(2)}$ is a positive 
definite operator we can obtain a cutoff which leads to a ``correct'' mode 
suppression. Since we expect the $R^2$-truncation anyhow to be reliable only 
for large $k$, it is actually sufficient if $\Gamma_k^{(2)}$ and ${\cal R}_k$
are positive definite for sufficiently large momenta $p^2\equiv -\bar{D}^2$. 
The reason is that, due to the factor $\partial_t{\cal R}_k(p^2)$ which 
emphasizes the region $p^2\approx k^2$, the traces on the RHS of the RG 
equation (\ref{truncflow}) receive the dominant contributions from modes whose
$p^2$ is close to $k^2$.

In general $\Gamma_k^{(2)}[g,\bar{g}]$ depends on both $g_{\mu\nu}$ and the
background metric $\bar{g}_{\mu\nu}$. Here we concentrate on 
$\Gamma_k^{(2)}[g,g]\equiv\Gamma_k^{(2)}$ with the two metrics identified.
Furthermore, we assume that $g_{\mu\nu}=\bar{g}_{\mu\nu}$ is the metric of a
$d$-sphere with radius $r$ since our projection 
technique requires these backgrounds only. In this case the eigenvalues 
$p^2=\Lambda_l(d,s)$ depend on the discrete quantum number $l$. The explicit 
expressions for $\Lambda_l(d,s)$ are tabulated in appendix \ref{harm}. They 
are strictly monotonically increasing functions of $l$ with 
$\lim_{l\rightarrow\infty}\Lambda_l(d,s)=\infty$.

In the following we show that the operator $\Gamma_k^{(2)}$ with $k$
very large indeed becomes
positive definite if it is restricted to the subspace spanned by the
$-D^2$-eigenfunctions with sufficiently large eigenvalues, certain assumptions
on the couplings being made. The spherical harmonics $T^{lm}_{\mu\nu}$,
$T^{lm}_\mu$, and $T^{lm}$ with $l$ larger than a certain minimum value 
$l_{\rm min}$ provide a basis of this subspace. We shall concentrate on the 
conditions implied by the leading large-$l$ behavior.

The Hessian  $\Gamma_k^{(2)}[g,g]$ as given by the quadratic form 
(\ref{gravquad}) is a symmetric blockdiagonal matrix. Therefore, according to
the Jacobi criterion, the condition
for positivity takes the simple form  
$\left(\Gamma_k^{(2)}\right)_{\bar{h}^T\bar{h}^T}>0$, 
$\left(\Gamma_k^{(2)}\right)_{\bar{\xi}\bar{\xi}}>0$, 
$\left(\Gamma_k^{(2)}\right)_{\bar{\phi}\bar{\phi}}>0$, and 
$\left(\Gamma_k^{(2)}\right)_{\bar{\sigma}\bar{\sigma}}\,
\left(\Gamma_k^{(2)}\right)_{\bar{\phi}\bar{\phi}}-\left(\Gamma_k^{(2)}
\right)_{\bar{\sigma}\bar{\phi}}^2>0$.
For sufficiently large values of $l$ the leading $l$-powers of $\Lambda_l(d,s)$
are the dominating contributions to the entries of $\Gamma_k^{(2)}$ in eq. 
(\ref{gravquad}) so that, 
in this limit, the above condition boils down to
\begin{eqnarray}
\label{noname}
0&<&\left(\Gamma_k^{(2)}\right)_{\bar{h}^T\bar{h}^T}
\stackrel{l\rightarrow\infty}
{\longrightarrow}\left(Z_{Nk}\kappa^2-\bar{\beta}_k\,R\right)\,\Lambda_l(d,2)
\hspace{0.5cm}\Longrightarrow\hspace{0.5cm}Z_{Nk}\kappa^2-\bar{\beta}_k\,R>0\\
0&<&\left(\Gamma_k^{(2)}\right)_{\bar{\xi}\bar{\xi}}\stackrel{l\rightarrow
\infty}{\longrightarrow} 2\frac{Z_{Nk}\kappa^2}{\alpha}\,\Lambda_l(d,1)
\hspace{0.5cm}\Longrightarrow\hspace{0.5cm}\frac{Z_{Nk}\kappa^2}{\alpha}>0\\
0&<&\left(\Gamma_k^{(2)}\right)_{\bar{\phi}\bar{\phi}}
\stackrel{l\rightarrow\infty}{\longrightarrow}
2\left(\frac{d-1}{d}\right)^2\,\bar{\beta}_k\,(\Lambda_l(d,0))^2\hspace{0.5cm}
\Longrightarrow\hspace{0.5cm}\bar{\beta}_k>0\\
0&<&\left(\Gamma_k^{(2)}\right)_{\bar{\sigma}\bar{\sigma}}\,
\left(\Gamma_k^{(2)}\right)_{\bar{\phi}\bar{\phi}}-\left(\Gamma_k^{(2)}
\right)_{\bar{\sigma}\bar{\phi}}^2\nonumber\\
& &
\stackrel{l\rightarrow\infty}{\longrightarrow}
\left(\frac{d-1}{d}
\right)^2\,\frac{Z_{Nk}\kappa^2\,\bar{\beta}_k}{\alpha}\,(\Lambda_l(d,0))^3>0
\hspace{0.5cm}\Longrightarrow\hspace{0.5cm} 
\frac{Z_{Nk}\kappa^2\,\bar{\beta}_k}{\alpha}>0
\end{eqnarray}
For nonnegative values of the gauge parameter, $\alpha\ge 0$, this leads to
the following restrictions on the dimensionless couplings:
\begin{eqnarray}
\label{rs}
g_k>0\;,\;\;\;\beta_k>0\;,\;\;\; k^2/(32\pi\,g_k\beta_k)>R\;.
\end{eqnarray}

In the UV fixed point regime of the $d=4$-dimensional case we have 
$g_k\approx g_*$ and $\beta_k\approx \beta_*$ with $g_*,\beta_*>0$. 
Hence, close to the non-Gaussian fixed point, the first two conditions of eq.
(\ref{rs}) are obviously satisfied. Furthermore, the third condition then takes
the form $R<k^2/(32\pi g_*\beta_*)$. For $R$ fixed this condition is satisfied
as well provided $k$ is sufficiently large. Thus, for $k$ large and on modes
with large eigenvalues of $-D^2$, the restricted operator $\Gamma_k^{(2)}$ is
positive. The cutoff should have the desired suppression properties therefore.

The above argument treats $R$ as a constant parameter. Recalling the derivation
of the projected flow equation where we compared powers of the radius $r\propto
R^{-1/2}$ it is indeed clear that in this context $r$ and $R$ should be 
regarded as fixed, $k$-independent quantities. 

It is instructive to look also at the operator $\Gamma_k^{(2)}[g^{\rm os}(k),
g^{\rm os}(k)]$ where $g^{\rm os}(k)$ is the $k$-dependent ``on-shell'' 
$S^d$-metric which solves the equation of motion $\delta\Gamma_k/\delta g_{\mu
\nu}=0$ for $g_{\mu\nu}=\bar{g}_{\mu\nu}$. The difference to the situation 
discussed before is that $R$ is a
function of $k$ now, to be computed from $g^{\rm os}(k)$. (The operator 
$\Gamma_k^{(2)}[g^{\rm os}(k),g^{\rm os}(k)]$ would appear in a standard 
one-loop (saddle point) calculation based upon the ``classical'' action
$\Gamma_k$.)

For the truncated action functional $\Gamma_k$ of eq. (\ref{trunc}) with 
$\bar{g}_{\mu\nu}=g_{\mu\nu}$ the field equation takes the form
\begin{eqnarray}
\label{fieldeqn}
2Z_{Nk}\kappa^2\left[G_{\mu\nu}+g_{\mu\nu}\,\bar{\lambda}_k\right]
+\bar{\beta}_k\left[-\left(G_{\mu\nu}+R_{\mu\nu}\right)R+2D_\mu D_\nu R
-2g_{\mu\nu}D^2R\right]=0
\end{eqnarray} 
with $G_{\mu\nu}=R_{\mu\nu}-g_{\mu\nu}R/2$ the Einstein tensor. Precisely for
$d=4$, the maximally symmetric solutions to eq. (\ref{fieldeqn}) 
satisfy Einstein's equation $G_{\mu\nu}=-g_{\mu\nu}\bar{\lambda}_k$, they are
not affected by the $R^2$-term. 
Inserting the contracted equation $R=4\bar{\lambda}_k$ into the third condition
of (\ref{rs}) leads to
\begin{eqnarray}
\label{oscond}
128\pi\,g_k\lambda_k\beta_k<1\;.
\end{eqnarray}
Remarkably, this condition is satisfied precisely if $\Gamma_k[g,g]$ is 
{\it not} a manifestly positive functional of $g_{\mu\nu}$, as follows from 
(\ref{posit}).
This implies that for $d=4$ the $S^4$ solution of eq. (\ref{fieldeqn}) cannot 
correspond to the absolute minimum of $\Gamma_k[g,g]$ if this functional is 
manifestly positive.

In the UV fixed point regime, the condition (\ref{oscond}) becomes 
$128\pi\,g_*\lambda_*\beta_*<1$. For all cutoffs employed we found that 
$0.17\lesssim 128\pi\,g_*\lambda_*\beta_*\lesssim 0.22$ so that this condition
is indeed satisfied. It is reassuring that also, upon inserting 
the $S^4$ solution of eq. (\ref{fieldeqn}), the Hessian $\Gamma_k^{(2)}$ 
becomes a positive operator for sufficiently large
values of $l$ and $k$, independently of the cutoff. 

Furthermore, the concomitant violation
of (\ref{posit}) implies that in the vicinity of the fixed point the functional
$\Gamma_k[g,g]$ is bounded below but not positive. By adding an appropriate
constant it is trivial though to turn it into a manifestly positive functional.

\subsection{Positivity of the cutoff}
\label{3S6B}
The cutoff $\Delta_kS$ is expected to be positive definite under the same 
conditions as found for the Hessian $\Gamma_k^{(2)}$ in the previous 
subsection. In order to obtain more quantitative information about 
$l_{\rm min}$ and the momentum
regime where $\Delta_k S$ is positive we continue our analysis with an explicit
investigation of the cutoff operator ${\cal R}_k$.
For simplicity we again restrict our considerations to the most interesting 
case of $d=4$ and to spherical backgrounds. 

After setting 
$\left({\cal R}_k\right)_{\bar{h}^T\bar{h}^T}^{\mu\nu\alpha\beta}
\equiv1/2\left(g^{\mu\alpha}g^{\nu\beta}+g^{\mu\beta}g^{\nu\alpha}\right)
\left({\cal R}_k\right)_{\bar{h}^T\bar{h}^T}$ and 
$\left({\cal R}_k\right)^{\mu\nu}_{\bar{\xi}\bar{\xi}}\equiv g^{\mu\nu}
\left({\cal R}_k\right)_{\bar{\xi}\bar{\xi}}$, and inserting the eigenvalues
of the covariant Laplacians, the entries of the cutoff matrix (\ref{cutoff}) 
assume the form
\begin{eqnarray}
\label{posdef}
\left({\cal R}_k\right)_{\bar{h}^T\bar{h}^T}
&=&\frac{k^4R^{(0)}(\Lambda_l(4,2)/k^2)}
{32\pi\,g_k}\left\{1-32\pi\,g_k\beta_k \,\frac{R}{k^2}\right\}\;,\nonumber\\
\left({\cal R}_k\right)_{\bar{\xi}\bar{\xi}}&=&
\frac{k^4R^{(0)}(\Lambda_l(4,1)/k^2)}
{16\pi\,g_k\alpha}\;,\nonumber\\
\left({\cal R}_k\right)_{\bar{\sigma}\bar{\sigma}}
&=&\frac{k^4R^{(0)}(\Lambda_l(4,0)/k^2)}
{32\pi\,g_k}\left\{
36\pi\,g_k\beta_k\left(2\Lambda_l(4,0)/
k^2+R^{(0)}(\Lambda_l(4,0)/k^2)\right)
+\frac{3}{4}\right\}\;,\nonumber\\
\left({\cal R}_k\right)_{\bar{\phi}\bar{\sigma}}
&=&\left({\cal R}_k\right)_{\bar{\sigma}\bar{\phi}_1}^\dagger
=\frac{9}{8}\beta_k\,k^4\Bigg\{
\left[\Lambda_l(4,0)/k^2+R^{(0)}
(\Lambda_l(4,0)/k^2)\right]^{\frac{3}{2}}\nonumber\\
& &\times\sqrt{\Lambda_l(4,0)/k^2+R^{(0)}
(\Lambda_l(4,0)/k^2)-\frac{R}{3k^2}}\nonumber\\
& &-\left[\Lambda_l(4,0)/k^2\right]^{\frac{3}{2}}\sqrt{\Lambda_l(4,0)/k^2
-\frac{R}{3k^2}}\Bigg]\Bigg\}\;,\\
\left({\cal R}_k\right)_{\bar{\phi}\bar{\phi}}&=&
\frac{k^4R^{(0)}(\Lambda_l(4,0)/k^2)}
{32\pi\,g_k}\left\{
36\pi\,g_k\beta_k
\left(2\Lambda_l(4,0)/k^2
+R^{(0)}(\Lambda_l(4,0)/k^2)\right)-\frac{1}{4}\right\}\;.\nonumber
\end{eqnarray} 
As compared to eq. (\ref{cutoff}) which was written in terms of the 
dimensionful
quantities $Z_{Nk}\kappa^2$, $\bar{\lambda}_k$, and $\bar{\beta}_k$, we 
switched here to a description in terms of the dimensionless couplings.

In analogy with the Hessian $\Gamma^{(2)}_k$ the condition for the cutoff 
matrix ${\cal R}_k$ to be positive definite reads:
$\left({\cal R}_k\right)_{\bar{h}^T\bar{h}^T}>0$, 
$\left({\cal R}_k\right)_{\bar{\xi}\bar{\xi}}>0$, 
$\left({\cal R}_k\right)_{\bar{\phi}\bar{\phi}}>0$, and
$\left({\cal R}_k\right)_{\bar{\phi}\bar{\phi}}\,
\left({\cal R}_k\right)_{\bar{\sigma}\bar{\sigma}}
-\left({\cal R}_k\right)_{\bar{\phi}\bar{\sigma}}^2>0$.
These conditions indeed reproduce the restrictions on the couplings obtained 
from $\Gamma_k^{(2)}$ for sufficiently large momenta $p^2=\Lambda_l(d,s)$. 
Provided that $\alpha\ge 0$, they take the form 
$g_k>0$, $\beta_k>0$ and $k^2/(32\pi\,g_k\beta_k)>R$, which coincides with
eq. (\ref{rs}).

Given an arbitrary set of parameters $(R,k,g_k,\beta_k)$ satisfying these 
three inequalities, we have $\left({\cal R}_k\right)_{\bar{h}^T\bar{h}^T}>0$ 
and $\left({\cal R}_k\right)_{\bar{\xi}\bar{\xi}}>0$ (and also 
$\left({\cal R}_k\right)_{\bar{\sigma}\bar{\sigma}}>0$) for {\it any} allowed
value of $l$. This is not the case for the other two conditions which stem from
the scalar sector of the cutoff. Clearly 
$\left({\cal R}_k\right)_{\bar{\phi}\bar{\phi}}$ of eq. (\ref{posdef}) can 
assume negative values for sufficiently small values of $l$, provided $g_k$, 
$\beta_k$ and $R/k^2$ are small enough. Since 
$\left({\cal R}_k\right)_{\bar{\sigma}\bar{\sigma}}>0$, a negative
$\left({\cal R}_k\right)_{\bar{\phi}\bar{\phi}}$ implies that also 
$\left({\cal R}_k\right)_{\bar{\phi}\bar{\phi}}\,\left({\cal R}_k
\right)_{\bar{\sigma}\bar{\sigma}}-\left({\cal R}_k
\right)_{\bar{\phi}\bar{\sigma}}^2<0$. 

The $l$-values for which 
$\left({\cal R}_k\right)_{\bar{\phi}\bar{\phi}}>0$  
satisfy 
\begin{eqnarray}
\label{phiphicond} 
2\Lambda_l(4,0)/k^2+R^{(0)}\Big(\Lambda_l(4,0)/k^2\Big)>
\frac{1}{144\pi\,g_k\beta_k}\;.
\end{eqnarray}
In appendix \ref{estimation} we derive a similar inequality involving the 
$\bar{\phi}$-$\bar{\sigma}$ cross term. There we find that
$\left({\cal R}_k\right)_{\bar{\phi}\bar{\phi}}\,\left({\cal R}_k
\right)_{\bar{\sigma}\bar{\sigma}}-\left({\cal R}_k
\right)_{\bar{\phi}\bar{\sigma}}^2>0$ at least for all values of $l$ 
satisfying
\begin{eqnarray}
\label{phisigma}
2\Lambda_l(4,0)/k^2+R^{(0)}\Big(\Lambda_l(4,0)/k^2\Big)
>\frac{1}{96\pi\,g_k\beta_k}\;.
\end{eqnarray}

Both inequalities, (\ref{phiphicond}) and (\ref{phisigma}), depend on 
$g_k\beta_k$, on $p^2/k^2$ with $p^2=\Lambda_l(4,0)=l(l+3)R/12$, and on the 
shape function $R^{(0)}$. Given a specific set $(R,k,g_k,\beta_k;R^{(0)};
T^{lm})$ with a certain scalar eigenmode $T^{lm}$, they 
tell us whether the contribution from this mode is suppressed correctly or 
not. The more restrictive inequality (\ref{phisigma}) applies to the scalar 
eigenmodes $\{T^{lm}\}$ with $l\ge 2$, while (\ref{phiphicond}) can be used 
for the constant mode $T^{l=0,m=1}$ and the PCKV's $\{T^{l=1,m}\}$ only.

Let us now focus on RG trajectories which run into the
non-Gaussian fixed point as $k\rightarrow\infty$. Furthermore, we assume that 
$R$ is either kept fixed or that $R=a_k\,k^2$ with a constant 
$a_k<(32\pi\,g_k\beta_k)^{-1}$. Then, for large enough values of $k$, we have 
$g_k\approx g_*>0$, $\beta_k\approx\beta_*>0$ and $k^2/(32\pi\,g_*\beta_*)>R$ 
so that the conditions (\ref{rs}) for the positivity of $\Gamma_k^{(2)}$ and
${\cal R}_k$ are satisfied. Moreover, the 
RHS of (\ref{phiphicond}) and (\ref{phisigma}) may be expressed as
$(144\pi\,g_*\beta_*)^{-1}$ and $(96\pi\,g_*\beta_*)^{-1}$, respectively.

Now we are in a position to determine the $l$-regime for which the
cutoff is manifestly positive definite. Using the family of exponential shape 
functions (\ref{expshape}) with $1\le s\le 30$, a numerical 
analysis reveals that {\it any} value of the ratio 
$x\equiv p^2/k^2=\Lambda_l(4,0)/k^2$ satisfies 
(\ref{phiphicond}) or (\ref{phisigma}) provided $s\gtrsim 2$ or 
$s\gtrsim 7$, respectively. Hence, under the above conditions, all cutoffs
employing an exponential shape function with $s\gtrsim 7$ 
are manifestly positive definite for {\it all} momenta, i.e. for all quantum
numbers $l$. 

This is a rather intriguing result. It might indicate that cutoffs with 
$s>7$ are particularly reliable.

\begin{figure}[ht]
\begin{minipage}{7.9cm}
        \epsfxsize=7.9cm
        \epsfysize=5.2cm
        \centerline{\epsffile{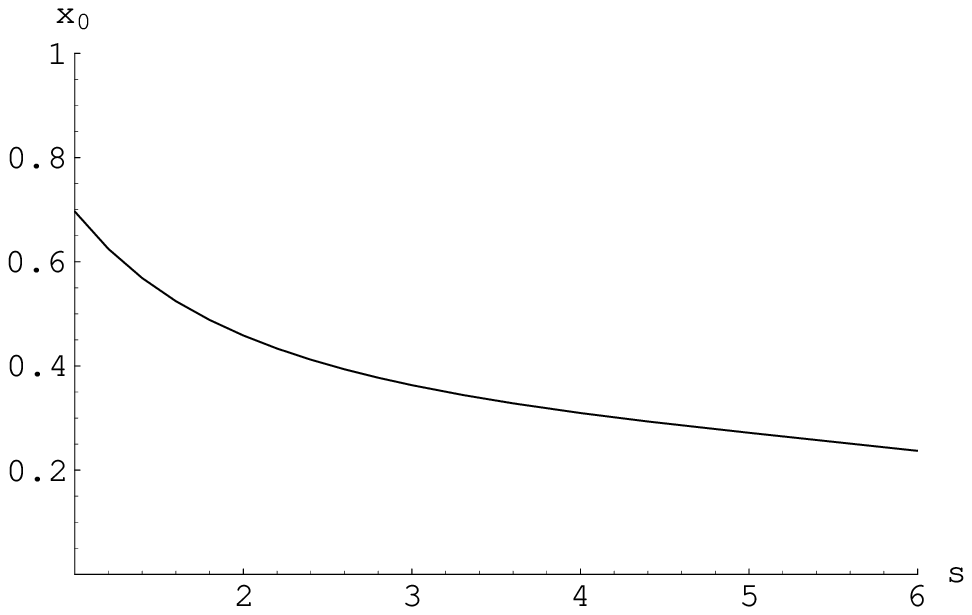}}
\centerline{(a)}
\end{minipage}
\hfill
\begin{minipage}{7.9cm}
        \epsfxsize=7.9cm
        \epsfysize=5.2cm
        \centerline{\epsffile{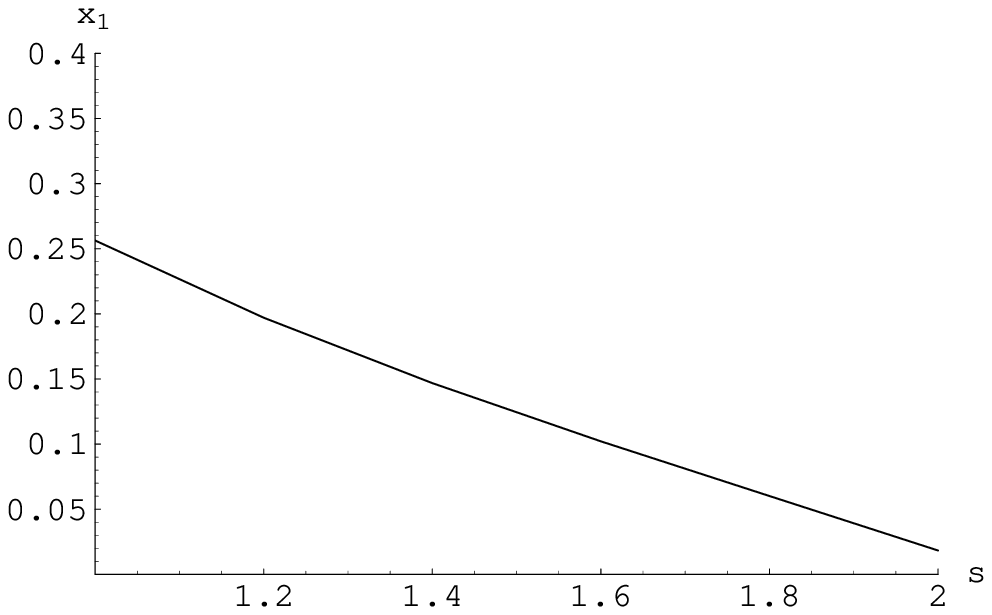}}
\centerline{(b)}
\end{minipage}
\caption{(a) $x_0$, and (b) $x_1$ as functions of $s$, using the family of 
exponential shape functions.}  
\label{plotA}
\end{figure}

Conversely, for any $s\lesssim 7$
there exists a specific value $x_0(s)$ such that all $x$ with $x\le x_0(s)$ 
violate (\ref{phisigma}). Furthermore, there exists a specific $x_1(s)$
for any $s\lesssim 2$ such that any $x\le x_1(s)$ leads to a violation of
(\ref{phiphicond}). In FIG. \ref{plotA} we show $x_0(s)$ and $x_1(s)$ in the 
ranges $1\le s\le 6$ and $1\le s\le 2$, respectively. It is important to note
that $x_0(s)<0.7$ and $x_1(s)<0.26$ for any value of $s$ considered. This 
implies
that in the UV fixed point regime the cutoff has the desired suppression 
properties for all modes with momenta ranging from infinity down to values well
below $k$.

\renewcommand{\baselinestretch}{1}
\small\normalsize
\renewcommand{\baselinestretch}{1.5}
\small\normalsize

To complete the analysis let us study the inequalities (\ref{phiphicond}) and 
(\ref{phisigma}) also at the
spherically symmetric stationary point $g^{\rm os}(k)$ of $\Gamma_k[g,g]$ which
we discussed in the previous subsection. In the vicinity of the fixed point 
the on-shell value of the curvature is $R\approx4\lambda_* k^2$. Hence, we 
obtain from (\ref{phiphicond}) and 
(\ref{phisigma}), respectively,
\begin{eqnarray}
\label{phiphi2}
f_1[R^{(0)};l]\equiv\frac{2l(l+3)\lambda_*}{3}+R^{(0)}(l(l+3)\lambda_*/3)-
\frac{1}{144\pi\,g_*\beta_*}>0\;,\;\;l=0,1
\end{eqnarray}
and  
\begin{eqnarray}
\label{phisigma2}
f_0[R^{(0)};l]\equiv\frac{2l(l+3)\lambda_*}{3}+R^{(0)}(l(l+3)\lambda_*/3)-
\frac{1}{96\pi\,g_*\beta_*}>0\;,\;\;l\ge 2\;.
\end{eqnarray}
The first inequality stems from the scalar eigenmodes $T^{lm}$ with $l=0,1$, 
and the second from those with $l\ge 2$. Both (\ref{phiphi2}) and
(\ref{phisigma2}) depend on $l$ and $R^{(0)}$.

\begin{figure}[ht]
\begin{minipage}{7.9cm}
        \epsfxsize=7.9cm
        \epsfysize=5.2cm
        \centerline{\epsffile{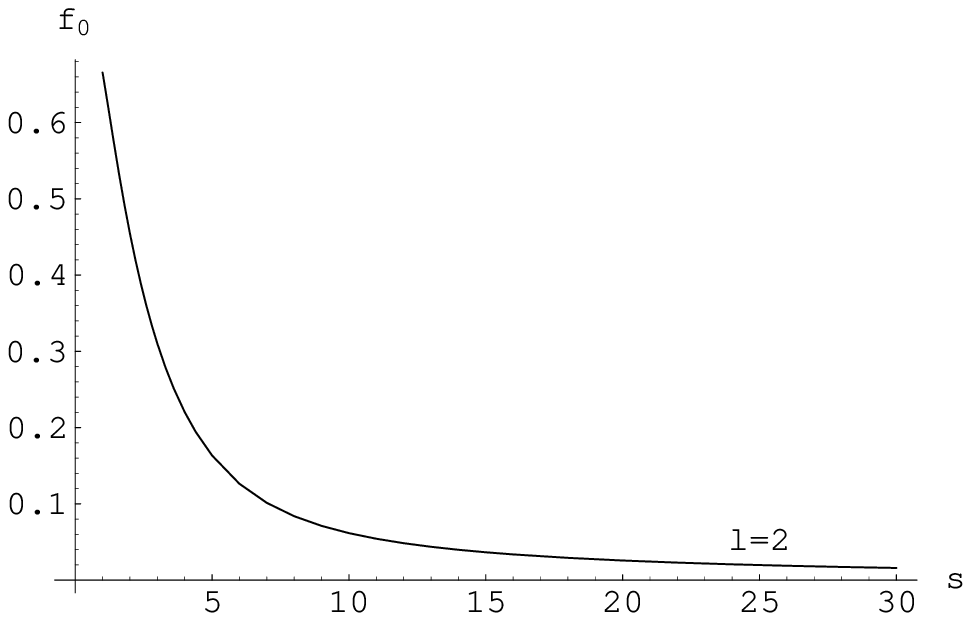}}
\centerline{(a)}
\end{minipage}
\hfill
\begin{minipage}{7.9cm}
        \epsfxsize=7.9cm
        \epsfysize=5.2cm
        \centerline{\epsffile{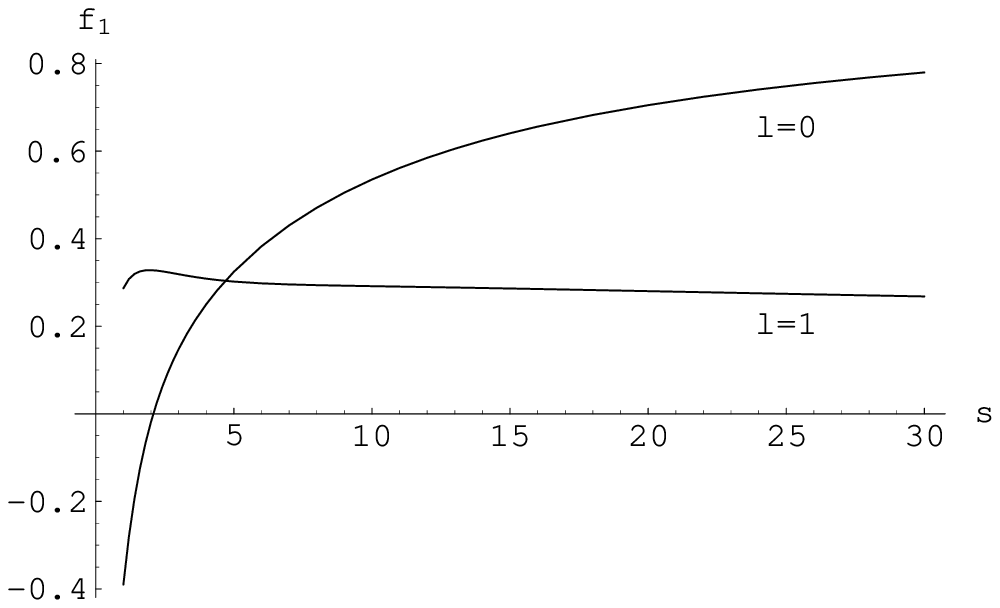}}
\centerline{(b)}
\end{minipage}
\caption{(a) $f_0(s,l=2)$, and (b) $f_1(s,l=0)$ and $f_1(s,l=1)$ as functions 
of $s$, using the family of exponential shape functions.}  
\label{plotB}
\end{figure}

Again we restrict our investigation to the family of exponential shape 
functions with 
$1\le s\le 30$. Then the LHS of (\ref{phiphi2}) and (\ref{phisigma2}) are 
functions of $s$ and $l$: $f_1[R^{(0)};l]\equiv f_1(s,l)$, $f_0[R^{(0)};l]
\equiv f_0(s,l)$. For $l\ge 2$ we have $f_0(s;l)\ge f_0(s;l=2)$ independently 
of the shape parameter. Numerically we find that $f_0(s;l=2)$ is always 
positive. Hence, {\it any} momentum with $l\ge 2$ satisfies the condition 
(\ref{phisigma2}) for all values of $s$ considered. Furthermore, our numerical
analysis shows that also $f_1(s,l=1)>0$ for all cutoffs employed. However, 
$f_1(s,l=0)$ is not always positive. We obtain $f_1(s,l=0)>0$ for $s\gtrsim 2$ 
and $f_1(s,l=0)<0$ for $s\lesssim 2$. Our results are illustrated in FIG. 
\ref{plotB}.

\renewcommand{\baselinestretch}{1}
\small\normalsize
\renewcommand{\baselinestretch}{1.5}
\small\normalsize

The above results have to be interpreted as follows. Assume that $k$
lies in the UV scaling regime, and consider the cutoff operator at the 
spherically symmetric stationary point, ${\cal R}_k[g^{\rm os}(k)]$. Then this 
operator is strictly positive on the space spanned by {\it all} 
spherical harmonics and would correctly suppress all 
modes in a path integral containing this ``on-shell'' cutoff, provided 
we choose an exponential shape function with $s\gtrsim 2$. For 
$s\lesssim 2$, only the contributions from the constant mode with $p^2=0$ are
not suppressed correctly. For any other mode the cutoff term is positive 
even in this case.

{\bf To summarize:} In this section we found that at least in the asymptotic
domain relevant in our investigation of the UV fixed point  the cutoff which
is adapted to  the $R^2$-truncation is positive definite and therefore has all
the required  mode suppression properties. No conformal factor problem and no
growing exponentials produced by $\exp(-\Delta_kS)$ are encountered.

\section{Summary and Conclusion}
\renewcommand{\theequation}{7.\arabic{equation}}
\setcounter{equation}{0}
\label{conclusio}
In this paper we evaluated the exact RG equation of Quantum Einstein Gravity
in a truncation which generalizes the Einstein-Hilbert approximation used so
far by the inclusion of a higher-derivative term. We derived the beta-functions
of the resulting $\lambda$-$g$-$\beta-$system which turned out to be by far 
more complicated than the old $\lambda$-$g-$system. We used these 
beta-functions in order to investigate how the two fixed points known to exist
in the Einstein-Hilbert truncation manifest themselves in the enlarged theory
space.

We found that the Gaussian fixed point of the Einstein-Hilbert truncation does
not generalize to a corresponding fixed point of the $R^2$-truncation. 
Nevertheless, the 2-dimensional projection of the $\lambda$-$g$-$\beta-$flow
onto the $\lambda$-$g-$plane at $\beta=0$, near the origin $\lambda=g=\beta=0$,
is well approximated by the flow resulting from the Einstein-Hilbert 
truncation. The projected flow does indeed have a fixed point at $\lambda_*
=g_*=0$. In the Einstein-Hilbert truncation there exists a distinguished RG
trajectory, the ``separatrix'' \cite{frank}, which gives rise to a vanishing
renormalized cosmological constant, $\lim_{k\rightarrow 0}\bar{\lambda}_k=0$.
In $d=4$, it turned out that this trajectory possesses a 3-dimensional ``lift''
which is characterized by a logarithmic running of the $R^2$-coupling 
$\beta_k$. For $d\neq 4$ its running is power-like, and there exists a 
``quasi-Gaussian'' fixed point at $\lambda_*=g_*=0$, $\beta_*\neq 0$. This
picture puts the older perturbative calculations in $R^2$-gravity into a
broader context.

Quite differently, the non-Gaussian fixed point $(\lambda_*,g_*)\neq(0,0)$ 
implied by the Einstein-Hilbert truncation does ``lift'' to a corresponding 
fixed point of the 3-dimensional flow, with a tiny but nonzero third component 
$\beta_*>0$. It is UV attractive in all 3 directions of 
$\lambda$-$g$-$\beta-$space. We demonstrated in detail that close to the fixed
point the flow on the extended theory space is essentially 2-dimensional, and
that the 2-dimensional projected flow is very well approximated by the 
Einstein-Hilbert flow. For the $\lambda_*$- and $g_*$-coordinates both 
truncations yield virtually identical values, and the same is true for the
critical exponents pertaining to the 2-dimensional subspace.  For universal
quantities the differences between the two truncations are typically smaller 
than their weak residual scheme dependence. 

This stability of the Einstein-Hilbert truncation against the inclusion of
a further invariant, together with the other pieces of evidence which we
summarized in Sections \ref{3S5.3.1} and \ref{3S5.3.2} strongly support the
conjecture that this approximation is at least qualitatively reliable in the 
UV. Hence it appears increasingly unlikely that the very existence of the
non-Gaussian fixed point is an artifact of the truncation. We believe that 
QEG has indeed very good chances of being nonperturbatively renormalizable.

A notorious difficulty of Euclidean quantum gravity is the conformal factor
problem. In the exact RG approach, it appears in the Einstein-Hilbert 
truncation, but not in the $R^2$-truncation provided $k$ is large enough.
When this complication occurs the construction of an appropriate cutoff 
operator is rather subtle. However, it was possible to show that our 
investigation of the non-Gaussian fixed point in the $R^2$-truncation is not
affected by this problem, and that a straightforward positive definite cutoff
can be employed. The numerical agreement of the results with those from the
Einstein-Hilbert truncation indicates that the rule for constructing an 
adapted cutoff in presence of the conformal factor problem which was proposed
in \cite{Reu96} (``${\cal Z}_k=z_k$-rule'') should indeed be correct.\\[24pt]
Acknowledgments: We would like to thank C. Wetterich for many helpful 
discussions. We are also grateful to A. Ashtekar, A. Bonanno, W. Dittrich,
P. Forg\'{a}cs, H. Gies, E. Gozzi, D. Litim, M. Niedermaier, H. B. Nielsen, R.
Percacci, J. B. Pitts, M. Salmhofer, F. Saueressig, L. Smolin, T. Thiemann, and
G. Veneziano for interesting conversations and communications.
\newpage
\begin{appendix}
\section{Evaluating the RHS of the truncated flow equation}
\renewcommand{\theequation}{A\arabic{equation}}
\setcounter{equation}{0}
\label{eval}
In this section we present several rather lengthy calculations needed for the 
discussion of the $R^2$-truncation in section \ref{$R^2$-trunc}. In the 
following, all calculations are performed with $g_{\mu\nu}=\bar{g}_{\mu\nu}$ 
where $\bar{g}_{\mu\nu}$ is assumed to correspond to a $S^d$ background. 
For simplicity the bars are omitted from the metric, the curvature and the 
operators.
\subsection{Computation of $\left(\Gamma_k^{(2)}
+{\cal R}_k\right)^{-1}$ and $\left(S_{\rm gh}^{(2)}+{\cal R}_k\right)^{-1}$}
\label{inv}
In section \ref{$R^2$-trunc} we derived explicit expressions for the kinetic 
operators $\widetilde{\Gamma}_k^{(2)}\equiv\Gamma_k^{(2)}+{\cal R}_k$ and
$\widetilde{S}_{\rm gh}^{(2)}\equiv S_{\rm gh}^{(2)}+{\cal R}_k$. They may be
represented as matrix differential operators acting on the column vectors
$(\bar{h}^T,\bar{\xi},\bar{\phi}_0,\bar{\sigma},\bar{\phi}_1)^{\bf T}$ and
$(\bar{v}^T,v^T,\bar{\varrho},\varrho)^{\bf T}$, respectively. In this
representation they take the form
\begin{eqnarray}
\label{83}
\widetilde{\Gamma}_k^{(2)}[g,g]=
\left(\begin{array}{cccc}
\left(\widetilde{\Gamma}_k^{(2)}[g,g]\right)_{\bar{h}^T\bar{h}^T} & 0 & 0 &
0_{1\times 2}\\0 & \left(\widetilde{\Gamma}_k^{(2)}[g,g]\right)_{\bar{\xi}
\bar{\xi}} & 0 & 0_{1\times 2}\\
0 & 0 & \left(\widetilde{\Gamma}_k^{(2)}[g,g]\right)_{\bar{\phi}_0
\bar{\phi}_0} & 0_{1\times 2}\\
0_{2\times 1} & 0_{2\times 1} & 0_{2\times 1} & {\cal Q}_k
\end{array}\right)
\end{eqnarray}
and
\begin{eqnarray}
\label{84}
\widetilde{S}_{\rm gh}^{(2)}[g,g]=
\left(\begin{array}{cccc}
0 & \left(\widetilde{S}_{\rm gh}^{(2)}[g,g]\right)_{\bar{v}^Tv^T} & 0 & 0\\ 
\left(\widetilde{S}_{\rm gh}^{(2)}[g,g]\right)_{v^T\bar{v}^T} & 0 & 0 & 0\\
0 & 0 & 0 & \left(\widetilde{S}_{\rm gh}^{(2)}[g,g]\right)_{\bar{\varrho}
\varrho}\\
0 & 0 & \left(\widetilde{S}_{\rm gh}^{(2)}[g,g]\right)_{\varrho\bar{\varrho}}
& 0
\end{array}\right) 
\end{eqnarray}
where
\begin{eqnarray}
\label{85}
{\cal Q}_k\equiv\left(
\begin{array}{cc}
\left(\widetilde{\Gamma}_k^{(2)}[g,g]\right)_{\bar{\sigma}
\bar{\sigma}} & \left(\widetilde{\Gamma}_k^{(2)}[g,g]
\right)_{\bar{\phi}_1\bar{\sigma}}\\
\left(\widetilde{\Gamma}_k^{(2)}[g,g]\right)_{\bar{\phi}_1\bar{\sigma}} & 
\left(\widetilde{\Gamma}_k^{(2)}[g,g]\right)_{\bar{\phi}\bar{\phi}}
\end{array}\right)\;.
\end{eqnarray}
The entries of these matrices are given in eq. (\ref{entries}). On the RHS of 
the flow equation (\ref{truncflow}) the inverse operators 
$[\widetilde{\Gamma}_k^{(2)}]^{-1}$ and $[\widetilde{S}_{\rm gh}^{(2)}]^{-1}$ 
appear which we determine in the following. At this point it is important
to note that, because of the maximally symmetric background, all covariant 
derivatives 
contained in the operators (\ref{83}) and (\ref{84}) appear as covariant 
Laplacians, and that the various entries are $x$-independent otherwise. This 
implies that these entries are {\it commuting} differential operators which 
allows for particularly simple manipulations. Therefore it is not difficult to
verify that the inverse operators assume the form
\begin{eqnarray}
\label{86}
\lefteqn{\left(\widetilde{\Gamma}_k^{(2)}[g,g]\right)^{-1}=}\nonumber\\
& &\left(\begin{array}{cccc}
\left[\left(\widetilde{\Gamma}_k^{(2)}[g,g]\right)_{\bar{h}^T\bar{h}^T}
\right]^{-1} & 0 &0 & 0_{1\times 2}\\
0 & \left[\left(\widetilde{\Gamma}_k^{(2)}[g,g]\right)_{\bar{\xi}\bar{\xi}}
\right]^{-1} & 0 & 0_{1\times 2}\\
0 & 0 &\left[\left(\widetilde{\Gamma}_k^{(2)}[g,g]\right)_{\bar{\phi}_0
\bar{\phi}_0}\right]^{-1} & 0_{1\times 2}\\
0_{2\times 1} & 0_{2\times 1} & 0_{2\times 1} & {\cal Q}_k^{-1}
\end{array}\right)
\end{eqnarray}
and
\begin{eqnarray}
\label{87}
& &\!\!\!\!\!\left(\widetilde{S}_{\rm gh}^{(2)}[g,g]\right)^{-1}=\\
& &\!\!\!\!\!\left(\begin{array}{cccc}
0 & \left[\left(\widetilde{S}_{\rm gh}^{(2)}[g,g]\right)_{\bar{v}^Tv^T}
\right]^{-1} & 0 & 0\\ 
\left[\left(\widetilde{S}_{\rm gh}^{(2)}[g,g]\right)_{v^T\bar{v}^T}
\right]^{-1} & 0 & 0 & 0\\
0 & 0 & 0 & \left[\left(\widetilde{S}_{\rm gh}^{(2)}[g,g]\right)_{
\bar{\varrho}\varrho}\right]^{-1}\\
0 & 0 &\left[\left(\widetilde{S}_{\rm gh}^{(2)}[g,g]\right)_{\varrho
\bar{\varrho}}\right]^{-1} & 0
\end{array}\right)\nonumber
\end{eqnarray}
with
\begin{eqnarray}
\label{88}
{\cal Q}_k^{-1}&=&
\left[\left(\widetilde{\Gamma}_k^{(2)}[g,g]\right)_{\bar{\sigma}
\bar{\sigma}} 
\left(\widetilde{\Gamma}_k^{(2)}[g,g]\right)_{\bar{\phi}_1\bar{\phi}_1}
-\left(\widetilde{\Gamma}_k^{(2)}[g,g]\right)^2_{\bar{\phi}_1\bar{\sigma}}
\right]^{-1}\nonumber\\
& &\times\left(\begin{array}{cc}
\left(\widetilde{\Gamma}_k^{(2)}[g,g]\right)_{\bar{\phi}_1\bar{\phi}_1}
&-\left(\widetilde{\Gamma}_k^{(2)}[g,g]
\right)_{\bar{\phi}_1\bar{\sigma}}\\
-\left(\widetilde{\Gamma}_k^{(2)}[g,g]\right)_{\bar{\phi}_1\bar{\sigma}} & 
\left(\widetilde{\Gamma}_k^{(2)}[g,g]\right)_{\bar{\sigma}\bar{\sigma}} 
\end{array}\right)\;.
\end{eqnarray}
Inserting these expressions into the RHS of the flow equation (\ref{truncflow})
leads to
\begin{eqnarray}
\label{061}
\lefteqn{{\cal S}_k(R)
=\frac{1}{2}{\rm Tr}'\left[\sum\limits_{\zeta\in\{\bar{h}^T,
\bar{\xi},\bar{\phi}_0\}}
\left[\left(\Gamma_k^{(2)}[g,g]
+{\cal R}_k[g]\right)_{\zeta\zeta}\right]^{-1}
\partial_t\left({\cal R}_k[g]\right)_{\zeta\zeta}\right]}
\nonumber\\
& &+\frac{1}{2}{\rm Tr}'\left[
\left\{\left(\widetilde{\Gamma}_k^{(2)}[g,g]\right)_{\bar{\sigma}\bar{\sigma}} 
\left(\widetilde{\Gamma}_k^{(2)}[g,g]\right)_{\bar{\phi}_1\bar{\phi}_1}
-\left(\widetilde{\Gamma}_k^{(2)}[g,g]\right)^2_{\bar{\phi}_1\bar{\sigma}}
\right\}^{-1}\right.\nonumber\\
& &\times\left\{\left(\Gamma_k^{(2)}[g,g]
+{\cal R}_k[g]\right)_{\bar{\sigma}\bar{\sigma}}
\partial_t\left({\cal R}_k[g]\right)_{\bar{\phi}_1\bar{\phi}_1}
+\left(\Gamma_k^{(2)}[g,g]+{\cal R}_k[g]\right)_{\bar{\phi}_1\bar{\phi}_1}
\partial_t\left({\cal R}_k[g]\right)_{\bar{\sigma}
\bar{\sigma}}\right.\nonumber\\
& &\left.\left.-2\left(\Gamma_k^{(2)}[g,g]+{\cal R}_k[g]
\right)_{\bar{\phi}_1\bar{\sigma}}
\partial_t\left({\cal R}_k[g]\right)_{\bar{\phi}_1\bar{\sigma}}
\right\}\right]\nonumber\\
& &-{\rm Tr}'\left[\sum\limits_{\psi\in\{v^T,\varrho\}}
\left[\left(S_{\rm gh}^{(2)}[g,g]
+{\cal R}_k[g]\right)_{\bar{\psi}\psi}\right]^{-1}
\partial_t\left({\cal R}_k[g]\right)_{\bar{\psi}\psi}\right]
\end{eqnarray}
where we used the relations
\begin{eqnarray}
\label{060}
{\left[\left(S_{\rm gh}^{(2)}\right)_{\bar{v}^Tv^T}
\right]^{\mu x}}_{\nu y}
&=&-{\left[\left(S_{\rm gh}^{(2)}\right)_{v^T\bar{v}^T}
\right]_{\nu y}}^{\mu x}
=\frac{1}{\sqrt{g(y)}}\frac{\delta}{\delta v^{T\nu}(y)}
\frac{1}{\sqrt{g(x)}}\frac{\delta S_{\rm gh}}{\delta\bar{v}^T_\mu(x)}
\nonumber\\
{\left[\left(S_{\rm gh}^{(2)}\right)_{\bar{\varrho}\varrho}
\right]^x}_y
&=&-{\left[\left(S_{\rm gh}^{(2)}\right)_{\varrho\bar{\varrho}}
\right]_y}^x
=\frac{1}{\sqrt{g(y)}}\frac{\delta}{\delta\varrho(y)}
\frac{1}{\sqrt{g(x)}}\frac{\delta S_{\rm gh}}{\delta\bar{\varrho}(x)}\;.
\end{eqnarray}
The trace of the $\phi_0$-term appearing in the first line of eq. (\ref{061}) 
may be evaluated easily since only the scalar eigenmodes  $T^{01}$ and $T^{1m}$
contribute. We obtain
\begin{eqnarray}
\label{phi0term}
\lefteqn{\frac{1}{2}{\rm Tr}'\left[\left[\left(\Gamma_k^{(2)}[g,g]
+{\cal R}_k[g]\right)_{\phi_0\phi_0}\right]^{-1}
\partial_t\left({\cal R}_k[g]\right)_{\phi_0\phi_0}\right]}\nonumber\\
&=&\frac{1}{2Z_{Nk}\kappa^2}\sum\limits_{l=0}^{1}\sum\limits_{m=1}^{D_l(d,0)}
\int d^dx\,\sqrt{g(x)}\,T^{lm}(x)
\bigg[C_{S2}(d,\alpha)\,C_{S1}(d,\alpha)\,\Big(P_k+A_{S1}(d,\alpha)\,R
\nonumber\\
& &+B_{S1}(d,\alpha)\,\bar{\lambda}_k\Big)
+(Z_{Nk}\kappa^2)^{-1}\bar{\beta}_k\Big(H_S(d)\,P_k^2
+G_{S2}(d)\,R\,P_k+G_{S3}(d)\,R^2\Big)\bigg]^{-1}
\nonumber\\
& &\times\partial_t\bigg[
C_{S2}(d,\alpha)\,C_{S1}(d,\alpha)\,Z_{Nk}\kappa^2
\,k^2R^{(0)}(-D^2/k^2)
+\bar{\beta}_k\Big(H_S(d)\,\left(-2D^2\,R^{(0)}(-D^2/k^2)\right.\nonumber\\
& &+\left.k^4R^{(0)}(-D^2/k^2)^2\right)
+G_{S2}(d)\,R\,k^2R^{(0)}(-D^2/k^2)\Big)\bigg]
T^{lm}(x)\nonumber\\
&=&
\frac{1}{2Z_{Nk}\kappa^2}\sum\limits_{l=0}^{1}\Bigg[
D_l(d,0)\bigg\{C_{S2}(d,\alpha)\,C_{S1}(d,\alpha)\,\Big(\Lambda_l(d,0)+
k^2R^{(0)}(\Lambda_l(d,0)/k^2)\nonumber\\
& &+A_{S1}(d,\alpha)\,R
+B_{S1}(d,\alpha)\,\bar{\lambda}_k\Big)\nonumber\\
& &+(Z_{Nk}\kappa^2)^{-1}\bar{\beta}_k\Big(H_S(d)\,\left(\Lambda_l(d,0)+
k^2R^{(0)}(\Lambda_l(d,0)/k^2)\right)^2\nonumber\\
& &+G_{S2}(d)\,R\left(\Lambda_l(d,0)+
k^2R^{(0)}(\Lambda_l(d,0)/k^2)\right)+G_{S3}(d)\,R^2\Big)\bigg\}^{-1}
\nonumber\\
& &\times\partial_t\bigg\{
C_{S2}(d,\alpha)\,C_{S1}(d,\alpha)\,Z_{Nk}\kappa^2
\,k^2R^{(0)}(\Lambda_l(d,0)/k^2)\nonumber\\
& &+\bar{\beta}_k\Big(H_S(d)\,
\left(2\Lambda_l(d,0)\,k^2R^{(0)}(\Lambda_l(d,0)/k^2)
+k^4R^{(0)}(\Lambda_l(d,0)/k^2)^2\right)\nonumber\\
& &+G_{S2}(d)\,R\,k^2R^{(0)}(\Lambda_l(d,0)/k^2)\Big)\bigg\}
\Bigg]
\;.
\end{eqnarray}
Here $\Lambda_l(d,0)$ is the eigenvalue with respect to $-D^2$ corresponding
to $T^{lm}$. Inserting also the remaining entries given in eq. 
(\ref{entries}) into eq. (\ref{061}) finally leads to eq. (\ref{squareflow}).
\subsection{Heat kernel expansion and evaluation of the traces}
\label{trace}
In this part of the appendix we expand ${\cal S}_k(R)$ of eq. 
(\ref{squareflow}) with respect to $r$
and evaluate the traces appearing in the resulting equation (\ref{derivexp})
below by applying the 
heat kernel expansion. Thereby we extract the contributions
proportional to $\int d^dx\,\sqrt{g}$, $\int d^dx\,\sqrt{g}R$ and 
$\int d^dx\,\sqrt{g}R^2$. This puts us in a position to read off the
RG equations for the three couplings. For technical convenience we restrict our
considerations to the gauge $\alpha=1$. 

We start our evaluation of ${\cal S}_k(R)$, eq. (\ref{squareflow}), by
expanding it with respect to $R\propto r^{-2}$.
Since we are only interested in the contributions proportional to 
$\int d^dx\,\sqrt{g}\propto r^d$, $\int d^dx\,\sqrt{g}R\propto r^{d-2}$ and 
$\int d^dx\,\sqrt{g}R^2\propto r^{d-4}$, only terms of order $r^d$, $r^{d-2}$
and $r^{d-4}$ are needed. This leads to  
\begin{eqnarray}
\label{derivexp}
\lefteqn{{\cal S}_k(R)={\rm Tr}_{(2ST^2)}\left[{\cal A}_1^{-1}{\cal N}\right]
+{\rm Tr}_{(1T)}'\left[{\cal A}_1^{-1}{\cal N}\right]
-h_1(d)\,{\rm Tr}_{(0)}''\left[{\cal A}_2^{-1}{\cal N}\right]
+a_k\,{\rm Tr}_{(0)}''\left[{\cal A}_2^{-1}{\cal T}_2\right]}\nonumber\\
& &+a_k\,{\rm Tr}_{(0)}''\left[({\cal A}_1{\cal A}_2)^{-1}P_k^2{\cal N}
\right]
-2{\rm Tr}_{(1T)}\left[P_k^{-1}{\cal N}_0\right]
-2{\rm Tr'}_{(0)}\left[P_k^{-1}{\cal N}_0\right]\nonumber\\
%
%
& &+\bigg\{-a_k\,{\rm Tr}_{(2ST^2)}
\left[{\cal A}_1^{-1}{\cal T}_1\right]
+a_k\,{\rm Tr}_{(2ST^2)}\left[{\cal A}_1^{-2}P_k{\cal N}\right]
-A_T(d)\,{\rm Tr}_{(2ST^2)}\left[{\cal A}_1^{-2}{\cal N}\right]\nonumber\\
& &-A_V(d,1)\,{\rm Tr}_{(1T)}'\left[{\cal A}_1^{-2}{\cal N}\right]
+h_2(d)\,a_k\,{\rm Tr}_{(0)}''\left[({\cal A}_1{\cal A}_2)^{-1}{\cal T}_2
\right]\nonumber\\
& &+h_3(d)\,a_k\,{\rm Tr}_{(0)}''\left[({\cal A}_1{\cal A}_2)^{-1}P_k{\cal N}
\right]
+h_4(d)\,{\rm Tr}_{(0)}''\left[({\cal A}_1{\cal A}_2)^{-1}{\cal N}\right]
+h_3(d)\,a_k\,{\rm Tr}_{(0)}''\left[{\cal A}_2^{-1}{\cal T}_1\right]\nonumber\\
& &-h_2(d)\,a_k^2\,{\rm Tr}_{(0)}''\left[{\cal A}_1^{-1}{\cal A}_2^{-2}P_k^2
{\cal T}_2\right]
-h_3(d)\,a_k^2\,{\rm Tr}_{(0)}''\left[{\cal A}_2^{-2}P_k{\cal T}_2\right]
-h_4(d)\,a_k\,{\rm Tr}_{(0)}''\left[{\cal A}_2^{-2}{\cal T}_2\right]
\nonumber\\
& &-h_2(d)\,a_k^2\,{\rm Tr}_{(0)}''\left[({\cal A}_1{\cal A}_2)^{-2}
P_k^4{\cal N}\right]
-h_3(d)\,a_k^2\,{\rm Tr}_{(0)}''\left[{\cal A}_1^{-1}{\cal A}_2^{-2}
P_k^3{\cal N}\right]\nonumber\\
& &-2h_4(d)\,a_k\,
{\rm Tr}_{(0)}''\left[{\cal A}_1^{-1}{\cal A}_2^{-2}P_k^2{\cal N}\right]
+h_1(d)\,h_3(d)\,a_k\,{\rm Tr}_{(0)}''\left[{\cal A}_2^{-2}P_k{\cal N}
\right]\nonumber\\
& &+h_1(d)\,h_4(d)\,{\rm Tr}_{(0)}''\left[{\cal A}_2^{-2}{\cal N}\right]
-\frac{2}{d}\,{\rm Tr}_{(1T)}\left[P_k^{-2}{\cal N}_0\right]
-\frac{4}{d}\,{\rm Tr}_{(0)}'\left[P_k^{-2}{\cal N}_0\right]\nonumber\\
& &+\frac{\delta_{d,2}}{4\pi}\left[(\bar{\beta}_k\,k^4)^{-1}
\,\partial_t(\bar{\beta}_k\,k^4)\right]\int d^dx\,\sqrt{g}\bigg\}\,R\nonumber\\
%
%
& &+\bigg\{
-a_k^2\,{\rm Tr}_{(2ST^2)}\left[{\cal A}_1^{-2}P_k{\cal T}_1\right]
+a_k\,A_T(d)\,{\rm Tr}_{(2ST^2)}\left[{\cal A}_1^{-2}{\cal T}_1\right]
-G_T(d)\,a_k\,
{\rm Tr}_{(2ST^2)}\left[{\cal A}_1^{-2}{\cal N}\right]\nonumber\\
& &+a_k^2\,{\rm Tr}_{(2ST^2)}\left[{\cal A}_1^{-3}P_k^2{\cal N}\right]
-2 A_T(d)\,a_k\,{\rm Tr}_{(2ST^2)}\left[{\cal A}_1^{-3}P_k{\cal N}\right]
+A_T(d)^2\,{\rm Tr}_{(2ST^2)}\left[{\cal A}_1^{-3}{\cal N}\right]\nonumber\\
& &-G_V(d)\,a_k\,{\rm Tr}_{(1T)}'\left[{\cal A}_1^{-2}{\cal N}\right]
+A_V(d,1)^2\,{\rm Tr}_{(1T)}'\left[{\cal A}_1^{-3}{\cal N}\right]
-h_5(d)\,a_k^2\,{\rm Tr}_{(0)}''\left[({\cal A}_1{\cal A}_2)^{-1}{\cal T}_2
\right]\nonumber\\
& &+h_6(d)\,a_k\,{\rm Tr}_{(0)}''\left[({\cal A}_1{\cal A}_2)^{-1}{\cal N}
\right]
-\frac{2}{d^2}\,a_k^2\,{\rm Tr}_{(0)}''\left[({\cal A}_1{\cal A}_2)^{-1}P_k
{\cal T}_1\right]\nonumber\\
& &+h_2(d)\,h_3(d)\,a_k\,
{\rm Tr}_{(0)}''\left[({\cal A}_1{\cal A}_2)^{-1}{\cal T}_1\right]
-h_2(d)^2\,a_k^2\,{\rm Tr}_{(0)}''\left[({\cal A}_1{\cal A}_2)^{-2}P_k^2
{\cal T}_2\right]\nonumber\\
& &+\frac{1}{2}h_2(d)\,a_k^3\,
{\rm Tr}_{(0)}''\left[{\cal A}_1^{-1}{\cal A}_2^{-2}P_k^2{\cal T}_2\right]
-2h_2(d)\,h_3(d)\,a_k^2\,
{\rm Tr}_{(0)}''\left[{\cal A}_1^{-1}{\cal A}_2^{-2}P_k{\cal T}_2\right]
\nonumber\\
& &-\frac{3}{2}h_2(d)\,h_4(d)\,a_k\,
{\rm Tr}_{(0)}''\left[{\cal A}_1^{-1}{\cal A}_2^{-2}{\cal T}_2\right]
-h_6(d)\,a_k^2\,{\rm Tr}_{(0)}''\left[{\cal A}_2^{-2}{\cal T}_2\right]
\nonumber\\
& &+\frac{1}{2}h_2(d)\,a_k^3\,
{\rm Tr}_{(0)}''\left[({\cal A}_1{\cal A}_2)^{-2}P_k^4{\cal N}
\right]
-2h_2(d)\,h_3(d)\,a_k^2\,
{\rm Tr}_{(0)}''\left[({\cal A}_1{\cal A}_2)^{-2}P_k^3{\cal N}
\right]\nonumber\\
& &-\frac{3}{2}h_2(d)\,h_4(d)\,a_k\,
{\rm Tr}_{(0)}''\left[({\cal A}_1{\cal A}_2)^{-2}P_k^2{\cal N}\right]
-h_7(d)\,a_k^2\,
{\rm Tr}_{(0)}''\left[{\cal A}_1^{-1}{\cal A}_2^{-2}P_k^2{\cal N}
\right]\nonumber\\
& &-3h_3(d)\,h_4(d)\,a_k\,{\rm Tr}_{(0)}''\left[{\cal A}_1^{-1}{\cal A}_2^{-2}
P_k{\cal N}\right]
-\frac{3}{2}h_4(d)^2\,
{\rm Tr}_{(0)}''\left[{\cal A}_1^{-1}{\cal A}_2^{-2}{\cal N}\right]\nonumber\\
& &+h_1(d)\,h_6(d)\,a_k\,{\rm Tr}_{(0)}''\left[{\cal A}_2^{-2}{\cal N}\right]
-h_2(d)\,h_3(d)\,a_k^2\,{\rm Tr}_{(0)}''\left[{\cal A}_1^{-1}
{\cal A}_2^{-2}P_k^2{\cal T}_1\right]\nonumber\\
& &-h_3(d)^2\,a_k^2\,  
{\rm Tr}_{(0)}''\left[{\cal A}_2^{-2}P_k{\cal T}_1\right]
-h_3(d)\,h_4(d)\,a_k\,{\rm Tr}_{(0)}''\left[{\cal A}_2^{-2}{\cal T}_1\right]
\nonumber\\
& &+h_2(d)^2\,a_k^3\,{\rm Tr}_{(0)}''\left[{\cal A}_1^{-2}
{\cal A}_2^{-3}P_k^4{\cal T}_2\right]
+2h_2(d)\,h_3(d)\,a_k^3\,
{\rm Tr}_{(0)}''\left[{\cal A}_1^{-1}{\cal A}_2^{-3}P_k^3{\cal T}_2\right]
\nonumber\\
& &+2h_2(d)\,h_4(d)\,a_k^2\,
{\rm Tr}_{(0)}''\left[{\cal A}_1^{-1}{\cal A}_2^{-3}P_k^2{\cal T}_2\right]
+h_3(d)^2\,a_k^3\,{\rm Tr}_{(0)}''
\left[{\cal A}_2^{-3}P_k^2{\cal T}_2\right]\nonumber\\
& &+2h_3(d)\,h_4(d)\,a_k^2\,{\rm Tr}_{(0)}''
\left[{\cal A}_2^{-3}P_k{\cal T}_2\right]
+h_4(d)^2\,a_k\,{\rm Tr}_{(0)}''
\left[{\cal A}_2^{-3}{\cal T}_2\right]\nonumber\\
& &+h_2(d)^2\,a_k^3\,{\rm Tr}_{(0)}''\left[({\cal A}_1
{\cal A}_2)^{-3}P_k^6{\cal N}\right]
+2h_2(d)\,h_3(d)\,a_k^3\,{\rm Tr}_{(0)}''\left[{\cal A}_1^{-2}{\cal A}_2^{-3}
P_k^5{\cal N}\right]\nonumber\\
& &+3h_2(d)\,h_4(d)\,a_k^2\,  
{\rm Tr}_{(0)}''\left[{\cal A}_1^{-2}{\cal A}_2^{-3}P_k^4{\cal N}\right]
+h_3(d)^2\,a_k^3\,{\rm Tr}_{(0)}''\left[{\cal A}_1^{-1}{\cal A}_2^{-3}P_k^4
{\cal N}\right]\nonumber\\
& &+4h_3(d)\,h_4(d)\,a_k^2\,{\rm Tr}_{(0)}''\left[{\cal A}_1^{-1}
{\cal A}_2^{-3}P_k^3{\cal N}\right]
+3h_4(d)^2\,a_k\,{\rm Tr}_{(0)}''\left[{\cal A}_1^{-1}{\cal A}_2^{-3}
P_k^2{\cal N}\right]\nonumber\\
& &-h_1(d)\,h_3(d)^2\,a_k^2\,{\rm Tr}_{(0)}''
\left[{\cal A}_2^{-3}P_k^2{\cal N}\right]
-2h_1(d)\,h_3(d)\,h_4(d)\,a_k\,{\rm Tr}_{(0)}''
\left[{\cal A}_2^{-3}P_k{\cal N}\right]\nonumber\\
& &-h_1(d)\,h_4(d)^2\,{\rm Tr}_{(0)}''\left[{\cal A}_2^{-3}{\cal N}\right]
-\frac{2}{d^2}\,{\rm Tr}_{(1T)}\left[P_k^{-3}{\cal N}_0\right]
-\frac{8}{d^2}\,{\rm Tr}_{(0)}'\left[P_k^{-3}{\cal N}_0\right]\nonumber\\
& &+\int d^dx\,\sqrt{g}\bigg[
\frac{\delta_{d,2}}{8\pi\,k^4}\left(1-3R^{(0)'}(0)\right)
\left((\bar{\beta}_k\,k^2)^{-1}\partial_t(\bar{\beta}_k\,k^4)
-\bar{\beta}_k^{-1}\partial_t(\bar{\beta}_k\,k^2)\right)
\nonumber\\
& &+\frac{\delta_{d,4}}{8(4\pi)^2}\Big(9a_k\,k^4-2(k^2-2\bar{\lambda}_k)
\Big)^{-1}\Big(9a_k\,\bar{\beta}_k^{-1}\partial_t(\bar{\beta}_k\,k^4)
-2Z_{Nk}^{-1}\partial_t(Z_{Nk}\,k^2)\Big)
\bigg]\bigg\}\,R^2\nonumber\\
& &+{\cal O}(r^{<d-4})\;.
\end{eqnarray}
Here we set
\begin{eqnarray}
\label{kalas}
{\cal A}_1&\equiv& P_k-2\bar{\lambda}_k\;,\nonumber\\
{\cal A}_2&\equiv& a_k\,P_k^2
-\frac{1}{2}h_1(d)\left(P_k-2\bar{\lambda}_k\right)
\end{eqnarray}
and 
\begin{eqnarray}
\label{a_k}
a_k\equiv(Z_{Nk}\kappa^2)^{-1}\bar{\beta}_k\;.
\end{eqnarray}
The quantities ${\cal N}$, ${\cal N}_0$, ${\cal T}_1$ and ${\cal T}_2$
are defined as in eq. (\ref{ntop}). Furthermore, ${\cal O}(r^{<d-4})$ means 
that terms $\propto r^n$ with powers $n<d-4$ are neglected. 

The terms in eq. (\ref{derivexp}) proportional to $\delta_{d,2}$ and 
$\delta_{d,4}$ arise from the last term in eq. (\ref{squareflow}). Contrary to
the other terms of eq. (\ref{squareflow}), its expansion does not contain 
$d$-dependent powers of $r$, but is of the form $\sum \limits_{m=0}^\infty 
b_{2m}r^{-2m}$ with $\{b_{2m}\}$ a set of $r$-independent coefficients. As for
comparing powers of $r$, this has the following consequence. Since, for all 
$m\ge 0$ and $d>0$, $-2m=d-4$ or $-2m=d-2$ are satisfied only if 
$(m,d)\in\{(0,2),(1,2),(0,4)\}$, and since $-2m=d$ cannot be satisfied at 
all, this term contributes to the evolution equation only in the 
two- and the four-dimensional case. Using eq. (\ref{spheres}) the 
pieces contributing, i.e. $b_{2m=0}\;r^0$ in $d=2$ and $d=4$, and 
$b_{2m=2}\;r^{-2}$ in $d=2$, may be expressed in terms of the operators 
$\int d^2x\sqrt{g}R$, $\int d^2x\sqrt{g}R^2$ or $\int d^4x\sqrt{g}R^2$. 
This yields the terms in eq. (\ref{derivexp}) which are proportional to the
$\delta$'s.

As the next step we evaluate the $r$-expansion of the traces appearing in eq.
(\ref{derivexp}) by applying the heat kernel expansion. In its 
original form it has often been used to compute traces of operators acting on 
unconstrained fields. For our purposes we need the {\bf heat kernel
expansions for operators acting on constrained fields}, i.e. fields satisfying
appropriate transversality conditions. In ref. \cite{LR1} these expansions
are derived in detail for Laplacians $D^2$ on $S^d$ backgrounds acting on 
symmetric transverse
traceless tensors, on transverse vectors and on scalars, with the following 
results:
\begin{eqnarray}
\label{kern1}
\lefteqn{{\rm Tr}_{(2ST^2)}\left[e^{-({\rm i} s-\varepsilon)D^2}\right]
=\left(\frac{{\rm i}}{4\pi (s+{\rm i}\varepsilon)}\right)^{d/2}\int d^dx\,
\sqrt{g}\Bigg\{
\frac{1}{2}(d-2)(d+1)}\nonumber\\
& &-\frac{(d+1)(d+2)(d-5+3\delta_{d,2})}{12(d-1)}
({\rm i} s-\varepsilon) R\\\
& &-\frac{(d+1)(5d^4-22d^3-83d^2-392d-228+1440\,\delta_{d,2}
+3240\,\delta_{d,4})}{720d(d-1)^2}(s+{\rm i}\varepsilon)^2 R^2\nonumber\\
& &+{\cal O}(R^3)\Bigg\}\;,\nonumber
\end{eqnarray}
\begin{eqnarray}
\label{kern2}
\lefteqn{{\rm Tr}_{(1T)}\left[e^{-({\rm i} s-\varepsilon)D^2}\right]
=\left(\frac{{\rm i}}{4\pi (s+{\rm i}\varepsilon)}\right)^{d/2}\int d^dx\,
\sqrt{g}\Bigg\{d-1}\nonumber\\
& &-\frac{(d+2)(d-3)+6\delta_{d,2}}{6d}({\rm i}s-\varepsilon) R\\
& &-\frac{5d^4-12d^3-47d^2-186d+180+360\,\delta_{d,2}
+720\,\delta_{d,4}}{360d^2(d-1)} (s+{\rm i}\varepsilon)^2 R^2
+{\cal O}(R^3)\Bigg\}\;,\nonumber
\end{eqnarray}
\begin{eqnarray}
\label{kern3}
\lefteqn{{\rm Tr}_{(0)}\left[e^{-({\rm i} s-\varepsilon)D^2}\right]
=\left(\frac{{\rm i}}{4\pi (s+{\rm i}\varepsilon)}\right)^{d/2}\int d^dx\,
\sqrt{g}}\nonumber\\
& &\times\Bigg\{1-\frac{1}{6}({\rm i} s-\varepsilon) R
-\frac{5d^2-7d+6}{360d(d-1)} (s+{\rm i}\varepsilon)^2 R^2
+{\cal O}(R^3)\Bigg\}\;.
\end{eqnarray}
Here the terms proportional to the $\delta$'s arise from the exclusion of 
unphysical modes of the type discussed in section \ref{3S2}.

Let us now consider an arbitrary function $W(z)$ with a Fourier
transform $\widetilde{W}(s)$. For such functions $W$, we may express the
trace of the operator $W(-D^2)$ that results from replacing the argument of $W$
with $-D^2$ in terms of $\widetilde{W}(s)$:
\begin{eqnarray}
\label{fourier}
{\rm Tr}\left[W(-D^2)\right]=\lim\limits_{\varepsilon\searrow 0}
\int\limits^{\infty}_{-\infty}ds\,\widetilde{W}(s)\,{\rm Tr}
\left[e^{-({\rm i} s-\varepsilon)D^2}\right]
\end{eqnarray}
We obtain the asymptotic expansion of ${\rm Tr}[W(-D^2)]$ by inserting 
the heat kernel expansion for ${\rm Tr}[e^{-({\rm i}s-\varepsilon)D^2}]$ into
eq. (\ref{fourier}). For Laplacians acting on the constrained fields considered
here they read as follows:
\begin{eqnarray}
\label{kern4}
\lefteqn{{\rm Tr}_{(2ST^2)}\left[W(-D^2)\right]=(4\pi)^{-d/2}\bigg\{
\frac{1}{2}(d-2)(d+1)\,Q_{d/2}[W]\,
\int d^dx\,\sqrt{g}}\nonumber\\
& &+\frac{(d+1)(d+2)(d-5+3\delta_{d,2})}{12(d-1)}
\,Q_{d/2-1}[W]\,\int d^dx\,\sqrt{g}R\nonumber\\
& &+\frac{(d+1)(5d^4-22d^3-83d^2-392d-228+1440\,\delta_{d,2}
+3240\,\delta_{d,4})}{720d(d-1)^2}\,Q_{d/2-2}[W]\nonumber\\
& &\times\int d^dx\,\sqrt{g}R^2
+{\cal O}(r^{<d-4})\bigg\}\;,
\end{eqnarray}
\begin{eqnarray}
\label{kern5}
\lefteqn{{\rm Tr}_{(1T)}\left[W(-D^2)\right]=(4\pi)^{-d/2}\bigg\{
(d-1)\,Q_{d/2}[W]\,\int d^dx\,\sqrt{g}}\nonumber\\
& &+\frac{(d+2)(d-3)+6\delta_{d,2}}{6d}
\,Q_{d/2-1}[W]\,\int d^dx\,\sqrt{g}R\nonumber\\
& &+\frac{5d^4-12d^3-47d^2-186d+180+360\,\delta_{d,2}
+720\,\delta_{d,4}}{360d^2(d-1)}\,Q_{d/2-2}[W]\,\int d^dx\,\sqrt{g}R^2
\nonumber\\
& &+{\cal O}(r^{<d-4})\bigg\}\;,
\end{eqnarray}
\begin{eqnarray}
\label{kern6}
\lefteqn{{\rm Tr}_{(0)}\left[W(-D^2)\right]=(4\pi)^{-d/2}\bigg\{Q_{d/2}[W]\,
\int d^dx\,\sqrt{g}}\\
& &+\frac{1}{6}\,Q_{d/2-1}[W]\,\int d^dx\,\sqrt{g}R
+\frac{5d^2-7d+6}{360d(d-1)}\,Q_{d/2-2}[W]\,\int d^dx\,\sqrt{g}R^2
+{\cal O}(r^{<d-4})\bigg\}\;.\nonumber
\end{eqnarray}
Here the set of functionals $Q_n[W]$ is defined as
\begin{eqnarray}
\label{Qn}
Q_n[W]\equiv\lim\limits_{\varepsilon\searrow 0}\int
\limits^{\infty}_{-\infty} ds\,(-{\rm i} s+\varepsilon)^{-n}\,
\widetilde{W}(s)\;.
\end{eqnarray}
By virtue of the Mellin transformation we may now reexpress $Q_n$ in terms of
$W$ so that
\begin{eqnarray}
\label{mellin}
Q_n[W]=\frac{(-1)^i}{\Gamma(n+i)}\int\limits_{0}^{\infty}dz\,z^{n+i-1}\,
\frac{d^i W(z)}{dz^i}\;,\;\,i>-n\;,\;\,i\in 
\mathbb{N}\cup\{0\}\;\;{\rm arbitrary}.
\end{eqnarray}
In particular we obtain $Q_0[W]=W(0)$. Furthermore, if $n>0$ we may choose 
$i=0$ for simplicity. As can be seen by an appropriate integration by parts,
$Q_n[W]$ does not depend on $i$.

At this point it is necessary to discuss the case where isolated eigenvalues 
have to be excluded from ${\rm Tr}[W(-D^2)]$. As we showed in \cite{LR1}, 
such traces can be expressed as 
\begin{eqnarray}
\label{modesubtract}
{\rm Tr}^{\prime\ldots\prime}[W(-D^2)]={\rm Tr}[W(-D^2)]
-\sum\limits_{l\in\{l_1,\ldots,l_n\}}D_l(d,s)W\left(\Lambda_l(d,s)\right)\;.
\end{eqnarray}
Here $n$ primes at ${\rm Tr}^{\prime\ldots\prime}$ symbolize the exclusion
of all eigenmodes $T^{lm}$ with $l\in\{l_1,\ldots,l_n\}$, 
and $\Lambda_l(d,s)$ and $D_l(d,s)$ denote the corresponding 
eigenvalues of $-D^2$ and their degrees of degeneracy, respectively. Since 
$\Lambda_l(d,s)\propto R$ we may view $W\left(\Lambda_l(d,s)\right)$ as a 
function of $R$. As outlined above, such a function 
contributes to the evolution of $\bar{\lambda}_k$, $Z_{Nk}$ and $\bar{\beta}_k$
only for $d=2$ and for $d=4$, with the contributions given by 
$W(0)+W'(0)\,\Lambda_l(2,s)$ and $W(0)$, respectively. Using the explicit 
expressions for $\Lambda_l(d,s)$ and$D_l(d,s)$ (see Table 2 in 
appendix \ref{harm}) and applying 
eq. (\ref{spheres}) we therefore obtain for the traces relevant to the flow 
equation:
\begin{eqnarray}
\label{subtract1}
\lefteqn{{\rm Tr}_{(1T)}'[W(-D^2)]={\rm Tr}_{(1T)}[W(-D^2)]
-\frac{\delta_{d,2}}{16\pi}\left\{6W(0)\,\int d^2x\,\sqrt{g}\,R\right.}
\nonumber\\
& &\left.+3W'(0)\,\int d^2x\,\sqrt{g}\,R^2\right\}
-\frac{5\delta_{d,4}}
{12(4\pi)^2}
\,W(0)\,\int d^4x\,\sqrt{g}\,R^2+{\cal O}\left(r^{<d-4}\right)\;,\nonumber\\
\end{eqnarray}
\begin{eqnarray}
\label{subtract2}
\lefteqn{{\rm Tr}_{(0)}''[W(-D^2)]={\rm Tr}_{(0)}[W(-D^2)]
-\frac{\delta_{d,2}}{16\pi}\left\{8W(0)\,\int d^2x\,\sqrt{g}\,R\right.}
\nonumber\\
& &\left.+3W'(0)\,\int d^2x\,\sqrt{g}\,R^2\right\}
-\frac{\delta_{d,4}}
{4(4\pi)^2}
\,W(0)\,\int d^4x\,\sqrt{g}\,R^2+{\cal O}\left(r^{<d-4}\right)\;,
\end{eqnarray}
\begin{eqnarray}
\label{subtract3}
\lefteqn{{\rm Tr}_{(0)}'[W(-D^2)]={\rm Tr}_{(0)}[W(-D^2)]}\nonumber\\
& &-\frac{\delta_{d,2}}{8\pi}\,W(0)\,\int d^2x\,\sqrt{g}\,R
-\frac{\delta_{d,4}}{24(4\pi)^2}
\,W(0)\,\int d^4x\,\sqrt{g}\,R^2+{\cal O}\left(r^{<d-4}\right)\;.
\end{eqnarray}
Here $W'$ denotes the derivative with respect to the argument:
$W'(z)=dW(z)/dz$ with $z=\Lambda_l(d,s)$.

The next step is to insert the expansions of the traces into ${\cal S}_k(R)$,
eq. (\ref{derivexp}), and to compare the coefficients of the operators
$\int d^dx\sqrt{g}$, $\int d^dx\sqrt{g}R$ and $\int d^dx\sqrt{g}R^2$ with 
those on the LHS, eq. (\ref{LHS}). This leads to the following differential 
equations:
\begin{eqnarray}
\label{q-level1}
\lefteqn{\partial_t\left(Z_{Nk}\bar{\lambda}_k\right)
=(4\kappa^2)^{-1}(4\pi)^{-d/2}
\Big\{h_8(d)\,Q_{d/2}\left[{\cal A}_1^{-1}{\cal N}\right]
-h_1(d)\,Q_{d/2}\left[{\cal A}_2^{-1}{\cal N}\right]}\nonumber\\
& &+a_k\,Q_{d/2}\left[{\cal A}_2^{-1}{\cal T}_2\right]
+a_k\,Q_{d/2}\left[({\cal A}_1{\cal A}_2)^{-1}P_k^2{\cal N}\right]
-2d\,Q_{d/2}\left[P_k^{-1}{\cal N}_0\right]\Big\}\;,
\end{eqnarray}
\begin{eqnarray}
\label{q-level2}
\lefteqn{\partial_t Z_{Nk}
=-(2\kappa^2)^{-1}(4\pi)^{-d/2}
\Big\{h_9(d)\,Q_{d/2-1}\left[{\cal A}_1^{-1}{\cal N}\right]
-\frac{1}{6}h_1(d)\,Q_{d/2-1}\left[{\cal A}_2^{-1}{\cal N}\right]}\nonumber\\
& &+\frac{1}{6}a_k\,Q_{d/2-1}\left[{\cal A}_2^{-1}{\cal T}_2\right]
+\frac{1}{6}a_k\,Q_{d/2-1}\left[({\cal A}_1{\cal A}_2)^{-1}P_k^2{\cal N}\right]
+h_{10}(d)\,Q_{d/2-1}\left[P_k^{-1}{\cal N}_0\right]\nonumber\\
& &-h_{11}(d)\,a_k\,Q_{d/2}\left[{\cal A}_1^{-1}{\cal T}_1\right]
+h_{11}(d)\,a_k\,Q_{d/2}\left[{\cal A}_1^{-2}P_k{\cal N}\right]
+h_{12}(d)\,Q_{d/2}\left[{\cal A}_1^{-2}{\cal N}\right]\nonumber\\
& &+h_2(d)\,a_k\,Q_{d/2}\left[({\cal A}_1{\cal A}_2)^{-1}{\cal T}_2
\right]
+h_3(d)\,a_k\,Q_{d/2}\left[({\cal A}_1{\cal A}_2)^{-1}P_k{\cal N}\right]
\nonumber\\
& &+h_4(d)\,Q_{d/2}\left[({\cal A}_1{\cal A}_2)^{-1}{\cal N}\right]
+h_3(d)\,a_k\,Q_{d/2}\left[{\cal A}_2^{-1}{\cal T}_1\right]
-h_2(d)\,a_k^2\,Q_{d/2}\left[{\cal A}_1^{-1}{\cal A}_2^{-2}P_k^2
{\cal T}_2\right]\nonumber\\
& &-h_3(d)\,a_k^2\,Q_{d/2}\left[{\cal A}_2^{-2}P_k{\cal T}_2\right]
-h_4(d)\,a_k\,Q_{d/2}\left[{\cal A}_2^{-2}{\cal T}_2\right]
-h_2(d)\,a_k^2\,Q_{d/2}\left[({\cal A}_1{\cal A}_2)^{-2}
P_k^4{\cal N}\right]\nonumber\\
& &-h_3(d)\,a_k^2\,Q_{d/2}\left[{\cal A}_1^{-1}{\cal A}_2^{-2}
P_k^3{\cal N}\right]
-2h_4(d)\,a_k\,
Q_{d/2}\left[{\cal A}_1^{-1}{\cal A}_2^{-2}P_k^2{\cal N}\right]\\
& &+h_1(d)\,h_3(d)\,a_k\,Q_{d/2}\left[{\cal A}_2^{-2}P_k{\cal N}
\right]
+h_1(d)\,h_4(d)\,Q_{d/2}\left[{\cal A}_2^{-2}{\cal N}\right]
+h_{13}(d)\,Q_{d/2}\left[P_k^{-2}{\cal N}_0\right]\Big\}\nonumber\;,
\end{eqnarray}
\begin{eqnarray}
\label{q-level3}
\lefteqn{\partial_t\bar{\beta}_k
=(4\kappa^2)^{-1}(4\pi)^{-d/2}
\Big\{h_{14}(d)\,Q_{d/2-2}\left[{\cal A}_1^{-1}{\cal N}\right]
-h_1(d)\,h_{15}(d)\,Q_{d/2-2}\left[{\cal A}_2^{-1}{\cal N}\right]}\nonumber\\
& &+h_{15}(d)\,a_k\,Q_{d/2-2}\left[{\cal A}_2^{-1}{\cal T}_2\right]
+h_{15}(d)\,a_k\,Q_{d/2-2}\left[({\cal A}_1{\cal A}_2)^{-1}P_k^2{\cal N}
\right]\nonumber\\
& &-h_{16}(d)\,Q_{d/2-2}\left[P_k^{-1}{\cal N}_0\right]
%
%
-h_{17}(d)\,a_k\,Q_{d/2-1}\left[{\cal A}_1^{-1}{\cal T}_1\right]
+h_{17}(d)\,a_k\,Q_{d/2-1}\left[{\cal A}_1^{-2}P_k{\cal N}\right]\nonumber\\
& &-h_{18}(d)\,Q_{d/2-1}\left[{\cal A}_1^{-2}{\cal N}\right]
+\frac{1}{6}h_2(d)\,a_k\,Q_{d/2-1}\left[({\cal A}_1{\cal A}_2)^{-1}{\cal T}_2
\right]\nonumber\\
& &+\frac{1}{6}h_3(d)\,a_k\,Q_{d/2-1}
\left[({\cal A}_1{\cal A}_2)^{-1}P_k{\cal N}\right]
+\frac{1}{6}h_4(d)\,Q_{d/2-1}\left[({\cal A}_1{\cal A}_2)^{-1}{\cal N}\right]
\nonumber\\
& &+\frac{1}{6}h_3(d)\,a_k\,Q_{d/2-1}\left[{\cal A}_2^{-1}{\cal T}_1\right]
-\frac{1}{6}h_2(d)\,a_k^2\,Q_{d/2-1}\left[{\cal A}_1^{-1}{\cal A}_2^{-2}
P_k^2{\cal T}_2\right]\nonumber\\
& &-\frac{1}{6}h_3(d)\,a_k^2\,Q_{d/2-1}\left[{\cal A}_2^{-2}P_k{\cal T}_2
\right]
-\frac{1}{6}h_4(d)\,a_k\,Q_{d/2-1}\left[{\cal A}_2^{-2}{\cal T}_2\right]
\nonumber\\
& &-\frac{1}{6}h_2(d)\,a_k^2\,Q_{d/2-1}\left[({\cal A}_1{\cal A}_2)^{-2}
P_k^4{\cal N}\right]
-\frac{1}{6}h_3(d)\,a_k^2\,Q_{d/2-1}\left[{\cal A}_1^{-1}{\cal A}_2^{-2}
P_k^3{\cal N}\right]\nonumber\\
& &-\frac{1}{3}h_4(d)\,a_k\,
Q_{d/2-1}\left[{\cal A}_1^{-1}{\cal A}_2^{-2}P_k^2{\cal N}\right]
+\frac{1}{6}h_1(d)\,h_{19}(d)\,a_k\,Q_{d/2-1}\left[{\cal A}_2^{-2}P_k{\cal N}
\right]\nonumber\\
& &+\frac{1}{6}h_1(d)\,h_4(d)\,Q_{d/2-1}\left[{\cal A}_2^{-2}{\cal N}\right]
-h_{20}(d)\,Q_{d/2-1}\left[P_k^{-2}{\cal N}_0\right]
%
%
-h_{11}(d)\,a_k^2\,Q_{d/2}\left[{\cal A}_1^{-2}P_k{\cal T}_1\right]\nonumber\\
& &+\frac{1}{2}h_{11}(d)\,h_{21}(d)\,a_k\,Q_{d/2}\left[{\cal A}_1^{-2}
{\cal T}_1\right]
+h_{22}(d)\,a_k\,Q_{d/2}\left[{\cal A}_1^{-2}{\cal N}\right]
\nonumber\\
& &+h_{11}(d)\,a_k^2\,Q_{d/2}\left[{\cal A}_1^{-3}P_k^2{\cal N}\right]
-h_{11}(d)\,h_{21}(d)\,a_k\,Q_{d/2}\left[{\cal A}_1^{-3}P_k{\cal N}\right]
\nonumber\\
& &+h_{23}(d)\,Q_{d/2}\left[{\cal A}_1^{-3}{\cal N}\right]
-h_5(d)\,a_k^2\,Q_{d/2}\left[({\cal A}_1{\cal A}_2)^{-1}{\cal T}_2
\right]\nonumber\\
& &+h_6(d)\,a_k\,Q_{d/2}\left[({\cal A}_1{\cal A}_2)^{-1}{\cal N}
\right]
-\frac{2}{d^2}\,a_k^2\,Q_{d/2}\left[({\cal A}_1{\cal A}_2)^{-1}P_k
{\cal T}_1\right]\nonumber\\
& &+h_2(d)\,h_3(d)\,a_k\,
Q_{d/2}\left[({\cal A}_1{\cal A}_2)^{-1}{\cal T}_1\right]
-h_2(d)^2\,a_k^2\,Q_{d/2}\left[({\cal A}_1{\cal A}_2)^{-2}P_k^2
{\cal T}_2\right]\nonumber\\
& &+\frac{1}{2}h_2(d)\,a_k^3\,
Q_{d/2}\left[{\cal A}_1^{-1}{\cal A}_2^{-2}P_k^2{\cal T}_2\right]
-2h_2(d)\,h_3(d)\,a_k^2\,
Q_{d/2}\left[{\cal A}_1^{-1}{\cal A}_2^{-2}P_k{\cal T}_2\right]
\nonumber\\
& &-\frac{3}{2}h_2(d)\,h_4(d)\,a_k\,
Q_{d/2}\left[{\cal A}_1^{-1}{\cal A}_2^{-2}{\cal T}_2\right]
-h_6(d)\,a_k^2\,Q_{d/2}\left[{\cal A}_2^{-2}{\cal T}_2\right]
\nonumber\\
& &+\frac{1}{2}h_2(d)\,a_k^3\,
Q_{d/2}\left[({\cal A}_1{\cal A}_2)^{-2}P_k^4{\cal N}\right]
-2h_2(d)\,h_3(d)\,a_k^2\,Q_{d/2}\left[({\cal A}_1{\cal A}_2)^{-2}P_k^3{\cal N}
\right]\nonumber\\
& &-\frac{3}{2}h_2(d)\,h_4(d)\,a_k\,
Q_{d/2}\left[({\cal A}_1{\cal A}_2)^{-2}P_k^2{\cal N}\right]
-h_7(d)\,a_k^2\,Q_{d/2}\left[{\cal A}_1^{-1}{\cal A}_2^{-2}P_k^2{\cal N}
\right]\nonumber\\
& &-3h_3(d)\,h_4(d)\,a_k\,Q_{d/2}\left[{\cal A}_1^{-1}{\cal A}_2^{-2}
P_k{\cal N}\right]
-\frac{3}{2}h_4(d)^2\,
Q_{d/2}\left[{\cal A}_1^{-1}{\cal A}_2^{-2}{\cal N}\right]\nonumber\\
& &+h_1(d)\,h_6(d)\,a_k\,Q_{d/2}\left[{\cal A}_2^{-2}{\cal N}\right]
-h_2(d)\,h_3(d)\,a_k^2\,Q_{d/2}\left[{\cal A}_1^{-1}
{\cal A}_2^{-2}P_k^2{\cal T}_1\right]\nonumber\\
& &-h_3(d)^2\,a_k^2\,Q_{d/2}\left[{\cal A}_2^{-2}P_k{\cal T}_1\right]
-h_3(d)\,h_4(d)\,a_k\,Q_{d/2}\left[{\cal A}_2^{-2}{\cal T}_1\right]
\nonumber\\
& &+h_2(d)^2\,a_k^3\,Q_{d/2}\left[{\cal A}_1^{-2}
{\cal A}_2^{-3}P_k^4{\cal T}_2\right]
+2h_2(d)\,h_3(d)\,a_k^3\,
Q_{d/2}\left[{\cal A}_1^{-1}{\cal A}_2^{-3}P_k^3{\cal T}_2\right]\nonumber\\
& &+2h_2(d)\,h_4(d)\,a_k^2\,
Q_{d/2}\left[{\cal A}_1^{-1}{\cal A}_2^{-3}P_k^2{\cal T}_2\right]
+h_3(d)^2\,a_k^3\,Q_{d/2}\left[{\cal A}_2^{-3}P_k^2{\cal T}_2\right]
\nonumber\\
& &+2h_3(d)\,h_4(d)\,a_k^2\,Q_{d/2}\left[{\cal A}_2^{-3}P_k{\cal T}_2\right]
+h_4(d)^2\,a_k\,Q_{d/2}\left[{\cal A}_2^{-3}{\cal T}_2\right]\nonumber\\
& &+h_2(d)^2\,a_k^3\,Q_{d/2}\left[({\cal A}_1
{\cal A}_2)^{-3}P_k^6{\cal N}\right]
+2h_2(d)\,h_3(d)\,a_k^3\,Q_{d/2}\left[{\cal A}_1^{-2}{\cal A}_2^{-3}
P_k^5{\cal N}\right]\nonumber\\
& &+3h_2(d)\,h_4(d)\,a_k^2\,
Q_{d/2}\left[{\cal A}_1^{-2}{\cal A}_2^{-3}P_k^4{\cal N}\right]
+h_3(d)^2\,a_k^3\,Q_{d/2}\left[{\cal A}_1^{-1}{\cal A}_2^{-3}P_k^4
{\cal N}\right]\nonumber\\
& &+4h_3(d)\,h_4(d)\,a_k^2\,Q_{d/2}\left[{\cal A}_1^{-1}
{\cal A}_2^{-3}P_k^3{\cal N}\right]
+3h_4(d)^2\,a_k\,Q_{d/2}\left[{\cal A}_1^{-1}{\cal A}_2^{-3}
P_k^2{\cal N}\right]\nonumber\\
& &-h_1(d)\,h_3(d)^2\,a_k^2\,Q_{d/2}\left[{\cal A}_2^{-3}P_k^2{\cal N}\right]
-2h_1(d)\,h_3(d)\,h_4(d)\,a_k\,
Q_{d/2}\left[{\cal A}_2^{-3}P_k{\cal N}\right]\nonumber\\
& &-h_1(d)\,h_4(d)^2\,Q_{d/2}\left[{\cal A}_2^{-3}{\cal N}\right]
+h_{24}(d)\,Q_{d/2}\left[P_k^{-3}{\cal N}_0\right]\nonumber\\
& &+\delta_{d,2}\Big[
-\frac{3}{2}a_k\,\frac{\partial_t\left[\bar{\beta}_k\,k^2\right]}
{\bar{\beta_k}\left(k^2-2\bar{\lambda}_k\right)}
+\left(\frac{11}{4}R^{(0)'}(0)+\frac{3}{2}a_k\,k^2\right)\,\frac{\partial_t
\left[Z_{Nk}\,k^2\right]}{Z_{Nk}\left(k^2-2\bar{\lambda}_k\right)^2}
\nonumber\\
& &-\frac{11}{4}R^{(0)'}(0)\frac{\partial_t\,Z_{Nk}}
{Z_{Nk}\left(k^2-2\bar{\lambda}_k\right)}-\frac{1}{2k^2}R^{(0)'}(0)
\Big]\nonumber\\
& &+\delta_{d,4}\Big[
\frac{1}{8}\,\frac{\partial_t\left[Z_{Nk}\,k^2\right]}
{Z_{Nk}\left(k^2-2\bar{\lambda}_k\right)}
+\frac{1}{4}\,\frac{\partial_t\left[Z_{Nk}\,k^2\right]}
{Z_{Nk}\left(3a_k\,k^4-\left(k^2-2\bar{\lambda}_k\right)\right)}\nonumber\\
& &-\frac{3}{8}a_k\,k^4\,\frac{\partial_t\left[Z_{Nk}\,k^2\right]}
{Z_{Nk}\left(k^2-2\bar{\lambda}_k\right)\left(3a_k\,k^4-\left(k^2
-2\bar{\lambda}_k\right)\right)}
-\frac{1}{4}\,\frac{\partial_t\left[Z_{Nk}\,k^2\right]}
{Z_{Nk}\left(9a_k\,k^4-2\left(k^2-2\bar{\lambda}_k\right)\right)}\nonumber\\
& &-\frac{3}{8}a_k\,\frac{\partial_t\left[\bar{\beta}_k\,k^4\right]}
{\bar{\beta}_k\left(3a_k\,k^4-\left(k^2-2\bar{\lambda}_k\right)\right)}
+\frac{9}{8}a_k\,\frac{\partial_t\left[\bar{\beta}_k\,k^4\right]}
{\bar{\beta}_k\left(9a_k\,k^4-2\left(k^2-2\bar{\lambda}_k\right)\right)}
\Big]\Big\}\;.
\end{eqnarray}
Here the coefficients $h_i$ are functions of the dimensionality $d$. They are
tabulated in  (\ref{coeff3}) of the appendix \ref{betacoeff}.

Now we introduce the cutoff-dependent {\bf generalized threshold functions}
\begin{eqnarray}
\label{psi}
\lefteqn{\Psi^{p;q}_{n;m}(v,w;d)\equiv
\frac{(-1)^i}{\Gamma(n+i)}\int\limits_{0}^{\infty} dy\,y^{n+i-1}
\frac{\partial^i}{\partial y^i}\Bigg[\left(y+R^{(0)}(y)\right)^m\,
\left(R^{(0)}(y)-yR^{(0)'}(y)\right)}\\
& &\times\left(y+R^{(0)}(y)+w\right)^{-p}
\left(32\pi\,v\left(y+R^{(0)}(y)\right)^2
-\frac{d-2}{2(d-1)}\left(y+R^{(0)}(y)+w\right)\right)^{-q}\Bigg]\nonumber
\end{eqnarray}
and
\begin{eqnarray}
\label{psitilde}
\lefteqn{\widetilde{\Psi}^{p;q}_{n;m;l}(v,w;d)\equiv 
\frac{(-1)^i}{\Gamma(n+i)}\int\limits_{0}^{\infty} dy\,y^{n+i-1}
\frac{\partial^i}{\partial y^i}\Bigg[
\left(y+R^{(0)}(y)\right)^m\,\left(2y+R^{(0)}(y)
\right)^l R^{(0)}(y)}\\
& &\times\left(y+R^{(0)}(y)+w\right)^{-p}
\left(32\pi\,v\left(y+R^{(0)}(y)\right)^2-\frac{d-2}{2(d-1)}
\left(y+R^{(0)}(y)+w\right)\right)^{-q}\Bigg]\;.\nonumber
\end{eqnarray}
In eqs. (\ref{psi}) and (\ref{psitilde}), $i$ is a nonnegative integer which 
satisfies $i>-n$, but which is arbitrary otherwise. The functions $\Psi$ and
$\widetilde{\Psi}$ are independent of $i$ which can be seen by an integration
by parts. Again, we may set $i=0$ if $n>0$.  Furthermore, noting that
\begin{eqnarray}
\label{phi}
\Phi^p_n(w)\equiv\Psi^{p;0}_{n;0}(v,w;d)\;,\;\;\;\widetilde{\Phi}^p_n(w)
\equiv\widetilde{\Psi}^{p;0}_{n;0;0}(v,w;d)\;\;\;\forall v\;\;\;\forall d
\end{eqnarray}
we recover the threshold functions $\Phi^p_n(w)$ and $\widetilde{\Phi}^p_n(w)$
originally defined in \cite{Reu96} in the context of the pure 
Einstein-Hilbert truncation.

Using the relations
\begin{eqnarray}
\label{70}
Q_n\left[{\cal A}_1^{-p}{\cal A}_2^{-q}P_k^m{\cal T}_1\right]
&=&k^{2(m+n-p-q+1)}\Psi^{p;q}_{n;m}(a_k\,k^2/(32\pi),\bar{\lambda}_k/k^2;d)
\nonumber\\
& &-\frac{1}{2}\eta_{\beta}(k)\,k^{2(m+n-p-q+1)}
\widetilde{\Psi}^{p;q}_{n;m;0}(a_k\,k^2/(32\pi),\bar{\lambda}_k/k^2;d)
\nonumber\\
Q_n\left[{\cal A}_1^{-p}{\cal A}_2^{-q}P_k^m{\cal T}_2\right]
&=&2k^{2(m+n-p-q+2)}\Psi^{p;q}_{n;m+1}(a_k\,k^2/(32\pi),\bar{\lambda}_k/k^2;d)
\nonumber\\
& &-\frac{1}{2}\eta_{\beta}(k)\,k^{2(m+n-p-q+2)}
\widetilde{\Psi}^{p;q}_{n;m;1}(a_k\,k^2/(32\pi),\bar{\lambda}_k/k^2;d)
\nonumber\\
Q_n\left[{\cal A}_1^{-p}{\cal A}_2^{-q}P_k^m{\cal N}\right]
&=&k^{2(m+n-p-q+1)}\Psi^{p;q}_{n;m}(a_k\,k^2/(32\pi),\bar{\lambda}_k/k^2;d)
\nonumber\\
& &-\frac{1}{2}\eta_N(k)\,k^{2(m+n-p-q+1)}
\widetilde{\Psi}^{p;q}_{n;m;0}(a_k\,k^2/(32\pi),\bar{\lambda}_k/k^2;d)
\nonumber\\
Q_n\left[P_k^{-p}{\cal N}_0\right]
&=&k^{2(n-p+1)}\Phi^p_n(0)
\end{eqnarray}
the differential equations (\ref{q-level1}), (\ref{q-level2}) and 
(\ref{q-level3}) may be rewritten in terms of the threshold functions 
$\Psi^{p;q}_{n;m}(v,w;d)$ and $\widetilde{\Psi}^{p;q}_{n;m;l}(v,w;d)$ instead 
of the $Q_n$. 

In order to make
the integrals in eq. (\ref{Qn}) convergent we have to demand that
$R^{(0)}(y)$ decreases rapidly as $y\rightarrow \pm\infty$. However, since
from now on its form for $y<0$ does not play a role any more we identify
$R^{(0)}(y)$ with its part for nonnegative arguments and assume that
$R^{(0)}(y)$ is a smooth function defined only for $y\ge 0$ and endowed with
the properties stated in subsection \ref{3S3B}.

Next we introduce the dimensionless couplings $\lambda_k$, 
$g_k$ and $\beta_k$ of eqs. (\ref{lambda})-(\ref{beta}). Inserting eq. 
(\ref{lambda}) into $\partial_t(Z_{Nk}\bar{\lambda}_k)$ leads to the relation
\begin{eqnarray}
\label{mix}
\partial_t\lambda_k=-\left(2-\eta_N(k)\right)\lambda_k+32\pi\,g_k\,
\kappa^2\,k^{-d}\,\partial_t\left(Z_{Nk}\bar{\lambda}_k\right)\;.
\end{eqnarray}
Then, by using eq. (\ref{q-level1}), we obtain the differential 
equation (\ref{del}) for the dimensionless cosmological constant. The 
corresponding differential equations for $g_k$ and $\beta_k$ may be 
determined as follows. Taking the scale derivative of eq. (\ref{g}) and 
(\ref{beta}) leads to eqs. (\ref{deg}) and (\ref{deb}), respectively. For the
anomalous dimensions $\eta_N$ and $\eta_{\beta}$ we obtain from eqs. 
(\ref{q-level2}) and (\ref{q-level3}), respectively,
\begin{eqnarray}
\label{eta0}
\eta_N=g_k\,B_1(\lambda_k,g_k,\beta_k;d)+\eta_Ng_k\,B_2(\lambda_k,g_k,
\beta_k;d)+\eta_{\beta}\,g_k\,B_3(\lambda_k,g_k,\beta_k;d)
\end{eqnarray}
and
\begin{eqnarray}
\label{rho0}
\eta_{\beta}=-\beta_k^{-1}\,C_1(\lambda_k,g_k,\beta_k;d)
-\eta_N\,\beta_k^{-1}\,C_2(\lambda_k,g_k,\beta_k;d)-\eta_{\beta}\,\beta_k^{-1}
\,C_3(\lambda_k,g_k,\beta_k;d)\;.
\end{eqnarray}
The set of equations (\ref{eta0}) and (\ref{rho0}) may now be solved for the
anomalous dimensions $\eta_N$ and $\eta_{\beta}$ in terms of $\lambda_k$, 
$g_k$, $\beta_k$ and $d$ which eventually leads to the expressions 
(\ref{eta}) and (\ref{rho}).

\section{Coefficient functions appearing in the 
$\mbox{\boldmath$\beta$}$-functions}
\renewcommand{\theequation}{B\arabic{equation}}
\setcounter{equation}{0}
\label{coefffnct}
In the following we list the coefficient functions $A_i$, $B_i$, $C_i$, 
$i=1,2,3$, which appear in
the $\mbox{\boldmath$\beta$}$-functions (\ref{del}), (\ref{deg}) and 
(\ref{deb}). The coefficients $h_i(d)$ contained in these expressions are
defined in subsection \ref{betacoeff}, and the generalized threshold functions
$\Psi$ and $\widetilde{\Psi}$ were introduced in eq. (\ref{psi}) and eq. 
(\ref{psitilde}), respectively. The other threshold functions, $\Phi$ and
$\widetilde{\Phi}$, are those already introduced in the context of the pure
Einstein-Hilbert truncation \cite{Reu96}.
\begin{eqnarray}
\label{coeffA1}
\lefteqn{A_1(\lambda_k,g_k,\beta_k;d)\equiv
-2\lambda_k+2(4\pi)^{1-\frac{d}{2}}\,g_k\Bigg(h_8(d)
\,\Phi^1_{d/2}(-2\lambda_k)-2d\,\Phi^1_{d/2}(0)}\nonumber\\
& &+64\pi\,g_k\beta_k\,
\Psi^{0;1}_{d/2;1}(g_k\beta_k,-2\lambda_k;d)
-h_1(d)\Psi^{0;1}_{d/2;0}(g_k\beta_k,-2\lambda_k;d)\nonumber\\
& &+32\pi\,g_k\beta_k\,
\Psi^{1;1}_{d/2;2}(g_k\beta_k,-2\lambda_k;d)\Bigg)
\end{eqnarray}
\begin{eqnarray}
\label{coeffA2}
\lefteqn{A_2(\lambda_k,g_k,\beta_k;d)\equiv
\lambda_k-(4\pi)^{1-\frac{d}{2}}\,g_k\Bigg(h_8(d)
\,\widetilde{\Phi}^1_{d/2}(-2\lambda_k)}\nonumber\\
& &+h_1(d)\widetilde{\Psi}^{0;1}_{d/2;0;0}
(g_k\beta_k,-2\lambda_k;d)+32\pi\,g_k\beta_k\,
\Psi^{1;1}_{d/2;2;0}(g_k\beta_k,-2\lambda_k;d)\Bigg)
\end{eqnarray}
\begin{eqnarray}
\label{coeffA3}
A_3(\lambda_k,g_k,\beta_k;d)\equiv
-8(4\pi)^{2-\frac{d}{2}}\,g_k^2\,\beta_k\,
\widetilde{\Psi}^{0;1}_{d/2;0;1}(g_k\beta_k,-2\lambda_k;d)
\end{eqnarray}
\begin{eqnarray}
\label{coeffB1}
\lefteqn{B_1(\lambda_k,g_k,\beta_k;d)\equiv
4(4\pi)^{1-\frac{d}{2}}\,\Bigg\{
h_9(d)\,\Phi^1_{d/2-1}(-2\lambda_k)+h_{10}(d)\,\Phi^1_{d/2-1}(0)}\nonumber\\
& &+\frac{32}{3}\pi\,g_k\beta_k\,
\Psi^{0;1}_{d/2-1;1}(g_k\beta_k,-2\lambda_k;d)
-\frac{1}{6}h_1(d)\,\Psi^{0;1}_{d/2-1;0}(g_k\beta_k,-2\lambda_k;d)\nonumber\\
& &+\frac{16}{3}\pi\,g_k\beta_k\,
\Psi^{1;1}_{d/2-1;2}(g_k\beta_k,-2\lambda_k;d)
-32h_{11}(d)\pi\,g_k\beta_k\,\Phi^1_{d/2}(-2\lambda_k)\nonumber\\
& &+32h_{11}(d)\pi\,g_k\beta_k\,\Psi^{2;0}_{d/2;1}(g_k\beta_k,-2\lambda_k;d)
+h_{12}(d)\,\Phi^2_{d/2}(-2\lambda_k)
+h_{13}(d)\,\Phi^2_{d/2}(0)\nonumber\\
& &+32h_{19}(d)\pi\,g_k\beta_k\,\Psi^{1;1}_{d/2;1}(g_k\beta_k,-2\lambda_k;d)
+h_4(d)\,\Psi^{1;1}_{d/2;0}(g_k\beta_k,-2\lambda_k;d)\nonumber\\
& &+32h_3(d)\pi\,g_k\beta_k\,\Psi^{0;1}_{d/2;0}(g_k\beta_k,-2\lambda_k;d)
-h_{19}(d)(32\pi\,g_k\beta_k)^2\,\Psi^{1;2}_{d/2;3}(g_k\beta_k,-2\lambda_k;d)
\nonumber\\
& &-64h_4(d)\pi\,g_k\beta_k\,\Psi^{1;2}_{d/2;2}(g_k\beta_k,-2\lambda_k;d)
-2h_3(d)(32\pi\,g_k\beta_k)^2\,\Psi^{0;2}_{d/2;2}(g_k\beta_k,-2\lambda_k;d)
\nonumber\\
& &+32h_1(d)h_{19}(d)\pi\,g_k\beta_k\,
\Psi^{0;2}_{d/2;1}(g_k\beta_k,-2\lambda_k;d)
+h_1(d)h_4(d)\,\Psi^{0;2}_{d/2;0}(g_k\beta_k,-2\lambda_k;d)\nonumber\\
& &-h_2(d)(32\pi\,g_k\beta_k)^2\,\Psi^{2;2}_{d/2;4}(g_k\beta_k,
-2\lambda_k;d)\Bigg\}
\end{eqnarray}
\begin{eqnarray}
\label{coeffB2}
\lefteqn{B_2(\lambda_k,g_k,\beta_k;d)\equiv
-2(4\pi)^{1-\frac{d}{2}}\,\Bigg\{
h_9(d)\,\widetilde{\Phi}^1_{d/2-1}(-2\lambda_k)
-\frac{1}{6}h_1(d)\,\widetilde{\Psi}^{0;1}_{d/2-1;0;0}
(g_k\beta_k,-2\lambda_k;d)}\nonumber\\
& &+\frac{16}{3}\pi\,g_k\beta_k\,
\widetilde{\Psi}^{1;1}_{d/2-1;2;0}(g_k\beta_k,-2\lambda_k;d)
+32h_{11}(d)\pi\,g_k\beta_k\,\widetilde{\Psi}^{2;0}_{d/2;1;0}
(g_k\beta_k,-2\lambda_k;d)
\nonumber\\
& &+h_{12}(d)\,\widetilde{\Phi}^2_{d/2}(-2\lambda_k)+32h_3(d)\pi\,g_k\beta_k\,
\widetilde{\Psi}^{1;1}_{d/2;1;0}(g_k\beta_k,-2\lambda_k;d)\nonumber\\
& &+h_4(d)\,\widetilde{\Psi}^{1;1}_{d/2;0;0}(g_k\beta_k,-2\lambda_k;d)
-h_3(d)(32\pi\,g_k\beta_k)^2\,
\widetilde{\Psi}^{1;2}_{d/2;3;0}(g_k\beta_k,-2\lambda_k;d)\nonumber\\
& &-64h_4(d)\pi\,g_k\beta_k\,
\widetilde{\Psi}^{1;2}_{d/2;2;0}(g_k\beta_k,-2\lambda_k;d)
+32h_1(d)h_3(d)\pi\,g_k\beta_k\,
\widetilde{\Psi}^{0;2}_{d/2;1;0}(g_k\beta_k,-2\lambda_k;d)\nonumber\\
& &+h_1(d)h_4(d)\,\widetilde{\Psi}^{0;2}_{d/2;0;0}(g_k\beta_k,-2\lambda_k;d)
-h_2(d)(32\pi\,g_k\beta_k)^2\,
\widetilde{\Psi}^{2;2}_{d/2;4;0}(g_k\beta_k,-2\lambda_k;d)\Bigg\}
\end{eqnarray}
\begin{eqnarray}
\label{coeffB3}
\lefteqn{B_3(\lambda_k,g_k,\beta_k;d)\equiv
-2(4\pi)^{1-\frac{d}{2}}\,\Bigg\{
\frac{16}{3}\pi\,g_k\beta_k\,
\widetilde{\Psi}^{0;1}_{d/2-1;0;1}(g_k\beta_k,-2\lambda_k;d)}\\
& &-32h_{11}(d)\pi\,g_k\beta_k\,\widetilde{\Phi}^1_{d/2}(-2\lambda_k)
+32h_2(d)\pi\,g_k\beta_k\,\widetilde{\Psi}^{1;1}_{d/2;0;1}
(g_k\beta_k,-2\lambda_k;d)\nonumber\\
& &+32h_3(d)\pi\,g_k\beta_k\,\widetilde{\Psi}^{0;1}_{d/2;0;0}
(g_k\beta_k,-2\lambda_k;d)
-h_2(d)(32\pi\,g_k\beta_k)^2\,\widetilde{\Psi}^{1;2}_{d/2;2;1}
(g_k\beta_k,-2\lambda_k;d)\nonumber\\
& &-h_3(d)(32\pi\,g_k\beta_k)^2\,\widetilde{\Psi}^{0;2}_{d/2;1;1}
(g_k\beta_k,-2\lambda_k;d)-32h_4(d)\pi\,g_k\beta_k\,
\widetilde{\Psi}^{0;2}_{d/2;0;1}(g_k\beta_k,-2\lambda_k;d)\Bigg\}\nonumber
\end{eqnarray}
\begin{eqnarray}
\label{coeffC1}
\lefteqn{C_1(\lambda_k,g_k,\beta_k;d)\equiv
(4\pi)^{-\frac{d}{2}}\,\Bigg\{h_{14}(d)
\,\Phi^1_{d/2-2}(-2\lambda_k)-h_{16}(d)\,\Phi^1_{d/2-2}(0)}\nonumber\\
& &+64h_{15}(d)\pi\,g_k\beta_k\,
\Psi^{0;1}_{d/2-2;1}(g_k\beta_k,-2\lambda_k;d)
-h_1(d)h_{15}(d)\Psi^{0;1}_{d/2-2;0}(g_k\beta_k,-2\lambda_k;d)\nonumber\\
& &+32h_{15}(d)\pi\,g_k\beta_k\,
\Psi^{1;1}_{d/2-2;2}(g_k\beta_k,-2\lambda_k;d)
-32h_{17}(d)\pi\,g_k\beta_k\,\Phi^1_{d/2-1}(-2\lambda_k)\nonumber\\
& &+\frac{16}{3}h_3(d)\pi\,g_k\beta_k\,
\Psi^{0;1}_{d/2-1;0}(g_k\beta_k,-2\lambda_k;d)
+\frac{16}{3}h_{19}(d)\pi\,g_k\beta_k\,
\Psi^{1;1}_{d/2-1;1}(g_k\beta_k,-2\lambda_k;d)\nonumber\\
& &+\frac{1}{6}h_4(d)\,\Psi^{1;1}_{d/2-1;0}(g_k\beta_k,-2\lambda_k;d)
+32h_{17}(d)\pi\,g_k\beta_k\,\Psi^{2;0}_{d/2-1;1}(g_k\beta_k,-2\lambda_k;d)
\nonumber\\
& &-h_{18}(d)\,\Phi^2_{d/2-1}(-2\lambda_k)
-h_{20}(d)\,\Phi^2_{d/2-1}(0)\nonumber\\
& &-\frac{1}{6}h_{19}(d)(32\pi\,g_k\beta_k)^2\,\Psi^{1;2}_{d/2-1;3}
(g_k\beta_k,-2\lambda_k;d)
\nonumber\\
& &-\frac{32}{3}h_4(d)\pi\,g_k\beta_k\,\Psi^{1;2}_{d/2-1;2}
(g_k\beta_k,-2\lambda_k;d)
-\frac{1}{3}h_3(d)(32\pi\,g_k\beta_k)^2\,\Psi^{0;2}_{d/2-1;2}
(g_k\beta_k,-2\lambda_k;d)\nonumber\\
& &+\frac{16}{3}h_1(d)h_{19}(d)\pi\,g_k\beta_k\,\Psi^{0;2}_{d/2-1;1}
(g_k\beta_k,-2\lambda_k;d)
+\frac{1}{6}h_1(d)h_4(d)\,\Psi^{0;2}_{d/2-1;0}(g_k\beta_k,-2\lambda_k;d)
\nonumber\\
& &-\frac{1}{6}h_2(d)(32\pi\,g_k\beta_k)^2\,\Psi^{2;2}_{d/2-1;4}
(g_k\beta_k,-2\lambda_k;d)
-h_{11}(d)(32\pi\,g_k\beta_k)^2\,\Psi^{2;0}_{d/2;1}(g_k\beta_k,-2\lambda_k;d)
\nonumber\\
& &+32h_{25}(d)\pi\,g_k\beta_k\,\Phi^2_{d/2}(-2\lambda_k)
+h_{11}(d)(32\pi\,g_k\beta_k)^2\,\Psi^{3;0}_{d/2;2}(g_k\beta_k,-2\lambda_k;d)
\nonumber\\
& &-32h_{11}(d)h_{21}(d)\pi\,g_k\beta_k\,\Psi^{3;0}_{d/2;1}
(g_k\beta_k,-2\lambda_k;d)
+h_{23}(d)\,\Phi^3_{d/2}(-2\lambda_k)+h_{24}(d)\,\Phi^3_{d/2}(0)
\nonumber\\
& &-h_2(d)(32\pi\,g_k\beta_k)^2\,\Psi^{1;1}_{d/2;1}
(g_k\beta_k,-2\lambda_k;d)
+32h_2(d)h_{26}(d)\pi\,g_k\beta_k\,\Psi^{1;1}_{d/2;0}
(g_k\beta_k,-2\lambda_k;d)\nonumber\\
& &+h_2(d)(32\pi\,g_k\beta_k)^3\,\Psi^{1;2}_{d/2;3}
(g_k\beta_k,-2\lambda_k;d)
-h_{27}(d)(32\pi\,g_k\beta_k)^2\,\Psi^{1;2}_{d/2;2}
(g_k\beta_k,-2\lambda_k;d)\nonumber\\
& &+32h_4(d)h_{28}(d)\pi\,g_k\beta_k\,\Psi^{1;2}_{d/2;1}
(g_k\beta_k,-2\lambda_k;d)
-\frac{3}{2}h_4(d)^2\,\Psi^{1;2}_{d/2;0}(g_k\beta_k,-2\lambda_k;d)
\nonumber\\
& &-h_{29}(d)(32\pi\,g_k\beta_k)^2\,\Psi^{0;2}_{d/2;1}
(g_k\beta_k,-2\lambda_k;d)
-32h_4(d)h_{26}(d)\pi\,g_k\beta_k\,\Psi^{0;2}_{d/2;0}
(g_k\beta_k,-2\lambda_k;d)\nonumber\\
& &+\frac{1}{2}h_2(d)(32\pi\,g_k\beta_k)^3\,\Psi^{2;2}_{d/2;4}(g_k\beta_k,
-2\lambda_k;d)\nonumber\\
& &+\frac{2}{3}h_2(d)h_{28}(d)
(32\pi\,g_k\beta_k)^2\,\Psi^{2;2}_{d/2;3}(g_k\beta_k,
-2\lambda_k;d)\nonumber\\
& &-48h_2(d)h_4(d)\pi\,g_k\beta_k\,\Psi^{2;2}_{d/2;2}(g_k\beta_k,
-2\lambda_k;d)
+2h_3(d)^2(32\pi\,g_k\beta_k)^3\,\Psi^{0;3}_{d/2;3}(g_k\beta_k,
-2\lambda_k;d)\nonumber\\
& &-3h_1(d)h_3(d)h_{30}(d)(32\pi\,g_k\beta_k)^2\,\Psi^{0;3}_{d/2;2}
(g_k\beta_k,-2\lambda_k;d)\nonumber\\
& &+\frac{64}{3}h_1(d)h_4(d)h_{28}(d)\pi\,g_k\beta_k\,\Psi^{0;3}_{d/2;1}
(g_k\beta_k,-2\lambda_k;d)
-h_1(d)h_4(d)^2\,\Psi^{0;3}_{d/2;0}(g_k\beta_k,
-2\lambda_k;d)\nonumber\\
& &+3h_3(d)h_{30}(d)(32\pi\,g_k\beta_k)^3\,\Psi^{1;3}_{d/2;4}(g_k\beta_k,
-2\lambda_k;d)\nonumber\\
& &-\frac{4}{3}h_4(d)h_{28}(d)(32\pi\,g_k\beta_k)^2\,\Psi^{1;3}_{d/2;3}
(g_k\beta_k,-2\lambda_k;d)
+96h_4(d)^2\pi\,g_k\beta_k\,\Psi^{1;3}_{d/2;2}
(g_k\beta_k,-2\lambda_k;d)\nonumber\\
& &-\frac{2}{3}h_2(d)h_{28}(d)(32\pi\,g_k\beta_k)^3\,\Psi^{2;3}_{d/2;5}
(g_k\beta_k,-2\lambda_k;d)\nonumber\\
& &+3h_2(d)h_4(d)(32\pi\,g_k\beta_k)^2\,\Psi^{2;3}_{d/2;4}
(g_k\beta_k,-2\lambda_k;d)\nonumber\\
& &+h_2(d)^2(32\pi\,g_k\beta_k)^3\,\Psi^{3;3}_{d/2;6}
(g_k\beta_k,-2\lambda_k;d)\nonumber\\
& &+\left[-\frac{1}{2}R^{(0)'}(0)-\frac{96\pi\,g_k\beta_k}{1-2\lambda_k}
+\frac{11R^{(0)'}(0)+192\pi\,g_k\beta_k}
{2(1-2\lambda_k)^2}\right]\delta_{d,2}\nonumber\\
& &+\left[\frac{288\pi\,g_k\,\beta_k-1}
{4(144\pi\,g_k\,\beta_k-(1-2\lambda_k))}
+\frac{1}{4(1-2\lambda_k)}\right.\nonumber\\
& &\left.-\frac{96\pi\,g_k\,\beta_k-1}{2(96\pi\,g_k\,\beta_k
-(1-2\lambda_k))}-\frac{24\pi\,g_k\,\beta_k}{(1
-2\lambda_k)\left(96\pi\,g_k\,\beta_k-(1-2\lambda_k\right)}
\right]\delta_{d,4}
\Bigg\}
\end{eqnarray}
\begin{eqnarray}
\label{coeffC2}
\lefteqn{C_2(\lambda_k,g_k,\beta_k;d)\equiv
-\frac{1}{2}(4\pi)^{-\frac{d}{2}}\,\Bigg\{h_{14}(d)
\,\widetilde{\Phi}^1_{d/2-2}(-2\lambda_k)
-h_1(d)h_{15}(d)\widetilde{\Psi}^{0;1}_{d/2-2;0;0}(g_k\beta_k,-2\lambda_k;d)}
\nonumber\\
& &+32h_{15}(d)\pi\,g_k\beta_k\,
\widetilde{\Psi}^{1;1}_{d/2-2;2;0}(g_k\beta_k,-2\lambda_k;d)
+\frac{16}{3}h_3(d)\pi\,g_k\beta_k\,
\widetilde{\Psi}^{1;1}_{d/2-1;1;0}(g_k\beta_k,-2\lambda_k;d)\nonumber\\
& &+\frac{1}{6}h_4(d)\,\widetilde{\Psi}^{1;1}_{d/2-1;0;0}
(g_k\beta_k,-2\lambda_k;d)
+32h_{17}(d)\pi\,g_k\beta_k\,\widetilde{\Psi}^{2;0}_{d/2-1;1;0}
(g_k\beta_k,-2\lambda_k;d)\nonumber\\
& &-h_{18}(d)\,\widetilde{\Phi}^2_{d/2-1}(-2\lambda_k)
-\frac{1}{6}h_3(d)(32\pi\,g_k\beta_k)^2\,\widetilde{\Psi}^{1;2}_{d/2-1;3;0}
(g_k\beta_k,-2\lambda_k;d)\nonumber\\
& &-\frac{32}{3}h_4(d)\pi\,g_k\beta_k\,\widetilde{\Psi}^{1;2}_{d/2-1;2;0}
(g_k\beta_k,-2\lambda_k;d)\nonumber\\
& &+\frac{16}{3}h_1(d)h_{19}(d)\pi\,g_k\beta_k\,
\widetilde{\Psi}^{0;2}_{d/2-1;1;0}
(g_k\beta_k,-2\lambda_k;d)\nonumber\\
& &+\frac{1}{6}h_1(d)h_4(d)\,\widetilde{\Psi}^{0;2}_{d/2-1;0;0}
(g_k\beta_k,-2\lambda_k;d)
-\frac{1}{6}h_2(d)(32\pi\,g_k\beta_k)^2\,\widetilde{\Psi}^{2;2}_{d/2-1;4;0}
(g_k\beta_k,-2\lambda_k;d)\nonumber\\
& &+32h_{22}(d)\pi\,g_k\beta_k\,\widetilde{\Phi}^2_{d/2}(-2\lambda_k)
+h_{11}(d)(32\pi\,g_k\beta_k)^2\,\widetilde{\Psi}^{3;0}_{d/2;2;0}
(g_k\beta_k,-2\lambda_k;d)\nonumber\\
& &-32h_{11}(d)h_{21}(d)\pi\,g_k\beta_k\,\widetilde{\Psi}^{3;0}_{d/2;1;0}
(g_k\beta_k,-2\lambda_k;d)
+h_{23}(d)\,\widetilde{\Phi}^3_{d/2}(-2\lambda_k)\nonumber\\
& &+32h_6(d)\pi\,g_k\beta_k\,\widetilde{\Psi}^{1;1}_{d/2;0;0}
(g_k\beta_k,-2\lambda_k;d)
-h_7(d)(32\pi\,g_k\beta_k)^2\,\widetilde{\Psi}^{1;2}_{d/2;2;0}
(g_k\beta_k,-2\lambda_k;d)\nonumber\\
& &-96h_3(d)h_4(d)\pi\,g_k\beta_k\,\widetilde{\Psi}^{1;2}_{d/2;1;0}
(g_k\beta_k,-2\lambda_k;d)
-\frac{3}{2}h_4(d)^2\,\widetilde{\Psi}^{1;2}_{d/2;0;0}
(g_k\beta_k,-2\lambda_k;d)\nonumber\\
& &+32h_1(d)h_6(d)\pi\,g_k\beta_k\,\widetilde{\Psi}^{0;2}_{d/2;0;0}
(g_k\beta_k,-2\lambda_k;d)\nonumber\\
& &+\frac{1}{2}h_2(d)(32\pi\,g_k\beta_k)^3\,\widetilde{\Psi}^{2;2}_{d/2;4;0}
(g_k\beta_k,-2\lambda_k;d)\nonumber\\
& &-2h_2(d)h_3(d)(32\pi\,g_k\beta_k)^2\,\widetilde{\Psi}^{2;2}_{d/2;3;0}
(g_k\beta_k,-2\lambda_k;d)\nonumber\\
& &-48h_2(d)h_4(d)\pi\,g_k\beta_k\,\widetilde{\Psi}^{2;2}_{d/2;2;0}
(g_k\beta_k,-2\lambda_k;d)\nonumber\\
& &-h_1(d)h_3(d)^2(32\pi\,g_k\beta_k)^2\,\widetilde{\Psi}^{0;3}_{d/2;2;0}
(g_k\beta_k,-2\lambda_k;d)\nonumber\\
& &-64h_1(d)h_3(d)h_4(d)\pi\,g_k\beta_k\,\widetilde{\Psi}^{0;3}_{d/2;1;0}
(g_k\beta_k,-2\lambda_k;d)\nonumber\\
& &-h_1(d)h_4(d)^2\,\widetilde{\Psi}^{0;3}_{d/2;0;0}
(g_k\beta_k,-2\lambda_k;d)\nonumber\\
& &+h_3(d)^2(32\pi\,g_k\beta_k)^3\,\widetilde{\Psi}^{1;3}_{d/2;4;0}
(g_k\beta_k,-2\lambda_k;d)\nonumber\\
& &+4h_3(d)h_4(d)(32\pi\,g_k\beta_k)^2\,\widetilde{\Psi}^{1;3}_{d/2;3;0}
(g_k\beta_k,-2\lambda_k;d)\nonumber\\
& &+96h_4(d)^2\pi\,g_k\beta_k\,\widetilde{\Psi}^{1;3}_{d/2;2;0}
(g_k\beta_k,-2\lambda_k;d)\nonumber\\
& &+2h_2(d)h_3(d)(32\pi\,g_k\beta_k)^3\,\widetilde{\Psi}^{2;3}_{d/2;5;0}
(g_k\beta_k,-2\lambda_k;d)\nonumber\\
& &+3h_2(d)h_4(d)(32\pi\,g_k\beta_k)^2\,\widetilde{\Psi}^{2;3}_{d/2;4;0}
(g_k\beta_k,-2\lambda_k;d)\nonumber\\
& &+h_2(d)^2(32\pi\,g_k\beta_k)^3\,\widetilde{\Psi}^{3;3}_{d/2;6;0}
(g_k\beta_k,-2\lambda_k;d)\nonumber\\
& &+\left[-\frac{11R^{(0)'}(0)}{2(1-2\lambda_k)}
+\frac{192\pi\,g_k\beta_k+11R^{(0)'}(0)}{2(1-2\lambda_k)^2}\right]
\delta_{d,2}\nonumber\\
& &+\left[-\frac{1}{4(144\pi\,g_k\,\beta_k-(1-2\lambda_k))}
+\frac{1}{4(1-2\lambda_k)}\right.\nonumber\\
& &\left.+\frac{1}{2(96\pi\,g_k\,\beta_k-(1-2\lambda_k))}
-\frac{24\pi\,g_k\,\beta_k}{(1
-2\lambda_k)\left(96\pi\,g_k\,\beta_k-(1-2\lambda_k)\right)}
\right]\delta_{d,4}
\Bigg\}
\end{eqnarray}
\begin{eqnarray}
\label{coeffC3}
\lefteqn{C_3(\lambda_k,g_k,\beta_k;d)\equiv
-\frac{1}{2}(4\pi)^{-\frac{d}{2}}\,\Bigg\{
32h_{15}(d)\pi\,g_k\beta_k\,
\widetilde{\Psi}^{0;1}_{d/2-2;0;1}(g_k\beta_k,-2\lambda_k;d)}\nonumber\\
& &-32h_{17}(d)\pi\,g_k\beta_k\,\widetilde{\Phi}^1_{d/2-1}(-2\lambda_k)
+\frac{16}{3}h_3(d)\pi\,g_k\beta_k\,
\widetilde{\Psi}^{0;1}_{d/2-1;0;0}(g_k\beta_k,-2\lambda_k;d)\nonumber\\
& &+\frac{16}{3}h_2(d)\pi\,g_k\beta_k\,
\widetilde{\Psi}^{1;1}_{d/2-1;0;1}(g_k\beta_k,-2\lambda_k;d)\nonumber\\
& &-\frac{1}{6}h_2(d)(32\pi\,g_k\beta_k)^2\,
\widetilde{\Psi}^{1;2}_{d/2-1;2;1}(g_k\beta_k,-2\lambda_k;d)\nonumber\\
& &-\frac{1}{6}h_3(d)(32\pi\,g_k\beta_k)^2\,
\widetilde{\Psi}^{0;2}_{d/2-1;1;1}(g_k\beta_k,-2\lambda_k;d)\nonumber\\
& &-\frac{16}{3}h_4(d)\pi\,g_k\beta_k\,
\widetilde{\Psi}^{0;2}_{d/2-1;0;1}(g_k\beta_k,-2\lambda_k;d)
-h_{11}(d)(32\pi\,g_k\beta_k)^2\,\widetilde{\Psi}^{2;0}_{d/2;1;0}(g_k\beta_k,-2
\lambda_k;d)\nonumber\\
& &+16h_{11}(d)h_{21}(d)\pi\,g_k\beta_k\,\widetilde{\Phi}^2_{d/2}(-2\lambda_k)
-h_5(d)(32\pi\,g_k\beta_k)^2\,\widetilde{\Psi}^{1;1}_{d/2;0;1}
(g_k\beta_k,-2\lambda_k;d)\nonumber\\
& &-\frac{2}{d^2}(32\pi\,g_k\beta_k)^2\,\widetilde{\Psi}^{1;1}_{d/2;1;0}
(g_k\beta_k,-2\lambda_k;d)
+32h_2(d)h_3(d)\pi\,g_k\beta_k\,\widetilde{\Psi}^{1;1}_{d/2;0;0}
(g_k\beta_k,-2\lambda_k;d)\nonumber\\
& &+\frac{1}{2}h_2(d)(32\pi\,g_k\beta_k)^3\,
\widetilde{\Psi}^{1;2}_{d/2;2;1}(g_k\beta_k,-2\lambda_k;d)\nonumber\\
& &-2h_2(d)h_3(d)(32\pi\,g_k\beta_k)^2\,\widetilde{\Psi}^{1;2}_{d/2;1;1}
(g_k\beta_k,-2\lambda_k;d)\nonumber\\
& &-48h_2(d)h_4(d)\pi\,g_k\beta_k\,\widetilde{\Psi}^{1;2}_{d/2;0;1}
(g_k\beta_k,-2\lambda_k;d)\nonumber\\
& &-h_2(d)h_3(d)(32\pi\,g_k\beta_k)^2\,\widetilde{\Psi}^{1;2}_{d/2;2;0}
(g_k\beta_k,-2\lambda_k;d)\nonumber\\
& &-h_6(d)(32\pi\,g_k\beta_k)^2\,\widetilde{\Psi}^{0;2}_{d/2;0;1}
(g_k\beta_k,-2\lambda_k;d)
-h_3(d)^2(32\pi\,g_k\beta_k)^2\,\widetilde{\Psi}^{0;2}_{d/2;1;0}
(g_k\beta_k,-2\lambda_k;d)\nonumber\\
& &-32h_3(d)h_4(d)\pi\,g_k\beta_k\,\widetilde{\Psi}^{0;2}_{d/2;0;0}
(g_k\beta_k,-2\lambda_k;d)\nonumber\\
& &-h_2(d)^2(32\pi\,g_k\beta_k)^2
\,\widetilde{\Psi}^{2;2}_{d/2;2;1}(g_k\beta_k,-2\lambda_k;d)
+h_3(d)^2(32\pi\,g_k\beta_k)^3\,
\widetilde{\Psi}^{0;3}_{d/2;2;1}(g_k\beta_k,-2\lambda_k;d)\nonumber\\
& &+2h_3(d)h_4(d)(32\pi\,g_k\beta_k)^2\,\widetilde{\Psi}^{0;3}_{d/2;1;1}
(g_k\beta_k,-2\lambda_k;d)\nonumber\\
& &+32h_4(d)^2\pi\,g_k\beta_k\,\widetilde{\Psi}^{0;3}_{d/2;0;1}
(g_k\beta_k,-2\lambda_k;d)\nonumber\\
& &+2h_2(d)h_3(d)(32\pi\,g_k\beta_k)^3\,\widetilde{\Psi}^{1;3}_{d/2;3;1}
(g_k\beta_k,-2\lambda_k;d)\nonumber\\
& &+2h_2(d)h_4(d)(32\pi\,g_k\beta_k)^2\,\widetilde{\Psi}^{1;3}_{d/2;2;1}
(g_k\beta_k,-2\lambda_k;d)\nonumber\\
& &+h_2(d)^2(32\pi\,g_k\beta_k)^3\,\widetilde{\Psi}^{2;3}_{d/2;4;1}
(g_k\beta_k,-2\lambda_k;d)\nonumber\\
& &-\frac{96\pi\,g_k\beta_k}{1-2\lambda_k}\,\delta_{d,2}
+\left[\frac{36\pi\,g_k\,\beta_k}{144\pi\,g_k\,\beta_k-(1
-2\lambda_k)}
-\frac{24\pi\,g_k\,\beta_k}{96\pi\,g_k\,\beta_k-(1-2\lambda_k)}\right]
\delta_{d,4}\Bigg\}
\end{eqnarray}

In eqs. (\ref{coeffC1}), (\ref{coeffC2}) and (\ref{coeffC3}) the terms 
proportional to $\delta_{d,2}$ or $\delta_{d,4}$ arise not only from the 
$\delta$-terms of eq. (\ref{derivexp}), but also by evaluating
the ``primed'' traces, i.e. by subtracting the contributions coming from 
unphysical modes, see appendix \ref{trace} for details.
All these contributions are obtained by expanding various functions $f(R)$ 
with respect to $R$ and retaining only the terms $f(0)+f'(0)R$ in $d=2$ 
and $f(0)$ in $d=4$. As we explained above, these are the only pieces of $f$ 
which may contribute to the evolution in the truncated parameter space. 
Furthermore, the heat kernel expansions of the traces corresponding to 
differentially constrained fields introduce additional contributions 
proportional to $\delta_{d,2}$ or $\delta_{d,4}$ into eqs. 
(\ref{coeffC1})-(\ref{coeffC3}).
\section{Tensor spherical harmonics on $S^d$}
\renewcommand{\theequation}{C\arabic{equation}}
\setcounter{equation}{0}
\label{harm}
The spherical harmonics $T^{lm}_{\mu\nu}$, 
$T^{lm}_\mu$ and $T^{lm}$ for symmetric transverse traceless $(ST^2)$ 
tensors $h^T_{\mu\nu}$, transverse $(T)$ vectors $\xi_\mu$, and scalars $\phi$
on $S^d$ form complete sets of orthogonal eigenfunctions with respect to the 
covariant Laplacians. They satisfy
\begin{eqnarray}
\label{094}
-\bar{D}^2\,T^{lm}_{\mu\nu}(x)&=&\Lambda_l(d,2)\,T^{lm}_{\mu\nu}(x)\;,
\nonumber\\
-\bar{D}^2\,T^{lm}_\mu(x)&=&\Lambda_l(d,1)\,T^{lm}_\mu(x)\;,\nonumber\\
-\bar{D}^2\,T^{lm}(x)&=&\Lambda_l(d,0)\,T^{lm}(x)
\end{eqnarray}
and, after proper normalization,
\begin{eqnarray}
\label{093}
\delta^{lk}\,\delta^{mn}&=&\int d^dx\,\sqrt{\bar{g}}\,
\left({\bf 1}_{(2ST^2)}\right)^{\mu\nu\rho\sigma}\,T^{lm}_{\mu\nu}\,
T^{kn}_{\rho\sigma}
=\int d^dx\,\sqrt{\bar{g}}\,\left({\bf 1}_{(1T)}\right)^{\mu\nu}
\,T^{lm}_\mu\,T^{kn}_\nu\nonumber\\
&=&\int d^dx\,\sqrt{\bar{g}}\,T^{lm}\,T^{kn}\;.
\end{eqnarray}
Here $\left({\bf 1}_{(2ST^2)}\right)^{\mu\nu\rho\sigma}=(d-2)/(2d)
\left(\bar{g}^{\mu\rho}\bar{g}^{\nu\sigma}+\bar{g}^{\mu\sigma}\bar{g}^{\nu
\rho}\right)$ and $\left({\bf 1}_{(1T)}\right)^{\mu\nu}=(d-1)/d\,
\bar{g}^{\mu\nu}$ are the unit matrices in the spaces of $ST^2$ tensors and
transverse vectors, respectively. The $\Lambda_l(d,s)$'s denote the eigenvalues
of $-\bar{D}^2$ where $s$ is the spin of the field under consideration
and $l$ takes the values $s,s+1,s+2,\cdots$. The index $m=1,\cdots,D_l(d,s)$
is a degeneracy index. 

In ref. \cite{OR} explicit expressions for $\Lambda_l(d,s)$ and the 
degeneracies $D_l(d,s)$ were derived
which are summarized in Table 2. The eigenvalues are expressed in terms of the
curvature scalar ${\bar R}=d(d-1)/r^2$ of the sphere with radius $r$. 

The spherical harmonics $T^{lm}_{\mu\nu}$, $T^{lm}_\mu$ and $T^{lm}$ span the 
spaces of $ST^2$ tensors, $T$ vectors, and scalars so that we may expand 
arbitrary functions $h^T_{\mu\nu}$, $\xi_\mu$ and $\phi$ according to
\begin{eqnarray}
\label{067}
h_{\mu\nu}^T(x)&=&
\sum\limits_{l=2}^{\infty}\sum\limits_{m=1}^{D_l(d,2)}h^T_{lm}\,
T^{lm}_{\mu\nu}(x)\;,\nonumber\\
\xi_\mu(x)&=&
\sum\limits_{l=1}^{\infty}\sum\limits_{m=1}^{D_l(d,1)}\xi_{lm}\,
T^{lm}_\mu(x)\;,\nonumber\\
\phi(x)&=&\sum\limits_{l=0}^{\infty}\sum\limits_{m=1}^{D_l(d,0)}\phi_{lm}\,
T^{lm}(x)\;.
\end{eqnarray}

Eqs. (\ref{067}) may now be used to expand also any 
symmetric non-$T^2$ tensor and nontransverse vector in terms of spherical 
harmonics
since they may be expressed in terms of $ST^2$ tensors, $T$ vectors and 
scalars by using the decompositions (\ref{TT}), (\ref{T}), see e.g.
\cite{OR,Al86,FT84,TV90}. 

Note that
the $D_1(d,1)=d(d+1)/2$ modes $\{T^{1,m}_\mu\}$ and the $D_1(d,0)=d+1$ modes 
$\{T^{1,m}\}$ satisfy the Killing equation (\ref{09}) and the scalar equation 
(\ref{010}), respectively, and that $T^{0,1}=\rm const$. 
Arbitrary symmetric rank-2 tensors receive no 
contribution from these modes. In the case of arbitrary vectors the 
constant scalar mode does not contribute. Such modes have no physical 
meaning and have to be omitted therefore. \vspace{0.5cm}  

\begin{center}
\begin{tabular}
{|c|c|c|c|c|}
\hline\multicolumn{5}{|c|}{Table 2: Eigenvalues of $-\bar{D}^2$ and their 
degeneracies on the $d$-sphere}\\
\hline Eigenfunction & Spin $s$ & Eigenvalue $\Lambda_l(d,s)$ & Degeneracy 
$D_l(d,s)$ & $l$ \\
\hline $T_{\mu\nu}^{lm}(x)$ & 2 & $\frac{l(l+d-1)-2}{d(d-1)}\bar{R}$
& $\frac{(d+1)(d-2)(l+d)(l-1)(2l+d-1)(l+d-3)!}{2(d-1)!(l+1)!}$ &
$2,3,\cdots$ \\
\hline $T_\mu^{lm}(x)$ & 1 & $\frac{l(l+d-1)-1}{d(d-1)}\bar{R}$ &
$\frac{l(l+d-1)(2l+d-1)(l+d-3)!}{(d-2)!(l+1)!}$ & $1,2,\cdots$
\\ \hline $T^{lm}(x)$ & 0 & $\frac{l(l+d-1)}{d(d-1)}\bar{R}$ & 
$\frac{(2l+d-1)(l+d-2)!}{l!(d-1)!}$ & $0,1,\cdots$ \\
\hline\end{tabular}
\end{center}
\vspace{0.5cm}

\section{Tables of coefficient functions}
\renewcommand{\theequation}{D\arabic{equation}}
\setcounter{equation}{0}
\label{somecoe}
\subsection{Coefficients introduced in \boldmath$\Gamma_k^{(2)}[g,g]$}
\label{S_kcoeff}
In this subsection we define the various $A$'s, $B$'s, $C$'s and $G$'s and 
$H_S(d)$ which appear in eqs. (\ref{gravquad})-(\ref{squareflow}) of subsection
\ref{3S4C} and in eqs. (\ref{phi0term}) and (\ref{derivexp}) of appendix 
\ref{eval}.
\begin{eqnarray}
\label{coeff2}
& &A_T(d)\equiv\frac{d(d-3)+4}{d(d-1)}\;,\;\;
G_T(d)\equiv-\frac{d(d-5)+8}{2d(d-1)}\;,\;\;
A_V(d,\alpha)\equiv\frac{\alpha(d-2)-1}{d}\;,\nonumber\\
& &G_V(d)\equiv-\frac{d-4}{2d}\;,\;\;
A_{S1}(d,\alpha)\equiv\frac{\alpha(d-4)}{2\alpha(d-1)-(d-2)}\;,\nonumber\\
& &A_{S2}(d,\alpha)\equiv-\frac{\alpha(d-2)-2}{\alpha(d-2)-2(d-1)}\;,\;\;
B_{S1}(d,\alpha)\equiv-\frac{2\alpha d}{2\alpha(d-1)-(d-2)}\;,\nonumber\\
& &B_{S2}(d,\alpha)\equiv\frac{2\alpha d}{\alpha(d-2)-2(d-1)}\;,\;\;
C_{S1}(d,\alpha)\equiv-\frac{2\alpha(d-1)-(d-2)}{4(d-1)-2\alpha (d-2)}
\frac{d-2}{d-1}\;,\nonumber\\
& &C_{S2}(d,\alpha)\equiv\frac{d-1}{d^2}\frac{2(d-1)-\alpha(d-2)}
{\alpha}\;,\;\;
C_{S3}(d,\alpha)\equiv\frac{(d-2)(\alpha-1)}{\alpha (d-2)-2(d-1)}\;,\nonumber\\
& &G_{S1}(d)\equiv\frac{(d-1)(d-4)}{2d^2}\;,\;\;
G_{S2}(d)\equiv\frac{(d-1)(d-6)}{d^2}\;,\nonumber\\
& &G_{S3}(d)\equiv\frac{(d-4)(d-6)}{4d^2}\;,\;\;
H_S(d)\equiv2\left(\frac{d-1}{d}\right)^2\;.
\end{eqnarray}
\subsection{Coefficients appearing in the $\mbox{\boldmath$\beta$}$-functions}
\label{betacoeff}
Next we define the coefficients $h_i(d)$ contained in the 
$\mbox{\boldmath$\beta$}$-functions (\ref{del}), (\ref{deg}) and (\ref{deb})
via the coefficient functions $A_i$, $B_i$, $C_i$, $i=1,2,3$, given in
appendix \ref{coefffnct}. They also appear in
the approximate solutions for the non-Gaussian fixed point of appendix 
\ref{approxngfp}.
\begin{eqnarray}
\label{coeff3}
& &h_1(d)\equiv\frac{d-2}{d-1}\;,\;\,h_2(d)\equiv\frac{d-4}{d}\;,\;\;
h_3(d)\equiv\frac{d^2-8d+4}{2d(d-1)}\;,\;\;
h_4(d)\equiv-\frac{(d-2)(d-4)}{d(d-1)}\;,\nonumber\\
& &h_5(d)\equiv\frac{d^2-4d-2}{2d^2}\,,\;
h_6(d)\equiv\frac{(d-4)^2}{2d(d-1)}\,,\;
h_7(d)\equiv\frac{5d^4-48d^3+148d^2-112d+16}{4d^2(d-1)^2}\;,\nonumber\\
& &h_8(d)\equiv\frac{d^2+d-4}{2}\;,\;\;
h_9(d)\equiv\frac{(d+3)(d+2)(d^2-5d+2)}{12d(d-1)}\;,\;\;
h_{10}(d)\equiv-\frac{d^2-6}{3d}\;,\nonumber\\
& &
h_{11}(d)\equiv\frac{(d+1)(d-2)}{2}\;,\;\;
h_{12}(d)\equiv-\frac{d^4-2d^3-5d^2+16d-14}{2d(d-1)}\;,\nonumber\\
& &h_{13}(d)\equiv-\frac{2(d+1)}{d}\;,\;\;
h_{14}(d)\equiv\frac{5d^6-7d^5-139d^4-545d^3-898d^2+504d-360}
{720d^2(d-1)^2}\;,\nonumber\\
& &h_{15}(d)\equiv\frac{5d^2-7d+6}{360d(d-1)}\;,\;\;
h_{16}(d)\equiv\frac{5d^4-7d^3-54d^2-180d+180}{180d^2(d-1)}\;,
\nonumber\\
& &h_{17}(d)\equiv\frac{(d+2)(d+1)(d-5)}{12(d-1)},\,
h_{18}(d)\equiv\frac{(d+2)(d^5-5d^4-5d^3+43d^2-68d+18)}{12d^2(d-1)^2},
\nonumber\\
& &h_{19}(d)\equiv\frac{5d^2-28d+20}{2d(d-1)}\;,\;\;
h_{20}(d)\equiv\frac{(d+3)(d-2)}{3d^2}\;,\nonumber\\
& &h_{21}(d)\equiv2\frac{d^2-3d+4}{d(d-1)}\;,\;\;
h_{22}(d)\equiv\frac{(d-3)(d^3-d^2-4d+8)}{4d(d-1)}\;,\nonumber\\
& &h_{23}(d)\equiv\frac{d^6-5d^5+3d^4+31d^3-86d^2+98d-50}{2d^2(d-1)^2}\;,\;\;
h_{24}(d)\equiv-2\frac{d+3}{d^2}\;,\nonumber\\
& &h_{25}(d)\equiv\frac{3d^4-12d^3+9d^2+24d-40}{4d(d-1)}\;,\;\;
h_{26}(d)\equiv\frac{d^2-6d+2}{d(d-1)}\;,\nonumber\\
& &h_{27}(d)\equiv\frac{15d^4-178d^3+628d^2-632d+176}{4d^2(d-1)^2}\;,\;\;
h_{28}(d)\equiv-\frac{9(d^2-6d+4)}{2d(d-1)}\;,\nonumber\\
& &h_{29}(d)\equiv\frac{5d^4-52d^3+168d^2-128d+16}{4d^2(d-1)^2}\;,\;\;
h_{30}(d)\equiv\frac{3d^2-16d+12}{2d(d-1)}\;,\nonumber\\
& &h_{31}(d)\equiv\frac{5d^6-27d^5-71d^4-405d^3-342d^2-960d+360}
{720d^2(d-1)^2}\;,\nonumber\\
& &h_{32}(d)\equiv-\frac{d^6-3d^5-7d^4+5d^3+26d^2-82d+12}
{12d^2(d-1)^2}\;,\nonumber\\
& &h_{33}(d)\equiv\frac{d^6-5d^5+7d^4-13d^3+42d^2-42d+2}
{2d^2(d-1)^2}\;,\nonumber\\
& &h_{34}(d)\equiv\frac{5d^6-7d^5-119d^4-593d^3-846d^2+480d-360}
{360d^2(d-1)^2}\;,\nonumber\\
& &h_{35}(d)\equiv-\frac{d^6-3d^5-11d^4+9d^3+54d^2-134d+36}
{3d^2(d-1)^2}\;,\nonumber\\
& &h_{36}(d)\equiv3\frac{d^6-5d^5+7d^4-9d^3+46d^2-62d+14}
{d^2(d-1)^2}\;,\nonumber\\
& &h_{37}(d)\equiv\frac{(d+2)(d^3-6d^2+3d-6)}{3d(d-1)}\;,\;\;
h_{38}(d)\equiv-2\frac{d^4-2d^3+3d^2-4d-2}{d(d-1)}\;,\nonumber\\
& &h_{39}(d)\equiv\frac{5d^2-7d+6}{45d(d-2)}\;,\;\;
h_{40}(d)\equiv\frac{30d^5-115d^4-362d^3+721d^2+182d+264}
{90d(d-1)(d-2)}\;,\nonumber\\
& &h_{41}(d)\equiv-2\frac{3d^6-17d^5+25d^4+39d^3-166d^2+224d-96}
{3d^2(d-1)(d-2)}\;,\nonumber\\
& &h_{42}(d)\equiv 4\frac{(d-1)(d-4)^2}{d(d-2)}\;,\;\;
h_{43}(d)\equiv-\frac{(d+2)(d^3-6d^2+3d-6)}{3d(d-1)(d-2)}\;,\nonumber\\
& &h_{44}(d)\equiv 2\frac{d^4-2d^3+3d^2-4d-2}{d(d-1)(d-2)}\;,\;\;
h_{45}(d)\equiv\frac{d^4-3d^3+32d-32}{d^2}\;,\nonumber\\
& &h_{46}(d)\equiv -\frac{d^4-13d^2-24d+12}{6d(d-1)}\;,\;\;
h_{47}(d)\equiv\frac{d^4-2d^3-d^2-4d+2}{d(d-1)}\;.
\end{eqnarray}
\section{Closed-form formulas for the fixed point location}
\renewcommand{\theequation}{E\arabic{equation}}
\setcounter{equation}{0}
\label{approxngfp}
In the following we derive the approximate formula for the position of the 
non-Gaussian fixed point discussed in subsection \ref{3S5C}. Here we restrict
our considerations to the case $d>2$. 

In a first approximation
we set $\lambda_k=\lambda_*=0$, $\beta_k=\beta_*=0$ and determine $g_*$ from 
the condition $\eta_{N*}=2-d$ alone. Since $\beta_*=0$, we may solve this 
equation for $g_*$ in closed form which leads to
\begin{eqnarray}
\label{gstar0}
g_*=\frac{2-d}{B_1(0,0,\lambda_*;d)-(d-2)B_2(0,0,\lambda_*;d)}\;.
\end{eqnarray}
As $\lambda_*=0$, it boils down to
\begin{eqnarray}
\label{gstar1}
g_*&=&(4\pi)^{\frac{d}{2}-1}\Bigg\{h_{43}(d)\,\Phi^1_{d/2-1}(0)
+h_{46}(d)\,\widetilde{\Phi}^1_{d/2-1}(0)+h_{44}(d)\,\Phi^2_{d/2}(0)\nonumber\\
& &+h_{47}(d)\,\widetilde{\Phi}^2_{d/2}(0)\Bigg\}^{-1}\;.
\end{eqnarray}
Here the $h_i(d)$ are again $d$-dependent coefficients which are defined in
subsection \ref{betacoeff} of appendix \ref{somecoe}. It is remarkable that 
the solution (\ref{gstar1}) coincides precisely with
the corresponding approximate solution (H2) of ref. \cite{LR1} with $\alpha=1$,
obtained in the framework of the Einstein-Hilbert truncation.

Employing the exponential shape function (\ref{expshape}) with $s=1$, and 
setting $d=4$, for instance, eq. (\ref{gstar1}) yields $g_*\approx 0.590$.
Here we used that, for this shape function, $\Phi^1_1(0)=\pi^2/6$,
$\Phi^2_2(0)=1$, $\widetilde{\Phi}^1_1(0)=1$,
$\widetilde{\Phi}^2_2(0)=1/2$, see appendix \ref{threshprop}.

In order to improve upon this approximation scheme, we determine 
$(\lambda_*,g_*,\beta_*)$ 
from a set of Taylor-expanded $\mbox{\boldmath $\beta$}$-functions. Using 
eqs. (\ref{psiexpansion1})-(\ref{phiexpansion}) we expand the 
$\mbox{\boldmath $\beta$}$-functions (\ref{del}), (\ref{deg}) and (\ref{deb})
about $\lambda_k=g_k=\beta_k=0$ and obtain
\begin{eqnarray}
\label{expandbeta}
\mbox{\boldmath $\beta$}_\lambda(\lambda_k,g_k,\beta_k;d)&=&-2\lambda_k+\nu_d
\,d\;g_k+{\cal O}\left({\rm g}^2\right)\;,\nonumber\\
\mbox{\boldmath $\beta$}_g(\lambda_k,g_k;\alpha,d)&=&(d-2)\,g_k-(d-2)
\omega_d\,g_k^2+{\cal O}\left({\rm g}^3\right)\;,\nonumber\\
\mbox{\boldmath $\beta$}_\beta(0,0,\beta_k;d)&=&\gamma_d+(4-d)\beta_k
+{\cal O}\left({\rm g}^2\right)\;.
\end{eqnarray}
Here $\gamma_d$, $\nu_d$, and $\omega_d$ are defined as in eqs. 
(\ref{gamma_d}), (\ref{nu_d}) and (\ref{omega_d}), respectively, and 
${\cal O}\left({\rm g}^n\right)$ stands for terms of $n$th and higher orders 
in the couplings ${\rm g}_1(k)=\lambda_k$, ${\rm g}_2(k)=g_k$
and ${\rm g}_3(k)=\beta_k$. Now $g_*$ is obtained as the nontrivial solution 
to $\mbox{\boldmath$\beta$}_g=0$, which reads
\begin{eqnarray}
\label{gstar2}
g_*=\omega_d^{-1}=(4\pi)^{\frac{d}{2}-1}\left\{h_{43}(d)\,
\Phi^1_{d/2-1}(0)+h_{44}(d)\,\Phi^2_{d/2}(0)\right\}^{-1}\;.
\end{eqnarray}
Inserting eq. (\ref{gstar2}) into $\mbox{\boldmath $\beta$}_\lambda=0$ leads to
\begin{eqnarray}
\label{lambdastar1}
\lambda_*=\frac{\nu_d\,d}{2\omega_d}=\frac{d(d-3)}{2}\,
\Phi^1_{d/2}(0)\left\{h_{43}(d)\,
\Phi^1_{d/2-1}(0)+h_{44}(d)\,\Phi^2_{d/2}(0)\right\}^{-1}\;.
\end{eqnarray}
Quite remarkably, also these results agree completely with those of ref. 
\cite{LR1} which follow
from the pure Einstein-Hilbert truncation. (See eqs. (H6) and (H7) of this 
reference.) 

Now we use $\mbox{\boldmath $\beta$}_\beta$ in order to determine $\beta_*$.
However, since the term linear in $\beta_k$ vanishes for $d=4$, the  
expanded $\mbox{\boldmath $\beta$}_\beta$ of eq. (\ref{expandbeta}) is not
sufficient in this case. Therefore we consider also those 
terms of second order in the couplings which are linear in $\beta_k$. For these
terms we find
\begin{eqnarray}
\label{additional}
\left.\frac{\partial^2\mbox{\boldmath $\beta$}_\beta}{\partial\lambda_k\partial
\beta_k}\right|_{\lambda_k=g_k=\beta_k=0}&=&0\;,\nonumber\\
\alpha_d\equiv\left.\frac{\partial^2\mbox{\boldmath $\beta$}_\beta}
{\partial g_k\partial
\beta_k}\right|_{\lambda_k=g_k=\beta_k=0}&=&-(4\pi)^{1-\frac{d}{2}}\left\{
2h_{39}(d)+2h_{45}(d)\,\Phi^2_{d/2}(0)-3\,\delta_{d,4}\right\}\;.
\end{eqnarray}
Taking the nonvanishing term of eq. (\ref{additional}) into account, and 
inserting $g_*$ of eq. (\ref{gstar2}) into $\mbox{\boldmath $\beta$}_\beta=0$ 
then leads to
\begin{eqnarray}
\label{betastar1}
\lefteqn{\beta_*=\frac{\gamma_d}{d-4-\alpha_d\,\omega_d^{-1}}}\\
&=&\frac{(4\pi)^{-\frac{d}{2}}\left\{h_{31}(d)\,\Phi^1_{d/2-2}(0)
+h_{32}(d)\,\Phi^2_{d/2-1}(0)+h_{33}(d)\,\Phi^3_{d/2}(0)\right\}}{d-4+\left\{
2h_{39}(d)+2h_{45}(d)\,\Phi^2_{d/2}(0)-3\,\delta_{d,4}\right\}
\left\{h_{43}(d)\,
\Phi^1_{d/2-1}(0)+h_{44}(d)\,\Phi^2_{d/2}(0)\right\}^{-1}}\,.\nonumber
\end{eqnarray}

Employing the shape function (\ref{expshape}) with $s=1$ we obtain from eqs. 
(\ref{gstar2}), (\ref{lambdastar1}) and (\ref{betastar1}) in $d=4$ dimensions
\begin{eqnarray}
\label{H7}
\lambda_*&=&\zeta(3)\left(\frac{13\pi^2}{144}+\frac{79}{24}\right)^{-1}
\approx 0.287\;,\nonumber\\
g_*&=&\left(\frac{13\pi}{144}+\frac{79}{24\pi}\right)^{-1}
\approx 0.751\;,\nonumber\\
\beta_*&=&\frac{419(13\pi^2+474)}{(4\pi)^2\,906768}\approx 0.0018\;.
\end{eqnarray}
Here we used the expressions for the threshold functions derived in 
appendix \ref{threshprop}. The numbers in (\ref{H7}) should be compared to the
exact result (\ref{R^2fixedpoint}).
\section{Properties of the threshold functions}
\renewcommand{\theequation}{F\arabic{equation}}
\setcounter{equation}{0}
\label{threshprop}
In this appendix we summarize various important properties of the threshold 
functions $\Psi^{p;q}_{n;m}$, $\widetilde{\Psi}^{p;q}_{n;m;l}$, $\Phi^p_n$ and
$\widetilde{\Phi}^p_n$ which are defined by eqs. (\ref{psi}), (\ref{psitilde}) 
and (\ref{phi}).

Expanding the generalized threshold functions $\Psi^{p;q}_{n;m}$, 
$\widetilde{\Psi}^{p;q}_{n;m;l}$ about vanishing couplings yields
\begin{eqnarray}
\label{psiexpansion1}
\lefteqn{\Psi^{p;q}_{n;m}(g_k\beta_k,-2\lambda_k;d)
=\left(-2\frac{d-1}{d-2}\right)^q
\,\Phi^{p+q-m}_n(0)+2(p+q)\left(-2\frac{d-1}{d-2}\right)^q}\nonumber\\
& &\times\Phi^{p+q-m+1}_n(0)\,\lambda_k
+2(p+q)(p+q+1)\left(-2\frac{d-1}{d-2}\right)^q
\,\Phi^{p+q-m+2}_n(0)\,\lambda_k^2\nonumber\\
& &-32\pi q\left(-2\frac{d-1}{d-2}\right)^{q+1}
\,\Phi^{p+q-m-1}_n(0)\,g_k\beta_k+{\cal O}({\rm g}^3)\;,
\end{eqnarray}
\begin{eqnarray}
\label{psiexpansion2}
\lefteqn{\widetilde{\Psi}^{p;q}_{n;m;0}(g_k\beta_k,-2\lambda_k;d)=
\left(-2\frac{d-1}{d-2}\right)^q\,\widetilde{\Phi}^{p+q-m}_n(0)
+2(p+q)\left(-2\frac{d-1}{d-2}\right)^q}\nonumber\\
& &\times\widetilde{\Phi}^{p+q-m+1}_n(0)\,\lambda_k
+2(p+q)(p+q+1)\left(-2\frac{d-1}{d-2}\right)^q
\,\widetilde{\Phi}^{p+q-m+2}_n(0)\,\lambda_k^2\nonumber\\
& &-32\pi q\left(-2\frac{d-1}{d-2}\right)^{q+1}
\,\widetilde{\Phi}^{p+q-m-1}_n(0)\,g_k\beta_k
+{\cal O}({\rm g}^3)\;,
\end{eqnarray}
\begin{eqnarray}
\label{psiexpansion3}
\lefteqn{\widetilde{\Psi}^{p;q}_{n;m;1}(g_k\beta_k,-2\lambda_k;d)=
\left(-2\frac{d-1}{d-2}\right)^q \left(\widetilde{\Phi}^{p+q-m-1}_n(0)
+n\,\widetilde{\Phi}^{p+q-m}_{n+1}(0)\right)}\nonumber\\
& &+2(p+q)\left(-2\frac{d-1}{d-2}\right)^q
\left(\widetilde{\Phi}^{p+q-m}_n(0)+n\,\widetilde{\Phi}^{p+q-m+1}_{n+1}(0)
\right)\,\lambda_k\nonumber\\
& &+2(p+q)(p+q+1)\left(-2\frac{d-1}{d-2}\right)^q
\left(\widetilde{\Phi}^{p+q-m+1}_n(0)+n\,\widetilde{\Phi}^{p+q-m+2}_{n+1}(0)
\right)\,\lambda_k^2\nonumber\\
& &-32\pi q\left(-2\frac{d-1}{d-2}\right)^{q+1}
\left(\widetilde{\Phi}^{p+q-m-2}_n(0)+n\,\widetilde{\Phi}^{p+q-m-1}_{n+1}(0)
\right)\,g_k\beta_k+{\cal O}({\rm g}^3)\;.
\end{eqnarray}
Here ${\cal O}\left({\rm g}^3\right)$ stands for terms of third and higher 
orders in the couplings ${\rm g}_1(k)=\lambda_k$, ${\rm g}_2(k)=g_k$
and ${\rm g}_3(k)=\beta_k$. Quite remarkably, every fixed order of these 
expansions depends only on the ``conventional'' threshold functions 
$\Phi^p_n$ and $\widetilde{\Phi}^p_n$ at vanishing arguments. 

By using eq. (\ref{phi}) the corresponding expansions of $\Phi^p_n$ and 
$\widetilde{\Phi}^p_n$ about vanishing argument can be read off directly from 
eqs. (\ref{psiexpansion1}) and (\ref{psiexpansion2}). They are given by
\begin{eqnarray}
\label{phiexpansion}
\Phi^p_n(-2\lambda_k)&=&
\Phi^p_n(0)+2p\,\Phi^{p+1}_n(0)\,\lambda_k
+2p(p+1)\,\Phi^{p+2}_n(0)\,\lambda_k^2+{\cal O}(\lambda_k^3)\nonumber\\
\widetilde{\Phi}^p_n(-2\lambda_k)&=&
\widetilde{\Phi}^p_n(0)+2p\,\widetilde{\Phi}^{p+1}_n(0)\,\lambda_k
+2p(p+1)\,\widetilde{\Phi}^{p+2}_n(0)\,\lambda_k^2
+{\cal O}(\lambda_k^3)
\end{eqnarray}

For $n=0$ the threshold functions are universal in the sense that they do not
depend on $R^{(0)}(y)$. In fact, setting
$n=0$ in $\Psi^{p;q}_{n;m}$, $\widetilde{\Psi}^{p;q}_{n;m;l}$, $\Phi^p_n$ and
$\widetilde{\Phi}^p_n$ leads to
\begin{eqnarray}
\label{genthresh}
\Psi_{0;m}^{p;q}(v,w;d)=\widetilde{\Psi}_{0;m;l}^{p;q}(v,w;d)
=\left(1+w\right)^{-p}\left(32\pi\,v-\frac{d-2}{2(d-1)}\left(1+w\right)
\right)^{-q}
\end{eqnarray}
and
\begin{eqnarray}
\label{thresh0}
\Phi_0^p(w)=\widetilde{\Phi}_0^p(w)=\left(1+w\right)^{-p}\;.
\end{eqnarray}

There exists a second class of universal values of certain threshold functions.
Using the boundary conditions for $R^{(0)}(y)$ one may easily verify that, for 
vanishing argument and for $n+1=p\ge 1$, $\Phi^p_n$ assumes the universal value
\begin{eqnarray}
\label{s=1,2}
\Phi^p_{p-1}(0)=\frac{1}{\Gamma(p)}\;.
\end{eqnarray}

Let us now be more specific and opt for the family of exponential cutoffs 
(\ref{expshape}). In this
case the integral that defines $\Phi^p_n$ can be carried out analytically
for vanishing argument. Using the integral representation of the polylogarithm
\cite{lewin}, 
\begin{eqnarray}
\label{polylog}
{\rm Li}_{n}(x)=\frac{1}{\Gamma(n)}\int\limits_0^\infty dz\,
\frac{x\,z^{n-1}}{e^z-x}\;,
\end{eqnarray}
one obtains for $p=0,\ldots,4$:
\begin{eqnarray}
\label{thresh}
\Phi^0_n(0)=n(n+1)\,s^{-n}\,\zeta(n+1)
\end{eqnarray}
\begin{eqnarray}
\label{thresh1}
\Phi^1_n(0)=n\,s^{-n}\left\{\zeta(n+1)-{\rm Li}_{n+1}(1-s)\right\}
\end{eqnarray}
\begin{eqnarray}
\label{thresh2}
\Phi^2_n(0)=\left\{\begin{array}{ll} s^{2-n}\,(1-s)^{-1}\,{\rm Li}_{n-1}(1-s)
& s\neq 1\\1 & s=1\end{array}\right.
\end{eqnarray}
\begin{eqnarray}
\label{thresh3}
\lefteqn{\Phi^3_n(0)=}\\
& &\left\{\begin{array}{llll} [2(n-1)(1-s)^2]^{-1}\,s^{3-n}\,\left\{
(2-s)\,{\rm Li}_{n-2}(1-s)-s\,{\rm Li}_{n-3}(1-s)\right\}
& n\neq 1\;,\;\;s\neq 1 \\ \left[2^{n-1}(n-1)\right]^{-1} (2^{n-1}-1) & 
n\neq 1\;,\;\;s=1 \\ - [2(1-s)^2]^{-1}\,s^{3-n}\,\left\{
(2-s)\,{\rm Li}_{-1}^{(1,0)}(1-s)
-s\,{\rm Li}_{-2}^{(1,0)}(1-s)\right\} & n=1\;,\;\;s\neq 1\\
\ln(2) & n=s=1 \end{array}\right.\nonumber
\end{eqnarray}
\begin{eqnarray}
\label{thresh4}
\lefteqn{\Phi^4_n(0)=}\\
& &\left\{\begin{array}{lllllll}\left[6(n-1)(n-2)(1-s)^3\right]^{-1}\,s^{4-n}
\,\left\{2(s^2-3s+3)\,{\rm Li}_{n-3}(1-s)\right. & \\
\hspace{0.5cm}\left.-3s(2-s)\,{\rm Li}_{n-4}(1-s)+s^2\,{\rm Li}_{n-5}(1-s)
\right\}
& n\neq 1,2\;,\;\,s\neq 1 \\ \left[(n-1)(n-2)\right]^{-1}(1-2^{3-n}+3^{2-n})
& n\neq 1,2\;,\;\,s=1 \\ \left[6(2n-3)(1-s)^3\right]^{-1}\,s^{4-n}\,\left\{
2(s^2-3s+3)\,{\rm Li}_{n-3}^{(1,0)}(1-s)\right. & \\
\hspace{0.5cm}\left.-3s(2-s)\,{\rm Li}_{n-4}^{(1,0)}(1-s)
+s^2\,{\rm Li}_{n-5}^{(1,0)}(1-s)\right\} & n\in\{1,2\}\;,\;\,s\neq 1\\
\ln(27/16) & n=s=1\\ \ln(4/3) & n=2\;,\;\,s=1 \end{array}\right.\nonumber
\end{eqnarray}
Here we defined 
\begin{eqnarray}
\label{polylogspecials1}
{\rm Li}_{n}^{(k,l)}(x)\equiv\frac{d^k}{dn^k}\frac{d^l}{dx^l}{\rm Li}_{n}(x)
\end{eqnarray}
and used the relations 
\begin{eqnarray}
\label{polylogspecials2}
{\rm Li}_{n}(1)=\zeta(n)\;,\;\;{\rm Li}_{n}^{(0,1)}(x)
=\frac{{\rm Li}_{n-1}(x)}{x}
\end{eqnarray}
with $\zeta$ denoting the Riemann zeta-function.  
For nonvanishing arguments an analytic solution to the integrals defining the
threshold functions is not known. 

For the exponential cutoff (\ref{expshape}) with $s=1$ there even exists a 
very useful relation among 
$\Phi^p_n(0)$ and $\widetilde{\Phi}^{p}_n(0)$. One may easily verify that 
\begin{eqnarray}
\label{s=1,1}
\Phi^p_n(0)=\widetilde{\Phi}^{p-1}_n(0)\;.
\end{eqnarray}
This relation allows us to calculate the $\widetilde{\Phi}^{p}_n(0)$-integrals 
analytically as well.

\section{Proof of the inequality (\ref{phisigma})}
\renewcommand{\theequation}{G\arabic{equation}}
\setcounter{equation}{0}
\label{estimation}
In this appendix we prove the inequality (\ref{phisigma}). 
As a first step we consider the function
\begin{eqnarray}
\label{estim}
f(a,y)=a\sqrt{a\left(a-y\right)}-\sqrt{1-y}
\end{eqnarray}
with $a>1$ and $0\le y\le 2/5$. (For $a=1$ this function vanishes identically:
$f(1,y)\equiv 0$.) An upper bound for this function may be obtained as follows.
For the first two derivatives of $f$ with respect to $y$ we obtain
\begin{eqnarray}
\label{estim2}
f^{(0,1)}(a,y)&\equiv&\frac{d}{dy}f(a,y)=-\frac{a^2}{2\sqrt{a\left(a-y\right)}}
+\frac{1}{2\sqrt{1-y}}\;,\nonumber\\
f^{(0,2)}(a,y)&\equiv&\frac{d^2}{dy^2}f(a,y)=
-\frac{a^3}{4\left[a\left(a-y\right)\right]^{\frac{3}{2}}}
+\frac{1}{4\left(1-y\right)^{\frac{3}{2}}}\;.
\end{eqnarray}
Solving $f^{(0,1)}(a,y)=0$ for $y$ leads to the {\it single} solution
\begin{eqnarray}
\label{solution}
y=y_0\equiv\frac{a(a+1)}{a^2+a+1}\;.
\end{eqnarray}
Since $f^{(0,2)}(a,y_0)=\left(a^2+a+1\right)^{\frac{3}{2}}\left(a^3-1\right)/
(4a^3)>0$, we have $f(a,y_0)\le f(a,y)$ for all $y\in[0,1]$ and
$a>1$. Hence, for $a>1$ fixed but arbitrary, $f$ monotonically decreases in
the interval $y\in[0,y_0]$ where $y_0<1$. 

Furthermore, $y_0=y_0(a)$ is a monotonically 
increasing function of $a$ for all $a\ge 1$. Therefore we have that
$y_0(a)\ge y_0(a=1)=2/3$. As a consequence, $f$ monotonically decreases in the
interval $0\le y\le 2/3$ for any value of $a>1$. Since we restricted 
our considerations to $y\in[0,2/5]$ we obtain $f(a,y)\le f(a,0)$.
 
$\left({\cal R}_k\right)_{\bar{\phi}
\bar{\sigma}}$ may be obtained from $f$ by replacing
\begin{eqnarray}
\label{replace2}
a\rightarrow\frac{\Lambda_l(4,0)/k^2+R^{(0)}
(\Lambda_l(4,0)/k^2)}{\Lambda_l(4,0)/k^2}\;,
\;\;\;\;\;\;y\rightarrow \frac{R}{3\Lambda_l(4,0)}=\frac{4}{l(l+3)}\;,\;\;
\end{eqnarray}
and multiplying the result by $9\beta_k\,k^4\left(\Lambda_l(4,0)/4k^2
\right)^{3/2}$. Note that for all $l\ge 2$ we have 
$\Lambda_l(4,0)\ge 5R/6$ so that $y\le 2/5$ is indeed satisfied. Moreover, 
$a>1$ is satisfied as long as $R^{(0)}(\Lambda_l(4,0)/k^2)>0$. If 
$R^{(0)}(\Lambda_l(4,0)/k^2)=0$, the cutoff in the scalar sector is zero 
anyway:
$\left({\cal R}_k\right)_{\bar{\sigma}\bar{\sigma}}
=\left({\cal R}_k\right)_{\bar{\phi}\bar{\sigma}}
=\left({\cal R}_k\right)_{\bar{\phi}\bar{\phi}}=0$. Hence, for positive values
of $\beta_k$, $f(a,y)\le f(a,0)$
leads to
\begin{eqnarray}
\label{glorious}
\left({\cal R}_k\right)_{\bar{\phi}
\bar{\sigma}}
\le\frac{9}{8}\beta_k\,k^4R^{(0)}
(\Lambda_l(4,0)/k^2)\left\{
2 \Lambda_l(4,0)/k^2+R^{(0)}
(\Lambda_l(4,0)/k^2)\right\}\;.
\end{eqnarray}

Next we insert $\left({\cal R}_k\right)_{\bar{\sigma}\bar{\sigma}}$ and
$\left({\cal R}_k\right)_{\bar{\phi}\bar{\phi}}$ of eq. (\ref{posdef}) into 
$\left({\cal R}_k\right)_{\bar{\phi}\bar{\phi}}\,
\left({\cal R}_k\right)_{\bar{\sigma}\bar{\sigma}}
-\left({\cal R}_k\right)_{\bar{\phi}\bar{\sigma}}^2$ and then use  
(\ref{glorious}). This yields 
\begin{eqnarray}
\label{hauptmincrit}
\left({\cal R}_k\right)_{\bar{\phi}\bar{\phi}}\,
\left({\cal R}_k\right)_{\bar{\sigma}\bar{\sigma}}
-\left({\cal R}_k\right)_{\bar{\phi}\bar{\sigma}}^2&\ge&
\left(\frac{k^4R^{(0)}(\Lambda_l(4,0)/k^2)}
{32\pi\,g_k}\right)^2\\
& &\times\left\{18\pi\,g_k\beta_k
\left(2\Lambda_l(4,0)/k^2+R^{(0)}
(\Lambda_l(4,0)/k^2)\right)
-\frac{3}{16}\right\}\;.\nonumber
\end{eqnarray}
Obviously the positivity condition $\left({\cal R}_k\right)_{\bar{\phi}
\bar{\phi}}\,\left({\cal R}_k\right)_{\bar{\sigma}\bar{\sigma}}
-\left({\cal R}_k\right)_{\bar{\phi}\bar{\sigma}}^2>0$ now boils down to
\begin{eqnarray}
\label{condpos}
v(k^2,l,R)+2\Lambda_l(4,0)/k^2
+R^{(0)}(\Lambda_l(4,0)/k^2)>\frac{1}{96\pi\,g_k\beta_k}\;.
\end{eqnarray}
Here $v$ is the nonnegative function of $k$, $l$ and $R$ which represents the 
contributions to $-\left({\cal R}_k\right)_{\bar{\phi}\bar{\sigma}}^2$ 
neglected on the RHS of (\ref{hauptmincrit}). We see that eq. (\ref{phisigma})
is a sufficient condition for the inequality (\ref{condpos}) to be valid, i.e.
for $\left({\cal R}_k\right)_{\bar{\phi}
\bar{\phi}}\,\left({\cal R}_k\right)_{\bar{\sigma}\bar{\sigma}}
-\left({\cal R}_k\right)_{\bar{\phi}\bar{\sigma}}^2$ to be positive. This is
what we wanted to proof.

\end{appendix}

\end{document}